\documentclass[twoside,twocolumn,9pt]{article}
\usepackage{extsizes}
\usepackage[super,sort&compress,comma]{natbib} 
\usepackage[version=3]{mhchem}
\usepackage[left=1.5cm, right=1.5cm, top=1.785cm, bottom=2.0cm]{geometry}
\usepackage{balance}
\usepackage{mathptmx}
\usepackage{sectsty}
\usepackage{graphicx} 
\usepackage{lastpage}
\usepackage[format=plain,justification=justified,singlelinecheck=false,font={stretch=1.125,small,sf},labelfont=bf,labelsep=space]{caption}
\usepackage{float}
\usepackage{fancyhdr}
\usepackage{fnpos}
\usepackage[english]{babel}
\addto{\captionsenglish}{%
  
}
\usepackage{array}
\usepackage{droidsans}
\usepackage{charter}
\usepackage[T1]{fontenc}
\usepackage[usenames,dvipsnames]{xcolor}
\usepackage{setspace}
\usepackage[compact]{titlesec}
\usepackage{hyperref}

\usepackage{jabbrv}

\usepackage{tikz}
\usetikzlibrary{arrows}
\usetikzlibrary{external}
\usepackage{layouts}

\definecolor{tol1}{HTML}{332288}
\definecolor{tol2}{HTML}{88CCEE}
\definecolor{tol3}{HTML}{44AA99}
\definecolor{tol4}{HTML}{117733}
\definecolor{tol7}{HTML}{CC6677}
\definecolor{col1}{rgb}{0,0.466667,0.733333}
\definecolor{col2}{rgb}{0.2,0.733333,0.933333}
\definecolor{col3}{rgb}{0,0.6,0.533333}
\definecolor{col4}{rgb}{0.933333,0.466667,0.2}
\definecolor{col5}{rgb}{0.8,0.2,0.0666667}
\definecolor{col6}{rgb}{0.933333,0.2,0.466667}
\definecolor{col7}{rgb}{0.733333,0.733333,0.733333}
\definecolor{col8}{rgb}{0.392157,0.392157,0.392157}

\newcommand{\etal}{{\it{et al.}}}

\usepackage{soul}

\definecolor{cream}{RGB}{222,217,201}

\begin{document}

\pagestyle{fancy}
\thispagestyle{plain}
\fancypagestyle{plain}{
\renewcommand{\headrulewidth}{0pt}
}

\makeFNbottom
\makeatletter
\renewcommand\LARGE{\@setfontsize\LARGE{15pt}{17}}
\renewcommand\Large{\@setfontsize\Large{12pt}{14}}
\renewcommand\large{\@setfontsize\large{10pt}{12}}
\renewcommand\footnotesize{\@setfontsize\footnotesize{7pt}{10}}
\makeatother

\renewcommand{\thefootnote}{\fnsymbol{footnote}}
\renewcommand\footnoterule{\vspace*{1pt}%
\color{cream}\hrule width 3.5in height 0.4pt \color{black}\vspace*{5pt}} 
\setcounter{secnumdepth}{5}

\makeatletter 
\renewcommand\@biblabel[1]{#1}            
\renewcommand\@makefntext[1]%
{\noindent\makebox[0pt][r]{\@thefnmark\,}#1}
\makeatother 
\renewcommand{\figurename}{\small{Fig.}~}
\sectionfont{\sffamily\Large}
\subsectionfont{\normalsize}
\subsubsectionfont{\bf}
\setstretch{1.125} 
\setlength{\skip\footins}{0.8cm}
\setlength{\footnotesep}{0.25cm}
\setlength{\jot}{10pt}
\titlespacing*{\section}{0pt}{4pt}{4pt}
\titlespacing*{\subsection}{0pt}{15pt}{1pt}

\fancyfoot{}
\fancyfoot[LO,RE]{\vspace{-7.1pt}\includegraphics[height=9pt]{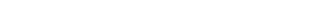}}
\fancyfoot[CO]{\vspace{-7.1pt}\hspace{11.9cm}\includegraphics{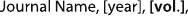}}
\fancyfoot[CE]{\vspace{-7.2pt}\hspace{-13.2cm}\includegraphics{head_foot/RF}}
\fancyfoot[RO]{\footnotesize{\sffamily{1--\pageref{LastPage} ~\textbar  \hspace{2pt}\thepage}}}
\fancyfoot[LE]{\footnotesize{\sffamily{\thepage~\textbar\hspace{4.65cm} 1--\pageref{LastPage}}}}
\fancyhead{}
\renewcommand{\headrulewidth}{0pt} 
\renewcommand{\footrulewidth}{0pt}
\setlength{\arrayrulewidth}{1pt}
\setlength{\columnsep}{6.5mm}
\setlength\bibsep{1pt}

\makeatletter 
\newlength{\figrulesep} 
\setlength{\figrulesep}{0.5\textfloatsep} 

\newcommand{\topfigrule}{\vspace*{-1pt}%
\noindent{\color{cream}\rule[-\figrulesep]{\columnwidth}{1.5pt}} }

\newcommand{\botfigrule}{\vspace*{-2pt}%
\noindent{\color{cream}\rule[\figrulesep]{\columnwidth}{1.5pt}} }

\newcommand{\dblfigrule}{\vspace*{-1pt}%
\noindent{\color{cream}\rule[-\figrulesep]{\textwidth}{1.5pt}} }

\makeatother

\twocolumn[
  \begin{@twocolumnfalse}
{\includegraphics[height=30pt]{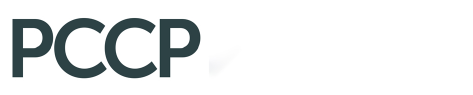}\hfill\raisebox{0pt}[0pt][0pt]{\includegraphics[height=55pt]{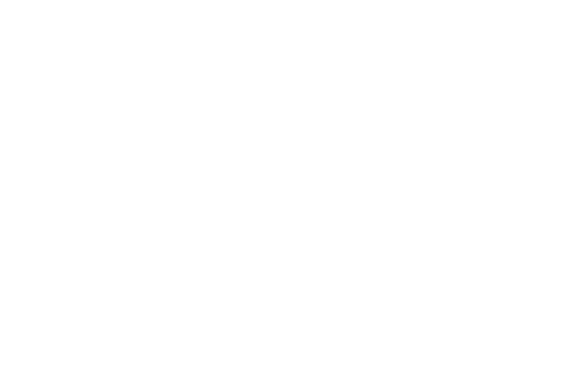}}\\[1ex]
\includegraphics[width=18.5cm]{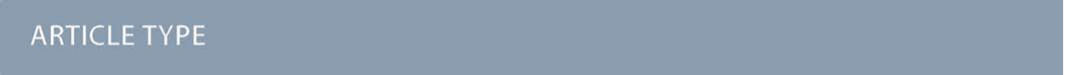}}\par
\vspace{1em}
\sffamily
\begin{tabular}{m{4.5cm} p{13.5cm} }

\includegraphics{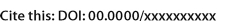} & \noindent\LARGE{\textbf{Electronic structure of norbornadiene and quadricyclane$^\dag$}} \\
\vspace{0.3cm} & \vspace{0.3cm} \\

 & \noindent\large{Joseph C. Cooper$^{\ast}$\textit{$^{a}$} and Adam Kirrander$^{\ddag}$\textit{$^{a}$}} \\

\includegraphics{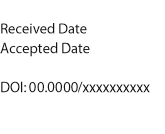} & \noindent\normalsize{The ground and excited state electronic structure of the molecular photoswitches quadricyclane and norbornadiene is examined qualitatively and quantitatively. A new custom basis set is introduced, optimised for efficient yet accurate calculations. A number of advanced multi-configurational and multi-reference electronic structure methods are evaluated, identifying those sufficiently accurate and efficient to be used in {\it{on-the-fly}} simulations of photoexcited dynamics. The key valence states participating in the isomerisation reaction are investigated, specifically mapping the important S$_1$/S$_0$ conical intersection that governs the non-radiative decay of the excited system. The powerful yet simple three-state valence model introduced here provides a suitable base for future computational exploration of the photodynamics of the substituted molecules suitable for \textit{e.g}.\ energy-storage applications.} 

\end{tabular}

 \end{@twocolumnfalse} \vspace{0.6cm}

  ]

\renewcommand*\rmdefault{bch}\normalfont\upshape
\rmfamily
\section*{}
\vspace{-1cm}


\footnotetext{\textit{$^{a}$~Physical and Theoretical Chemistry Laboratory, University of Oxford, South Parks Road, Oxford, OX1 3QZ, UK}}
\footnotetext{\textit{$^{\ast}$~E-mail: joseph.cooper@new.ox.ac.uk}}
\footnotetext{\textit{$^{\ddag}$~E-mail: adam.kirrander@chem.ox.ac.uk}}

\footnotetext{$^\dag$~Supplementary Information available: Images of orbitals, tabulated geometries and excitation energies, description of LIIC geometries, further detail about the conical intersections, basis sets and Rydberg states, and some further electronic structure calculations. See DOI: 10.1039/cXCP00000x/ }





\section{Introduction}

\begin{figure}[bh!]
    \centering
    \centering
    \begin{tikzpicture}[xscale=1, yscale=1]

    \begin{scope}
    \draw[very thick] (-0.5,0.5) -- (0.5, 0.35) -- (1.5,0.5);
    
    \filldraw[draw=white, fill=white] (-0.05, 0.) rectangle ++(0.1,1);
    \filldraw[draw=white, fill=white] (0.95, 0.) rectangle ++(0.1,1);
    
    \draw[very thick] (0,0) -- (1,0) -- (1.5,0.5) -- (1,1) -- (0,1) -- (-0.5, 0.5) -- (0,0)--(1,0);
    \draw[very thick] (0,0) -- (0,1);
    \draw[very thick] (1,0) -- (1,1);

    \node[anchor=north east] at (0,0)  {\scriptsize 1};
    \node[anchor=south east] at (0,1)  {\scriptsize 2};
    \node[anchor=south west] at (1,1)  {\scriptsize 3};
    \node[anchor=north west] at (1,0)  {\scriptsize 4};
     
    \node[anchor=east] at (-0.5,0.5)   {\scriptsize 5};
    \node[anchor=west] at (1.5, 0.5)   {\scriptsize 6};
    \node[anchor=south] at (0.5, 0.35) {\scriptsize 7};

    \end{scope}

    \draw[very thick, -left to] (2,0.55) -- (2.8,0.55);
    \draw[very thick, -left to] (2.8,0.45) -- (2,0.45);
    \node[anchor=south] at (2.4, 0.5) {UV};

   \draw[very thick,<->] (2.15,-0) -- (2.15,-0.5) -- (2.65,-0.5);
\node[anchor=east] at (2.15,0) {$x$};
\node[anchor=north] at (2.65,-0.5) {$y$}; 
    \node at (0.5, -1) {QC};
    \node at (4.3, -1) {NBD};

    \begin{scope}[xshift=3.8cm, yshift=-0.366925cm, yscale=1.73205]

    \draw[very thick] (0,0.92) -- (1,0.92);
    \draw[very thick] (0,0.08) -- (1,0.08);
    \filldraw[draw=white, fill=white] (-0.05, 0.) rectangle ++(0.1,1);
    \filldraw[draw=white, fill=white] (0.95, 0.) rectangle ++(0.1,1);
    \draw[very thick] (0,0) -- (1,0) -- (1.5,0.5) -- (1,1) -- (0,1) -- (-0.5, 0.5) -- (0,0)--(1,0);
    \draw[very thick] (-0.5,0.5) -- (0.5, 0.35) -- (1.5,0.5);
    
    \node[anchor=north east] at (0,0)  {\scriptsize 1};
    \node[anchor=south east] at (0,1)  {\scriptsize 2};
    \node[anchor=south west] at (1,1)  {\scriptsize 3};
    \node[anchor=north west] at (1,0)  {\scriptsize 4};
     
    \node[anchor=east] at (-0.5,0.5)   {\scriptsize 5};
    \node[anchor=west] at (1.5, 0.5)   {\scriptsize 6};
    \node[anchor=south] at (0.5, 0.35) {\scriptsize 7};
    \end{scope}
        
    \end{tikzpicture}
    \caption{Skeletal formulae of quadricyclane (QC, left) and norbornadiene (NBD, right), with the carbon atoms numbered. UV light absorption drives the reversible transformation between the two isomers. NBD has two doubly-bonded carbon atoms, \ce{C1=C4} and \ce{C2=C3}, referred to as `wings'. In QC, the wings come together, with electrons from the double bonds in NBD forming \ce{C1-C2} and \ce{C4-C3} single bonds, resulting in a strained four-membered ring. In both molecules, the \ce{C7} atom forms a bridge between the \ce{C5} and \ce{C6} atoms.}
    \label{fig:rxn}
\end{figure}
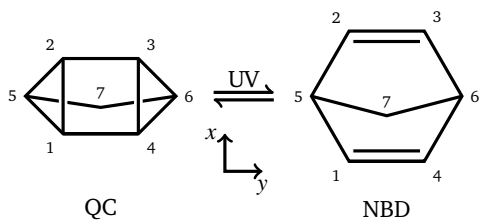

The isomers quadricyclane (QC) and norbornadiene (NBD) form a compact molecular photoswitch, capable of inter-converting upon photoabsorption via a [2+2] (retro)cycloaddition reaction,\cite{woodward_conservation_1969,bernardi_mechanism_1990} as depicted in Fig.\ \ref{fig:rxn}. The system has been the subject of intense theoretical\cite{jorner_unraveling_2017,qin_ab_2005,vessally_steric_2009,kuisma_comparative_2016,valentini_selective_2020,antol_photodeactivation_2013,coppola_norbornadienequadricyclane_2023,manso_establishing_2020,kilde_orthogonal_2020,borne_ultrafast_2024} and experimental\cite{fus_ultrafast_2002,rudakov_ultrafast_2012,palmer_high-level_2020,palmer_vacuum_2021,palmer_high-level_2023,borne_ultrafast_2024} scrutiny. It constitutes the central unit in molecular solar thermal (MOST) systems, capable of capture and storage of solar energy,\cite{kuisma_comparative_2016,orrego-hernandez_engineering_2020} and has been proposed for other applications that include information storage and optical devices.\cite{ergettetebikachew_turn-off_2018,kilde_orthogonal_2020,dreos_three-input_2018,kunz_112_2021} 

In applications, substituents are used to modify the basic QC/NBD system to maximise the absorption of incident solar light and to increase the quantum yield of interconversion between the isomers. This is often achieved by breaking the $C_{2v}$ symmetry, allowing stronger absorption into the reactive states, and using chromophoric groups conjugating into the NBD $\pi$-system.\cite{orrego-hernandez_engineering_2020} In the gas phase, especially at higher energies ($\geq 6$ eV), the Rydberg state manifold comes into play, complicating the isomerisation by introducing slower and less efficient decay channels.\cite{fus_ultrafast_2002, borne_ultrafast_2024} However, substitutions often lower the relative energy of the reactive valence states, separating the two manifolds. Moreover, in applications such as MOSTs, the issue is removed altogether since the Rydberg states are quenched in the condensed phase.\cite{robin_pressure_1970} The focus of the current work is thus on the valence electronic states that play a critical role in the decay and isomerisation dynamics.

The photochemistry of the QC/NBD system is of fundamental interest. As mentioned, the system provides an important example of a [2+2] (retro)cycloaddition.\cite{woodward_conservation_1969,bernardi_mechanism_1990} Furthermore, it has a single characteristic conical intersection (CI), which governs the decay from the first excited to the ground electronic state. This same CI is accessed whether QC or NBD is excited, providing an opportunity to study how the approach of wavepackets to CIs affects their transmission. Understanding the photoexcited dynamics of the QC/NBD system and interpreting a growing number of time-resolved experiments requires simulations of the excited dynamics and the decay process. A necessary prerequisite for these is accurate, computationally feasible, and globally valid electronic structure models, which is one of the goals of the current study.

The electronic structure of the excited states in the QC/NBD system is challenging. Rydberg states notwithstanding, the system exhibits strong multi-configurational character with dramatic changes as the molecule distorts between the QC and the NBD isomers during the dynamics. Consequently, different electronic structure methods can predict quite different results, and many suffer significant stability issues. Benchmarking the predictions against experimental data is non-trivial, given that standard spectroscopy only reveals information about the bright states of the two isomers, and then only in the Franck-Condon regions. Therefore, it is essential to assess the electronic structure methods via secondary properties such as quantum-yields and, as here, by carefully comparing different electronic structure methods.

In previous theoretical work on QC/NBD, one focal area has been the assignment of vibronic transitions in the highly excited Rydberg and ionic manifolds above the ground-state geometries.\cite{zgierski_franckcondon_1993,roos_theoretical_1994,palmer_high-level_2020,palmer_vacuum_2021,palmer_high-level_2023, cooper_valence_2024} One of the first studies to consider the role of the excited electronic states in the dynamics was carried out by Antol, who used an augmented CASSCF(4,4)+3s approach to predict how an excited 3s state may decay through a doubly and then a singly excited state to a conical intersection with the ground electronic state.\cite{antol_photodeactivation_2013} Coppola \etal\ used CASSCF(4,7) and CASPT2 to highlight the role of the doubly-excited state in non-adiabatic transfer,\cite{coppola_norbornadienequadricyclane_2023} in close accord with Antol. At about the same time, Hernandez \etal\ performed surface-hopping dynamics simulations (including on a substituted derivative),\cite{hernandez_multiconfigurational_2023} returning qualitative similar results to Antol and Coppola \etal. Finally, Valentini \etal\ used CASSCF(4,8)-level theory to model coherent control experiments in QC/NBD.\cite{valentini_selective_2020} Previous work by the current authors simulated the dynamics in photoexcited QC using RMS-CASPT2(2,6) electronic structure and compared the results to time-resolved photoelectron spectroscopy experiments.\cite{borne_ultrafast_2024} All the previous treatments listed above include the Rydberg states, differing mostly in how and which additional Rydberg orbitals are included, and most employ active spaces closely related to the (4,4)-active space. There are also previous studies that consider more systematically the role of substitutions on the QC/NBD system, and how they affect the (excited state) potentials.\cite{jorner_unraveling_2017,kuisma_comparative_2016}

Our main goal is to undertake a series of theoretical explorations of the QC/NBD system and to provide detailed benchmarks, first concentrating on the electronic structure of the unsubstituted molecule. The insights provided by systematically evaluating different electronic structure methods and basis sets allow us to make observations regarding static \textit{vs} dynamic correlation, the role of the doubly excited character in the wavefunction, assess the validity of the results away from the Franck-Condon region, and to discuss the different electronic structure methods considered. In doing so, we report extensive benchmarks using multi-configurational active space methods (CASSCF, CASPT2, and MRCI), selected configuration interaction (SHCI), and coupled cluster methods (LR-CC3 and LR-CCSD). These systematic comparisons allow us to identify electronic structure models suitable for dynamics simulations. Several recent publications have highlighted the importance of these models on non-adiabatic simulations,\cite{szymczak_influence_2011,janos_what_2023,papineau_which_2024,bellshaw_correspondence_2019} and the electronic structure models presented here can be exploited to investigate how subtle changes in the potential energy surfaces affect the photochemical dynamics.

\section{Qualitative photochemistry} \label{sec:active-space}

\begin{figure}
    \centering
    \begin{tikzpicture}[scale=1]
\begin{scope}[xshift = 0.5cm,yshift=0.5cm]

\draw[very thick,<->] (-1.5,-0.5) -- (-1.5,-1) -- (-1.,-1);
\node[anchor=east] at (-1.5,-0.5) {$x$};
\node[anchor=north] at (-1.,-1) {$y$};
\end{scope}
\begin{scope}[xshift=0cm, yshift=0cm]
    \draw[thick, black] (0,0) -- (1,0);
    \draw[thick, black] (0,0) -- (1,0);
    \filldraw[thick,fill=black] (0,0) .. controls (-0.5,0.8) and (0.5,0.8) .. (0,0);
    \filldraw[thick,fill=black] (1,0) .. controls (0.5,0.8) and (1.5,0.8) .. (1,0);
    \filldraw[thick,fill=white] (0,0) .. controls (-0.5,-0.8) and (0.5,-0.8) .. (0,0);
    \filldraw[thick,fill=white] (1,0) .. controls (0.5,-0.8) and (1.5,-0.8) .. (1,0);
\node at (0.5,-1) {$A_1(\pi\sigma)$};
\node[anchor=east] at (0.,0) {\scriptsize 1};
\node[anchor=west] at (1.,0) {\scriptsize 4};
\end{scope}

\begin{scope}[xshift=0cm, yshift=1.5cm]
    \draw[thick, black] (0,0) -- (1,0);
    \draw[thick, black] (0,0) -- (1,0);
    \filldraw[thick,fill=white] (0,0) .. controls (-0.5,0.8) and (0.5,0.8) .. (0,0);
    \filldraw[thick,fill=white] (1,0) .. controls (0.5,0.8) and (1.5,0.8) .. (1,0);
    \filldraw[thick,fill=black] (0,0) .. controls (-0.5,-0.8) and (0.5,-0.8) .. (0,0);
    \filldraw[thick,fill=black] (1,0) .. controls (0.5,-0.8) and (1.5,-0.8) .. (1,0);
\node[anchor=east] at (0.,0) {\scriptsize 2};
\node[anchor=west] at (1.,0) {\scriptsize 3};
\end{scope}

\begin{scope}[xshift=2.0cm, yshift=0cm]
    \draw[dashed,black] (-0.2,0.75) -- (1.2,0.75);
    \draw[thick, black] (0,0) -- (1,0);
    \draw[thick, black] (0,0) -- (1,0);
    \filldraw[thick,fill=black] (0,0) .. controls (-0.5,0.8) and (0.5,0.8) .. (0,0);
    \filldraw[thick,fill=black] (1,0) .. controls (0.5,0.8) and (1.5,0.8) .. (1,0);
    \filldraw[thick,fill=white] (0,0) .. controls (-0.5,-0.8) and (0.5,-0.8) .. (0,0);
    \filldraw[thick,fill=white] (1,0) .. controls (0.5,-0.8) and (1.5,-0.8) .. (1,0);
\node at (0.5,-1) {$B_1(\pi\sigma^*)$};
\node[anchor=east] at (0.,0) {\scriptsize 1};
\node[anchor=west] at (1.,0) {\scriptsize 4};
\end{scope}

\begin{scope}[xshift=2.0cm, yshift=1.5cm]
    \draw[thick, black] (0,0) -- (1,0);
    \draw[thick, black] (0,0) -- (1,0);
    \filldraw[thick,fill=black] (0,0) .. controls (-0.5,0.8) and (0.5,0.8) .. (0,0);
    \filldraw[thick,fill=black] (1,0) .. controls (0.5,0.8) and (1.5,0.8) .. (1,0);
    \filldraw[thick,fill=white] (0,0) .. controls (-0.5,-0.8) and (0.5,-0.8) .. (0,0);
    \filldraw[thick,fill=white] (1,0) .. controls (0.5,-0.8) and (1.5,-0.8) .. (1,0);
\node[anchor=east] at (0.,0) {\scriptsize 2};
\node[anchor=west] at (1.,0) {\scriptsize 3};
\end{scope}

\begin{scope}[xshift=4.0cm, yshift=0cm]
    \draw[dashed,black] (0.5,-0.7) -- (0.5,2.2);
    \draw[thick, black] (0,0) -- (1,0);
    \draw[thick, black] (0,0) -- (1,0);
    \filldraw[thick,fill=white] (0,0) .. controls (-0.5,0.8) and (0.5,0.8) .. (0,0);
    \filldraw[thick,fill=black] (1,0) .. controls (0.5,0.8) and (1.5,0.8) .. (1,0);
    \filldraw[thick,fill=black] (0,0) .. controls (-0.5,-0.8) and (0.5,-0.8) .. (0,0);
    \filldraw[thick,fill=white] (1,0) .. controls (0.5,-0.8) and (1.5,-0.8) .. (1,0);
\node at (0.5,-1) {$B_2(\pi^*\sigma)$};
\node[anchor=east] at (0.,0) {\scriptsize 1};
\node[anchor=west] at (1.,0) {\scriptsize 4};
\end{scope}

\begin{scope}[xshift=4.0cm, yshift=1.5cm]
    \draw[thick, black] (0,0) -- (1,0);
    \draw[thick, black] (0,0) -- (1,0);
    \filldraw[thick,fill=black] (0,0) .. controls (-0.5,0.8) and (0.5,0.8) .. (0,0);
    \filldraw[thick,fill=white] (1,0) .. controls (0.5,0.8) and (1.5,0.8) .. (1,0);
    \filldraw[thick,fill=white] (0,0) .. controls (-0.5,-0.8) and (0.5,-0.8) .. (0,0);
    \filldraw[thick,fill=black] (1,0) .. controls (0.5,-0.8) and (1.5,-0.8) .. (1,0);
\node[anchor=east] at (0.,0) {\scriptsize 2};
\node[anchor=west] at (1.,0) {\scriptsize 3};
\end{scope}

\begin{scope}[xshift=6.0cm, yshift=0cm]
    \draw[dashed,black] (0.5,-0.7) -- (0.5,2.2);
    \draw[dashed,black] (-0.2,0.75) -- (1.2,0.75);
    \draw[thick, black] (0,0) -- (1,0);
    \draw[thick, black] (0,0) -- (1,0);
    \filldraw[thick,fill=white] (0,0) .. controls (-0.5,0.8) and (0.5,0.8) .. (0,0);
    \filldraw[thick,fill=black] (1,0) .. controls (0.5,0.8) and (1.5,0.8) .. (1,0);
    \filldraw[thick,fill=black] (0,0) .. controls (-0.5,-0.8) and (0.5,-0.8) .. (0,0);
    \filldraw[thick,fill=white] (1,0) .. controls (0.5,-0.8) and (1.5,-0.8) .. (1,0);
\node at (0.5,-1) {$A_2(\pi^*\sigma^*)$};
\node[anchor=east] at (0.,0) {\scriptsize 1};
\node[anchor=west] at (1.,0) {\scriptsize 4};
\end{scope}

\begin{scope}[xshift=6.0cm, yshift=1.5cm]
    \draw[thick, black] (0,0) -- (1,0);
    \draw[thick, black] (0,0) -- (1,0);
    \filldraw[thick,fill=white] (0,0) .. controls (-0.5,0.8) and (0.5,0.8) .. (0,0);
    \filldraw[thick,fill=black] (1,0) .. controls (0.5,0.8) and (1.5,0.8) .. (1,0);
    \filldraw[thick,fill=black] (0,0) .. controls (-0.5,-0.8) and (0.5,-0.8) .. (0,0);
    \filldraw[thick,fill=white] (1,0) .. controls (0.5,-0.8) and (1.5,-0.8) .. (1,0);
\node[anchor=east] at (0.,0) {\scriptsize 2};
\node[anchor=west] at (1.,0) {\scriptsize 3};
\end{scope}

\draw[tol1,very thick, dashed] (1.5,-1.4) -- (1.5,+2.4) -- (5.55,+2.4) -- (5.5,-1.4) -- cycle;

    \end{tikzpicture}
     \caption{Schematic of the four principal orbitals, with only carbons 1-4 and their p$_x$ orbitals shown (cf fig. \ref{fig:rxn}).  In NBD, the $B_1$ and $B_2$ orbitals are the HOMO and LUMO, respectively, and vice versa in QC. Inter-nuclear nodes are drawn as dashed lines. All four orbitals constitute the (4,4) active space, while the dashed box denotes the (2,2) active space. Orbitals from electronic structure calculations are shown in Section 1 of the supplementary information$^\dag$.}
    
    \label{fig:AS-schematic}
\end{figure}
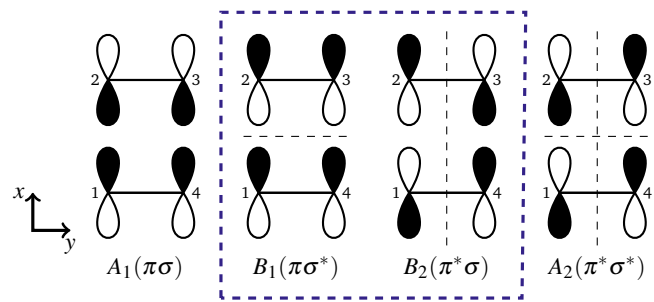

In this section, we discuss the general trends in the electronic structure of QC/NBD. These qualitative aspects manifest in all the electronic structure methods investigated here and provide general insights into the overall photochemical behaviour of this system. Notably, they constitute an essential background for analysing the efficacy of the various approaches to the electronic structure.

Consider first the two isomers shown schematically in Fig.\ \ref{fig:rxn}. NBD (right) has two double-bonded `wings' formed by atoms \ce{C1=C4} and \ce{C2=C3}, respectively. In QC (left), the wings move closer together as the molecule undergoes [2+2] retro-cycloaddition, whereby the double bonds break and cyclise to form the eponymous four-membered ring of QC. Given the central role played by the four carbons \ce{C1}-\ce{C4}, it is reasonable to start with a simple H\"uckel-like picture. The four symmetrised combinations of \ce{C}(p$_x$)-orbitals are shown in Fig.\ \ref{fig:AS-schematic}, with the $x$-axis aligned with the \ce{C1-C2} and \ce{C3-C4} bonds in QC and the $y$-axis with the \ce{C1=C4} and \ce{C2=C3} bonds in NBD. The former bonds have predominantly $\sigma$-character, while the latter are $\pi$-character. 

Each orbital has a symmetry label, which we refer to throughout the text, sometimes combined with a pithy label that indicates the $\pi$-bonding character for the wings (along $y$) and the $\sigma$-bonding character between the wings (along $x$). For example, in the $B_1(\pi\sigma^*)$ orbital the label in parenthesis signifies that the orbital has bonding character along the \ce{C1-C4} and \ce{C2-C3} $\pi$-bonds, \textit{but} anti-bonding character for the \ce{C1-C2} and \ce{C3-C4} $\sigma$-bonds. As each carbon contributes one electron to these orbitals, we have four electrons in these orbitals. The corresponding configuration state functions (CSFs) in this subsystem are labelled as $|\eta_{A_1}\eta_{B_1}\eta_{B_2}\eta_{A_2}\rangle$, where $\eta$ denotes the occupation for each specified orbital, either 0, 1 (u for up and d for down spin), or 2.

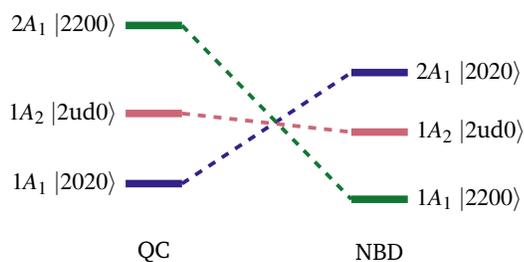
\begin{figure}
    \centering
    \centering
\begin{tikzpicture}[yscale=0.15,xscale=0.75]

    \def \offset {4.};
    \def \lw {0.1cm};

    \draw[tol1, line width = \lw] (0,0) -- (1,0);
    \draw[tol7, line width = \lw] (0,6.222) -- (1,6.222);
    \draw[tol4, line width = \lw] (0,14.0243) -- (1,14.0243);
    \draw[tol4, line width = \lw] (0+\offset,-1.38828)       -- (1+\offset,-1.38828);
    \draw[tol7, line width = \lw] (0+\offset,4.57042)   -- (1+\offset,4.57042);
    \draw[tol1, line width = \lw] (0+\offset,9.8496) -- (1+\offset,9.8496);

    \node[anchor=east] at (0,0) {$1A_1\;|2020\rangle$};
    \node[anchor=east] at (0,6.222) {$1A_2\;|2\text{ud}0\rangle$};
    \node[anchor=east] at (0,14.0243) {$2A_1\;|2200\rangle$};
    
    \node[anchor=west] at (1+\offset,-1.38828) {$1A_1\;|2200\rangle$};
    \node[anchor=west] at (1+\offset,4.57042) {$1A_2\;|2\text{ud}0\rangle$};
    \node[anchor=west] at (1+\offset,9.8496) {$2A_1\;|2020\rangle$};

    \draw[tol1, dashed,line width = 0.05cm] (1,0) -- (0+\offset,9.8496);
    \draw[tol7, dashed,line width = 0.05cm] (1,6.222) -- (0+\offset,4.57042);
    \draw[tol4, dashed,line width = 0.05cm] (1,14.0243) -- (0+\offset,-1.38828);

    \node at (0.5, -6.0) {QC};
    \node at (\offset+0.5, -6.0) {NBD};


\end{tikzpicture}
    \caption{State correlation diagram for the QC/NBD system, with energy increasing up the diagram. Figure drawn to scale for SA(3)-CASSCF(2,2)/p-cc-(p)VDZ energies.}
    \label{fig:state_correlation}
\end{figure}

We consider three states: the ground S$_0$ ($1A_1$) state, the first excited S$_1$ ($1A_2$) valence state, and the second, doubly excited S$_2$ ($2A_1$) valence state. The S$_1$ ($1A_2$)  state is the simplest, with $|2\text{ud}0\rangle$ the leading configuration at both QC and NBD geometries. The two $A_1$ states, S$_0$ and S$_2$, are more complicated. In NBD, S$_0$ has $|2200\rangle$ character while S$_2$ has $|2020\rangle$. In QC, this is reversed, with the ground state having $|2020\rangle$ character and S$_2$ $|2200\rangle$. The correlation between states is shown pictorially in Fig.\ \ref{fig:state_correlation}. Thus, the origin of the MOST and photoswitch nature of this system is apparent: the ground state isomerisation involves a change in character of the wavefunction, resulting in a large barrier, while the excited states couple to the other isomer's ground state, leading to efficient excited-state isomerisation. The reaction of the molecules can therefore be rationalised using the orbital characters in Fig.\ \ref{fig:AS-schematic}, and it is clear exciting either the S$_1$ or S$_2$ state from the NBD minimum will cause motion towards the QC minimum, or vice versa.

For non-radiative decay, the S$_1$/S$_0$ conical intersection (CI) plays a crucial role. The minimum energy conical intersection (MECI) geometry is included in Fig.\ \ref{fig:geoms}. The S$_1$ state has $\mathrm{A_2}$ symmetry and thus the MECI distorts to C$_2$ symmetry, with a distinctive rhombic arrangement of atom \ce{C1}-\ce{C4}. This distortion is akin to other [2+2] systems, such as the well-studied ethylene dimerisation.\cite{bernardi_mechanism_1990,celani_geometry_1995,serrano-perez_extended_2012} The $C_2$ symmetry indicates that left- and right-handed CI variants exist --- we shall not distinguish them. At the S$_1$/S$_0$ conical intersection, the wavefunction is exceptionally multi-configurational, with all three of the $|2020\rangle$, $|2\text{ud}0\rangle$ and $|2200\rangle$ configurations strongly occupied in the S$_0$ and S$_1$ states.

During dynamics, the molecule will break the $C_{2v}$ symmetry, and so we shall drop the state symmetry labels and use only the adiabatic S$_n$ labelling scheme, punctuated with a leading character label to keep track of the state character. Finally, we stress that the model introduced here does not include the Rydberg or other valence states that appear between the singly- and doubly-excited valence states in the gas phase. The current adiabatic labels S$_n$ are thus only valid in the context of this model. A complete analysis of the spectra of these molecules in the gas phase is given in our previous work.\cite{cooper_valence_2024}, and more details on why this approximation is valid are given in Section 6 of the supplementary information (SI)$^{\dag}$.

\section{Computational methods}
\subsection{Electronic structure calculations}
Electronic structure calculations are performed using OpenMolcas v23.02 \cite{aquilante_modern_2020} (CASSCF, XMS-CASPT2),  COLUMBUS 7.6\cite{lischka_generality_2020} (CASSCF, MRCI) and e$^T$ 1.9\cite{folkestad_et_2020} (LR-CC). For CASSCF calculations, the results from OpenMolcas and COLUMBUS are effectively identical. The XMS-CASPT2 calculations are performed with a level-shift of 0.2$i$, the minimum value required to remove all intruder states. The potential energies are relatively sensitive to this shift (see Fig.\ S20, SI$^\dag$). The MRCI calculations are performed using COLUMBUS with the uncontracted formalism, and the XMS-CASPT2 use the single-state single-reference formulation in OpenMolcas. Analysis shows the different formalisms in XMS-CASPT2 and MRCI affect the (4,4) active space but not (2,2) active space (see Fig.\ S15, SI$^\dag$). Cholesky decomposition\cite{folkestad_et_2020,aquilante_modern_2020} (CD) is used when possible, and the inclusion of this does not significantly change the results. All correlated methods (MRCI, CASPT2, LR-CC) use frozen carbon 1s orbitals. Compute times are reported for a single-threaded OpenMolcas calculation running on an Intel\textsuperscript{\textregistered} Xeon\textsuperscript{\textregistered} Silver 4314 CPU with a clock-speed of 2.40GHz.

The electronic structure calculations are not overly sensitive to the basis, provided the basis set is sufficiently diffuse. In view of future non-adiabatic dynamics simulations, which require a large number of electronic structure evaluations, we have developed a custom adapted basis that is both efficient and accurate for this specific system. The basis is an altered version of cc-pVDZ.\cite{dunning_gaussian_1989}, denoted p-cc-(p)VDZ. We remove the polarisation functions from the hydrogens, giving a [4s$|$2s] contraction equivalent in size to def2-SV(P) or 6-31G$^*$, and add only the additional p diffuse functions from the aug-cc-pVDZ basis\cite{kendall_electron_1992} to the carbons, resulting in a [9s5p1d$|$3s3p1d] contraction. Discussion and evaluation of this basis, including the full contraction, is given in Section \ref{sec:basis}, with further benchmarking provided in Section 10 of the SI$^\dag$. Finally, a summary of the individual methods used in this work is presented in Table S6 in Section 12 of the SI$^\dag$.


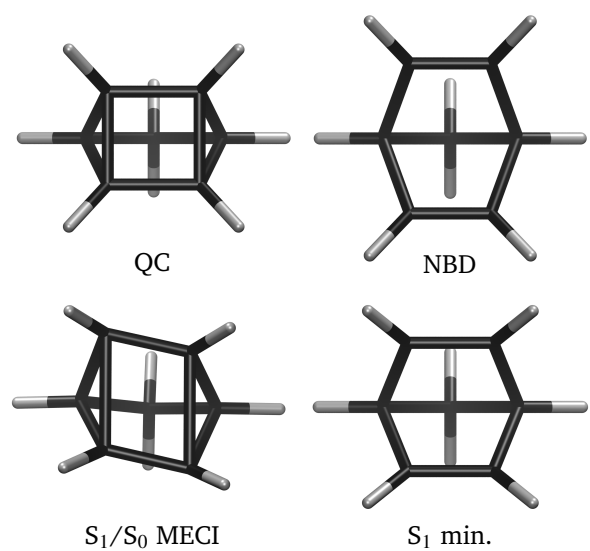
\begin{figure}
    \centering
    \begin{tikzpicture}[x=0.9\linewidth,y=0.9\linewidth]
    \node at (0,0) {\includegraphics[width=0.9\linewidth]{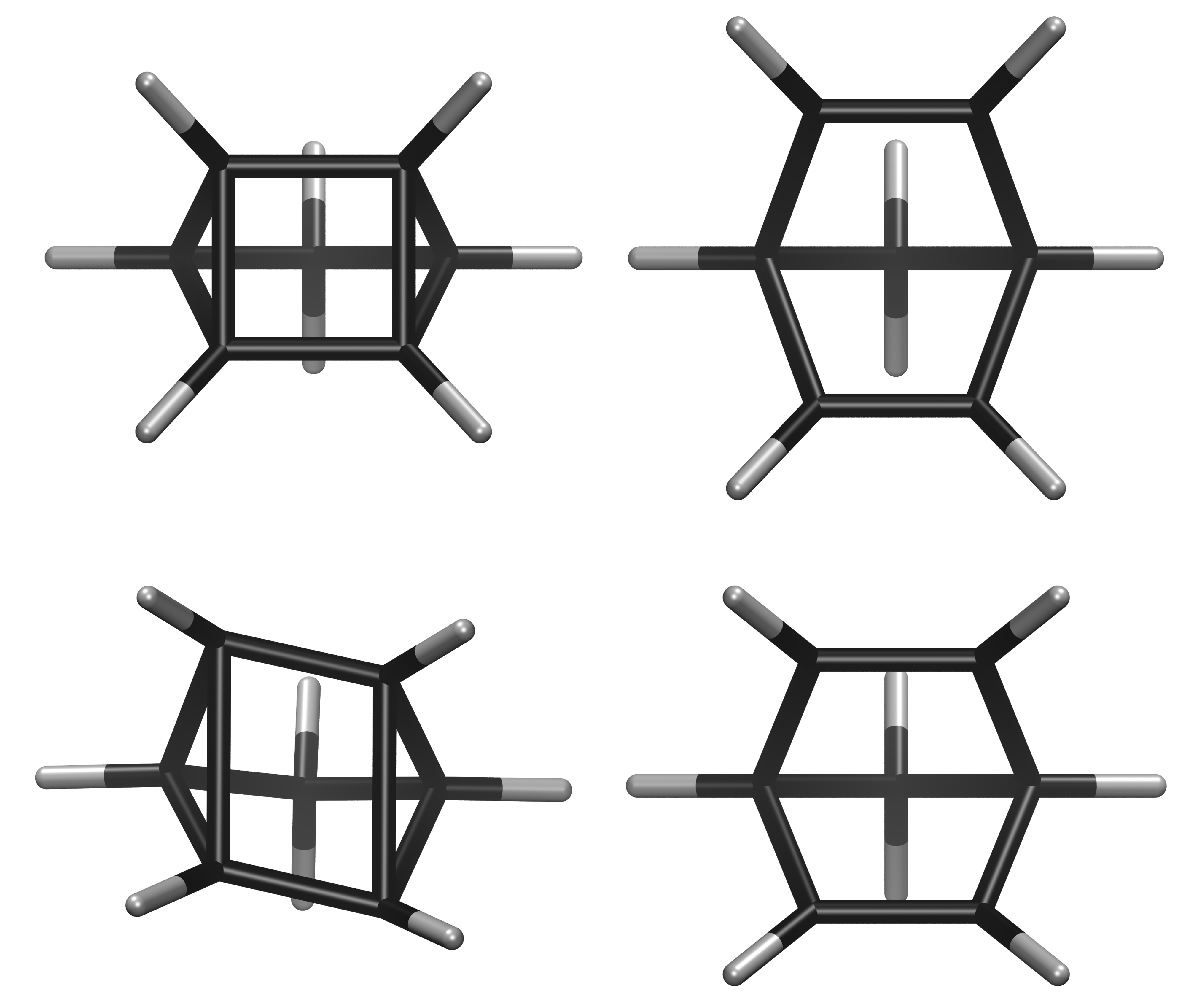}};

    \node at (-0.24,0.) {\large\centering QC};
    \node at (+0.25,0.) {\large\centering NBD};
    \node at (-0.24,-0.45) {\large\centering S$_1$/S$_0$ MECI};
    \node at (0.25,-0.45) {\large\centering S$_1$ min.};
    \end{tikzpicture}
    \caption{Representative optimised molecular geometries. The QC ground state (top left) has the characteristic four-membered ring, whereas the NBD (top right) has two separate double-bonded `wings'. The S$_1$/S$_0$ MECI (bottom left) has a distinct rhombic arrangement of the four-carbon moiety, while the S$_1$ minimum (bottom right), which appears in some of the calculations, is quite similar to the NBD ground state, with slightly closer wings. The optimised geometries shown are obtained using SA(3)-CASSCF(2,2)/p-cc-(p)VDZ.}
    \label{fig:geoms}
\end{figure}

\subsection{Interpolations of internal coordinates}\label{sec:liics_nbd}
To compare the different electronic structure methods, we calculate the energies along linear interpolations in internal coordinates (LIICs), which provide effective one-dimensional `reaction coordinates'. We use two different sets of LIICs. The first, which is similar to that used in Borne \etal.\cite{borne_ultrafast_2024}, connects the QC ground state equilibrium geometry to the S$_1$/S$_0$ MECI, and then proceeds to the NBD ground state equilibrium. We note that the combined LIIC involves a change in direction at the S$_1$/S$_0$ MECI. The second LIIC proceeds from the NBD ground state to the S$_1$ minimum, only present in some of the calculations, and then to the S$_1$/S$_0$ MECI. This approximates the path for dynamics that proceeds via the S$_1$ minimum. This second LIIC does not include the final path from the MECI towards QC, since this is already included in the first LIIC. The first LIIC (QC$\rightarrow$S$_1$/S$_0$$\rightarrow$NBD), used in Figs.\ \ref{fig:CASSCF}, \ref{fig:pt2cc3} and \ref{fig:basis}, is defined based on molecular geometries optimised using CASSCF(2,2)/p-cc-(p)VDZ for CASSCF calculations, while the LIIC is based on MRCI(2,2)/p-cc-(p)VDZ optimised geometries when comparing all other electronic structure methods. The second LIIC (S$_1$/S$_0$$\rightarrow$S$_1$-min$\rightarrow$NBD), used in Fig.\ \ref{fig:s1_liic}, exists in one version only, obtained using CASSCF(2,2)/p-cc-(p)VDZ geometries. For reference, the CASSCF(2,2)/p-cc-(p)VDZ optimised geometries are shown in Fig.\ \ref{fig:geoms}. As an aside, we note that all internal coordinates in the molecule change across each LIIC.

Finally, in Figs.\ \ref{fig:PES} and \ref{fig:PES_S1}, we show two-dimensional potential energy surfaces. These are calculated by using linear interpolations in Cartesian coordinates. Figure \ref{fig:PES} shows potential energy in the plane that contains the NBD and QC minima and the S$_1$/S$_0$ MECI, and Fig.\ \ref{fig:PES_S1} in the plane containing the NBD and S$_1$ minima and the S$_1$/S$_0$ MECI.

\section{Results}
\subsection{Static correlation}

\begin{figure}[h!]
    \centering
    \input{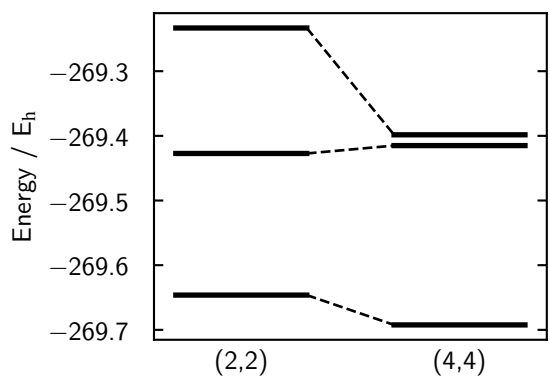}
    \caption{Absolute energies for the three states S$_0$, S$_1$, and S$_2$ at NBD equilibrium geometry for the (2,2) (left) and (4,4) (right) active spaces, calculated at the SA(3)-CASSCF/p-cc-(p)VDZ level. Note that although the state-averaged energy is lower for the (4,4) calculation, the S$_1$ state energy is  \textit{higher} for the (4,4) active space compared to the (2,2).}
    \label{fig:diff_corr}
\end{figure}

The multi-configurational and doubly-excited character of the NBD/QC system means that single-reference methods, such as ADC(2) and TDDFT, often give poor results, especially around the critical S$_1$/S$_0$ CI. The exception to this are high-order coupled cluster calculations, which are discussed later, but these are too expensive for dynamics in a system of this size. As such, we focus mainly on multi-configurational methods.

From the qualitative discussion in Section \ref{sec:active-space}, it is clear that a global representation of the electronic states requires that the active space includes the $|2200\rangle$, $|2020\rangle$, and $|2\text{ud}0\rangle$ configurations, which make leading contributions to the three states. A standard approach would be to utilise the complete set of the four orbitals and electrons as detailed in Fig.\ \ref{fig:AS-schematic}, \textit{i.e.}\ a CASSCF(4,4) approach. Almost all previous work on the excited states of these molecules used methods based on this active space, generally including Rydberg states.\cite{antol_photodeactivation_2013,valentini_selective_2020,hernandez_multiconfigurational_2023} We show how the addition of Rydberg states affects these potentials in Section 6 of the SI$^\dag$.

Alternatively, one could remove the $A_1$ and $A_2$ orbitals; these orbitals are fully occupied or unoccupied in the important configurations. This leads to a CASSCF(2,2) approach, which has particular computational advantage as state-averaging over three states leads to both orbitals always having a state-averaged occupation number of 1, leading to stable convergence.

This additional stability is evident in practice. The (4,4) active space has trouble converging to the same active space in QC-like geometries due to the formation of the $\sigma$-bond; the $A_1$ orbital, with $\sigma$-bonding character, drops significantly in energy, while the $A_2$ orbital, with $\sigma^*$-anti-bonding character, rises significantly, leading to those orbitals being replaced in the optimisation. A third choice is a (4,3) active space, which gives similar results to the (2,2) active space but exhibits similar instabilities as the (4,4) active space. 

We find that the (2,2) active space provides a better qualitative description of the potential energy surfaces, with the (4,4) active space biased against the S$_1$ $|2\text{ud}0\rangle$ state. This is seen in Fig.\ \ref{fig:diff_corr}, which shows the absolute energies for the (2,2) and (4,4) active spaces at the NBD geometry. The critical S$_1$ $|2\text{ud}0\rangle$ state has a \textit{higher} absolute energy in the larger (4,4) active space than the (2,2), While this may seem counter-intuitive, as the (4,4) state has more parameters and hence should give a lower variational energy, this only applies to the optimised \textit{state-averaged} energy rather than the energy of individual states.

The poor description of S$_1$ in CASSCF(4,4) can be rationalised using orbital occupations. The two $A_1$ states contain four orbitals with occupations significantly different from 2 or 0, indicating that all four orbitals contribute to the correlation. The $A_2$ state, on the other hand, only has the central two orbitals $B_1(\pi\sigma^*)$ and $B_2(\pi^*\sigma)$ with occupations not close to 2 or 0, and requires its correlation to come from excitations to the virtual space. CASSCF(4,4), therefore, describes the two $A_1$ states better than the $A_2$ state, leading to qualitatively incorrect energy gaps, while CASSCF(2,2) gives a more balanced description of the individual states.
This pattern is seen in other molecules with two $\pi$-bonds, such as cyclopentadiene\cite{bertram_mapping_2023} and 1,3-cyclohexadiene\cite{polyak_ultrafast_2019}, where the doubly-excited state is \textit{lower} in energy than the singly-excited state in CASSCF. Only by including more correlation, in those cases \textit{via} XMS-CASPT2, does one retrieve the correct state ordering.

\begin{figure}
    \centering
    \input{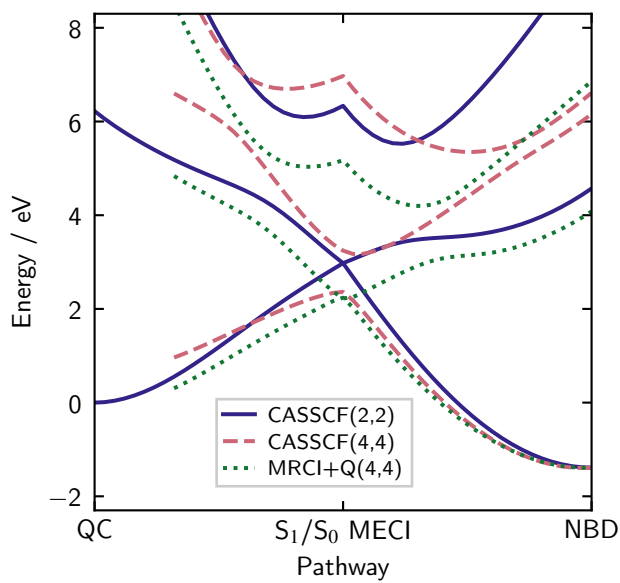}
    \caption{Energies for S$_0$, S$_1$, and S$_2$ calculated using SA(3)-CASSCF(2,2) (solid indigo lines),  SA(3)-CASSCF(4,4) (dashed rose lines), and MRCI+Q(4,4) (dotted green lines) with the p-cc-(p)VDZ basis set. The (4,4) active space fails to converge around QC and is thus not shown in that region. CASSCF(2,2) and MRCI+Q(4,4) agree on the shape of the potential but not on the excitation energy at the NBD ground state equilibrium. Details of the LIIC pathway are given in Section \ref{sec:liics_nbd}}
    \label{fig:CASSCF}
\end{figure}

Here, we note that the CASSCF(2,2) does not describe the S$_2$ state well in either NBD or QC. Specifically in the region around the NBD ground state equilibrium, the CASSCF(2,2) method does not include the $|\mathrm{uudd}\rangle$, $|2002\rangle$, and $|0220\rangle$ configurations, which all contribute $\approx10\%$ to the doubly excited state. This leads to a significant increase in energy of the S$_2$ state for CASSCF(2,2), as seen in Table \ref{tab:energies}. Fortunately, both the S$_1$ and S$_0$ states are well described, even in regions with significant doubly excited character. Additionally, we do not expect S$_2$ to be populated to any notable degree during photoexcited dynamics (see the discussion below and in Section 5 of the SI$^{\dag}$), so the poor description should not affect simulations.

The analysis so far only concerns the NBD geometry. A better overview is gained by making the comparison using potential energy cuts (PECs) along the LIICs introduced earlier (see Methods for details). All calculations use the optimised p-cc-(p)VDZ basis, with the results shown in Fig.\ \ref{fig:CASSCF}. 

The indigo solid lines are the relative energies for the three states in the SA(3)-CASSCF(2,2) calculation.
We start on the right of the diagram, corresponding to NBD. The ground state, which has primary $|2200\rangle$ character, begins to rise as we move towards the centre of the plot, the S$_1$/S$_0$ MECI. Correspondingly, the S$_1$ state, with primary $|2\text{ud}0\rangle$ character, comes down in energy to meet the ground state at the MECI. In the CI region, in the middle of the plot, these two states are of mixed character, containing strong contributions from the $|2200\rangle$, $|2\text{ud}0\rangle$ and $|2020\rangle$ configurations. The S$_2$ state starts at much higher energy but then descends, mixing with the other two states. We note that in CASSCF(2,2), the doubly excited state is not well described, as mentioned earlier, but the two dynamically important states --- S$_0$ and S$_1$ --- are well described across the LIIC. Finally, as we continue left in the plot, towards QC, the lowest two states separate again. The ground state, now with primary $|2020\rangle$ character, comes down to a value approximately 1 eV above the NBD ground state minimum. This difference between QC and NBD ground state energies is the difference exploited for energy storage in MOST systems.\cite{jorner_unraveling_2017,orrego-hernandez_engineering_2020} Correspondingly, the S$_1$ state again acquires primary $|2\text{ud}0\rangle$ character, and the S$_2$ state rises very high in energy, where it is of primary $|2200\rangle$ character.

The rose dashed lines in the same plot show the relative energies for SA(3)-CASSCF(4,4). At the NBD equilibrium geometry, the S$_0$ obtained by (4,4) is quite similar to (2,2), but as observed above, the S$_1$ $|2\text{ud}0\rangle$ state appears at around 7.5 eV, far higher than the experimental value of 5.25 eV.\cite{cooper_valence_2024,xing_198-225-nm_1994,roos_theoretical_1994,palmer_vacuum_2021,frueholz_excited_1979, doering_electron_1981,allan_study_1989,zgierski_franckcondon_1993,lightner_electronic_1980} In terms of dynamics, this would have a severe impact, both due to the energy shift and changes in gradients. Additionally, this pushes it closer to the S$_2$ $|2020\rangle$ state, leading to their interaction region appearing closer to the NBD geometry. Moving towards the MECI, the two states do not quite meet, an artefact of using the CASSCF(2,2) geometries for the LIIC. When the states separate as we move to QC, we can see a notably steeper S$_1$ state, again reflecting the poor description of the $|2\text{ud}0\rangle$ state. As mentioned earlier, the active space is unstable in QC, so we do not show the energies.

In summary, the CASSCF(2,2) calculations, while exceptionally simple, give a balanced description of the potential energy surfaces. On the other hand, the CASSCF(4,4) calculations specifically bias \textit{against} the S$_1$ $|2\text{ud}0\rangle$ state, increasing its energy. This is further confirmed in the next section, when we add dynamical correlation.

\subsection{Dynamic correlation}
Methods beyond CASSCF are required to recover the dynamical correlation. To evaluate the methods, we use a very similar LIIC, except calculated with a correlated method (MRCI(2,2)/p-cc-(p)VDZ, see Section \ref{sec:liics_nbd}). We now discuss MRCI, XMS-CASPT2, and finally methods that do not define an active space, such as linear response theories.

\subsubsection{MRCI}

\begin{figure}
    \centering
    \input{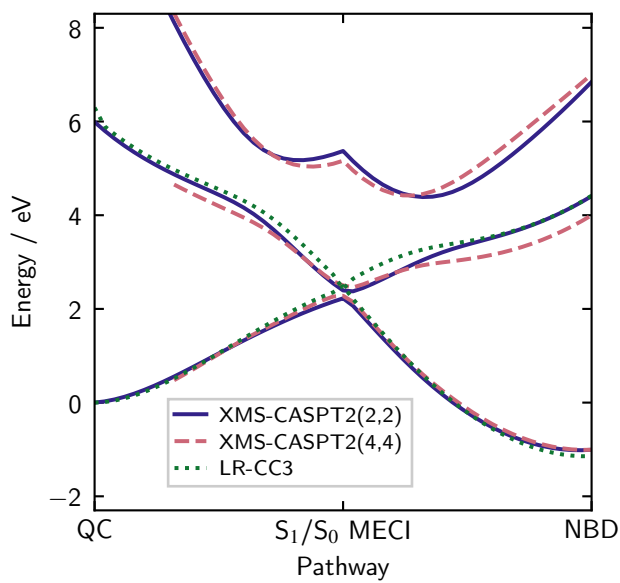}
    \caption{Energies for S$_0$, S$_1$, and S$_2$ calculated using XMS-CASPT2(2,2) (solid, purple),  XMS-CASPT2(4,4) (rose, dashed), and LR-CC3 (green, dotted) with the p-cc-(p)VDZ basis set. All three methods agree well and also with MRCI+Q(4,4) shown in Fig. \ref{fig:CASSCF}. XMS-CASPT2(4,4) shows a notably lower and different shape potential around the NBD geometry. Details of the LIIC pathway are given in Section \ref{sec:liics_nbd}}
    \label{fig:pt2cc3}
\end{figure}

We first focus on MRCI, which considers excitations from reference configurations taken from CASSCF. This method gives high-quality energies and wavefunctions, but lacks size-extensivity and is computationally expensive compared to perturbative and density-functional-based methods. The calculations shown here are MRCI+Q, which uses only single and double excitations\footnote[6]{The use of MRCI($m$,$n$) denotes MRCISD performed on a CASSCF($m$,$n$) reference.} and adds a size-consistency correction (here the renormalised Davidson correction,\cite{davidson_size_1977,siegbahn_multiple_1978} see Section 7 of the SI$^\dag$).

For the (2,2) active space, the MRCI reference weights for all three roots are approximately the same, indicating that the quality of all three states in the original CASSCF calculation is similar. Indeed, the CASSCF(2,2) calculations show nice overall agreement with the MRCI+Q(2,2) calculations, as shown in Fig.\ S13 in the SI$^\dag$. The highest S$_2$ state comes down in energy in the MRCI compared to CASSCF(2,2), but in dynamics at reasonably low energies, say $<8$ eV, this state is not expected to be populated.

For the (4,4) active space, it is crucial to include the Davidson correction (see Section 7 of the SI$^\dag$). Interestingly, MRCI+Q(4,4) gives a similar potential energy surface to CASSCF(2,2), as shown in Fig.\ \ref{fig:CASSCF} (green dotted lines). Notably, the two states are relatively parallel on the right-hand side of the pathway, only beginning to diverge around the conical intersection. The lack of agreement between the CASSCF(4,4) and MRCI+Q(4,4) calculations reflects the comments about the S$_1$ $|2\text{ud}0\rangle$ state above; the reference weight of this state is much lower than the other two states, confirming its poor description in the CASSCF picture.

Overall, the MRCI+Q calculations give excellent quality potential energy surfaces, but the computational expense means that they are not suitable for \textit{on-the-fly} dynamics. This is especially true since the Davidson correction only corrects the energy and not the underlying wavefunction, meaning that it is not possible to calculate analytical gradients and couplings.

\subsubsection{XMS-CASPT2}

An alternative to MRCI is CASPT2, with XMS-CASPT2\cite{shiozaki_communication_2011} the most popular dynamically-correlated multi-reference electronic structure method for dynamics. This is a variant of the MS-CASPT2 method\cite{finley_multi-state_1998} that gives exceptionally high-quality, smooth wavefunctions and potential energy surfaces,\cite{sen_comprehensive_2018,park_single-state_2019} yet is computationally less expensive than its variational cousin MRCI.\cite{finley_multi-state_1998} Crucially, analytical gradients and non-adiabatic coupling vectors have been implemented in multiple software packages.\cite{shiozaki_communication_2011, nishimoto_analytic_2022,aquilante_modern_2020,shiozaki_bagel_2018} Other forms of multi-reference perturbation theory, such as QD-NEVPT2, MS-CASPT2 and XMC-QDPT2, give similar results, as shown in Section 9 of the SI$^\dag$.

At first glance, the XMS-CASPT2 calculations with the (2,2) active space, shown as the solid purple lines in Fig.\ \ref{fig:pt2cc3}, agree well with the MRCI+Q(4,4) calculations (Fig.\ \ref{fig:CASSCF}) across all states. This agreement can be seen further in the excitation energies (shown in {Section 3 of the SI$^\dag$). Furthermore, XMS-CASPT2(4,4) calculations, also included in Fig.\ \ref{fig:pt2cc3}, agree well with \textit{both} these methods, with only minor differences in the excitation energy of S$_1$. The conclusion, so far, is that MRCI+Q(4,4) produces credible reference potential energy surfaces but is not suitable for \textit{on-the-fly} dynamics. The three methods that \textit{are} feasible for dynamics, namely CASSCF(2,2), XMS-CASPT2(2,2) and (4,4), all agree on the overall topography of the potential energy surfaces. In the next section, we turn to non-active space methods as a final arbiter.

\subsubsection{Non-active space methods}

As a final check, we perform LR-CC3,\cite{paul_new_2021} which generates excitation energies from a high-quality coupled-cluster ground state wavefunction. Although this method is unstable around the conical intersection, it provides highly accurate energies elsewhere.\cite{loos_mountaineering_2018,paul_new_2021} Most importantly, LR-CC3 calculations are relatively free of human bias, as they have no active space. In Fig.\ \ref{fig:pt2cc3}, we show the calculations for LR-CC3 for the ground and first excited $|2\text{ud}0\rangle$ state, compared to XMS-CASPT2. The LR-CC3 agrees closely with the XMS-CASPT2 and MRCI+Q calculations for both active spaces, indicating that both methods do not show significant human bias. It agrees particularly well with XMS-CASPT2(4,4), with the two states being parallel across the majority of the potential energy surfaces. Additionally, this indicates a nice qualitative agreement with the CASSCF(2,2) calculations. Unfortunately, LR-CC3 is exceptionally expensive, which, combined with its instability, makes it unsuitable for dynamics simulations. Other coupled cluster methods, such as LR-CC2 and LR-CCSD, do not describe doubly-excited states well and so are not well-suited for this particular system. More information about LR-CCSD, as well as selected configuration interaction, is shown in Section 8 of the SI$^\dag$.


\subsection{Basis sets} \label{sec:basis}

\begin{figure}
    \centering
    \input{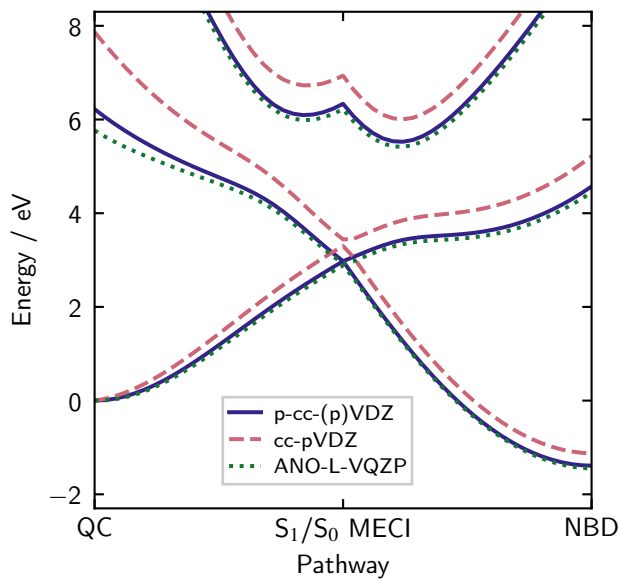}
    \caption{Basis set comparison of p-cc-(p)VDZ (solid purple),  cc-pVDZ (dashed rose), and ANO-L-VTQZ (dotted green), using the SA(3)-CASSCF(2,2) method.  The S$_0$, S$_1$, and S$_2$ energies agree very well for p-cc-(p)VDZ and ANO-L-VTQZ, which both have sufficiently diffuse character, while cc-pVDZ shows a significant increase in energy for the S$_1$ and S$_2$ excited states. Details of the LIIC pathway are given in Section \ref{sec:liics_nbd}.}
    \label{fig:basis}
\end{figure}

As mentioned earlier, the electronic structure calculations in QC/NBD are relatively insensitive to the choice of basis. However, accurate energies require that the basis is sufficiently diffuse. This is particularly notable for the S$_1$ $|2\text{ud}0\rangle$ state, which has a marked diffuse character due to strong mixing with the 3p$_x$ Rydberg state near the QC equilibrium geometry.\cite{borne_ultrafast_2024,cooper_valence_2024} To demonstrate, we compare SA(3)-CASSCF(2,2) energies using the p-cc-(p)VDZ and cc-pVDZ basis sets in Fig.\ \ref{fig:basis}. The p-cc-(p)VDZ basis, which lacks hydrogen polarisation functions but adds diffuse p functions, shows a significantly lower excitation energy for the valence state than cc-pVDZ, especially near the QC ground state equilibrium geometry. The large ANO-L-VTQZ basis agrees well with the p-cc-(p)VDZ, justifying the new contraction. Further discussion of basis sets, including diffuseness, can be found in Section 10 of the SI$^\dag$.

The p-cc-(p)VDZ basis uses only 135 spherical functions, reducing compute time for a single-point SA(3)-CASSCF(2,2) by two orders of magnitude compared to the ANO-L-VQZP basis. This computational efficiency is crucial for non-adiabatic dynamics simulations, which involve many repeated evaluations of the electronic structure. Finally, we note that if the Rydberg states are to be accounted for in the calculations, even more specifically adapted diffuse basis sets are required.\cite{borne_ultrafast_2024}

\section{Discussion}

\subsection{Critical points on the potential energy surface}

\begin{table*}
    \centering
    \caption{Selected carbon-carbon distances (see Eqs.\ \ref{eq:rcc}-\ref{eq:rdb}) and vertical excitation energies calculated for CASSCF(2,2) /p-cc-(p)VDZ optimised geometries (with XMS-CASPT2(2,2)/ANO-L-VQZP excitation energies in brackets) for the QC and NBD ground state equilibrium geometries, the S$_1$/S$_0$ MECI, and the S$_1$ minimum. Leading configurations are given in occupation number representation for each of the three states S$_0$-S$_2$ in the format $|\eta_{A_1}\eta_{B_1}\eta_{B_2}\eta_{A_2}\rangle$ (orbitals as in Fig.\ \ref{fig:AS-schematic}), with \textit{Mix} indicating that the states contain strong contributions from all of the $|2200\rangle$, $|2\text{ud}0\rangle$, and $|2020\rangle$ configurations.}
    \begin{tabular}{lcccccccc}
    \hline
        Geometry               &$r_{\mathrm{cc}}$ / \AA &$r_{\mathrm{rh}}$ / \AA & $r_{\mathrm{db}}$ / \AA & S$_1$ / eV & S$_2$ / eV & S$_0$ & S$_1$ & S$_2$ \\ \hline
        QC                     &1.55  &  0  & 1.53 &6.22 (5.75) & 14.02 (11.59) & $|2020\rangle$ & $|2\text{ud}0\rangle$ & $|2200\rangle$\\
        NBD                    &2.47  &  0 & 1.33 &5.96 (5.34) & 11.24 (7.85) & $|2200\rangle$ & $|2\text{ud}0\rangle$ & $|2020\rangle$\\
        S$_1$/S$_0$ MECI      &1.94 & $\pm$0.50 & 1.44 &0 (0.20) & 3.36 (3.11) & Mix & Mix & Mix\\
        S$_1$ min.\            &2.12  &  0  & 1.41 & 2.85 (2.85) & 4.56 (3.29) & $|2200\rangle$ & $|2\text{ud}0\rangle$ & $|2020\rangle$  \\
        \hline
    \end{tabular}
    \label{tab:energies}
\end{table*}

Purely for the sake of visualisation, we define three coordinates, all involving carbon atoms \ce{C1}-\ce{C4}, as
\begin{align}
r_{\mathrm{cc}} &= \frac{1}{2}\left( r_{12} + r_{34}\right)\label{eq:rcc}\\
r_{\mathrm{rh}} &= r_{13} - r_{24}\label{eq:rrh}\\
r_{\mathrm{db}} &= \frac{1}{2}\left( r_{14} + r_{23}\right),\label{eq:rdb}
\end{align}
where $r_{ij}=|\mathbf{R}_j-\mathbf{R}_i|$ indicates the distance between carbons $i$ and $j$ (see Fig.\ \ref{fig:rxn} for the numbering of the atoms). The $r_{\mathrm{cc}}$ is a measure of the \textit{wing-separation} and is the mean distance between the two ethylenic moieties, with a large value in NBD and a small in QC. The $r_{\mathrm{rh}}$ represents the \textit{rhombicity} and is the difference between the two diagonal distances across the four-carbon ring \ce{C1}-\ce{C4}. This is zero at the NBD and QC ground states and increases as the four-carbon ring distorts to rhombic or parallelogram-like geometries (as seen at the S$_1$/S$_0$ MECI in Fig.\ \ref{fig:geoms}). Finally, $r_{\mathrm{db}}$ corresponds to the mean length of the carbon bonds that acquire double bond character (and hence shorten) in NBD. We show LIIC pathways in the $(r_{\mathrm{cc}},r_{\mathrm{rh}})$-plane in Fig.\ S3 in the SI$^\dag$.

\begin{figure}
    \centering
    \includegraphics[width=\linewidth]{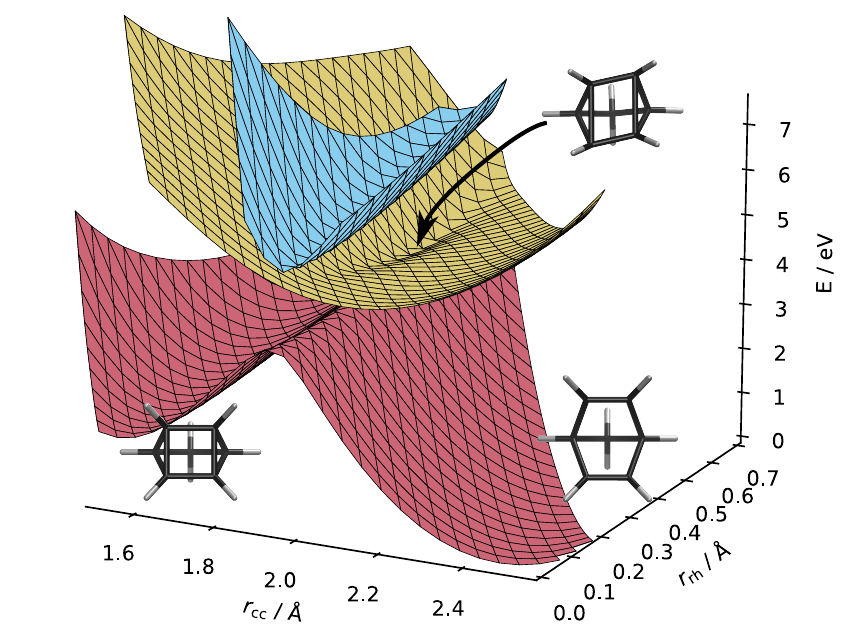}
    \caption{Potential energy surfaces in the plane defined by the NBD and QC  minima and the S$_1$/S$_0$ MECI, calculated at the CASSCF(2,2)/p-cc-(p)VDZ level. The surfaces are plotted as a function of $r_{\mathrm{cc}}$ and $r_{\mathrm{rh}}$ (see Eqs.\ \ref{eq:rcc}-  \ref{eq:rrh}). The S$_0$ surface (rose) has two clear minima corresponding to the equilibrium geometries of QC ($(r_{\mathrm{cc}},r_{\mathrm{rh}})\approx(1.5,0)$ \AA) and NBD ($(r_{\mathrm{cc}},r_{\mathrm{rh}})\approx(2.5,0)$ \AA), separated by a large barrier. The S$_1$ surface (yellow) connects with the S$_0$ surface at the rhombic conical intersection on top of this barrier, indicated by the arrow. The S$_2$ surface (light blue) interacts most strongly at $r_{\mathrm{cc}}\approx2.0$ \AA, halfway between the two minima. Molecular structures from Fig.\ \ref{fig:geoms} are included for reference.}
    \label{fig:PES}
\end{figure}

Figure \ref{fig:PES} shows the potential energy surfaces for the three important states in the plane defined by the QC and NBD minima and the S$_1$/S$_0$ MECI. The barrier between the two ground state minima is immediately apparent, with two distinct wells on the ground state surface corresponding to QC and NBD. The conical intersection linking to the ground state appears on top of this barrier, which explains the photoswitch nature of the system --- depending on the path through the intersection, a wavepacket can end up in either potential well. Finally, the S$_2$ state sits well above the S$_1$ state {at the geometries shown in Fig.\ \ref{fig:PES}, with a trough mirroring the potential barrier on the ground state. The change of character in the ground state wavefunction is evident, with strong coupling between the $|2020\rangle$ and $|2200\rangle$ configurations.

The NBD, QC, and S$_1$/S$_0$ MECI geometries remain relatively consistent across the different electronic structure methods (see \textit{e.g.}\ Table \ref{tab:MECIs} later and Table S1 in SI$^\dag$). However, in some methods, an additional S$_1$ minimum with $C_{2v}$ symmetry and $1^1A_2$ $|2\text{ud}0\rangle$ character is found. This local minimum appears at higher energies than the S$_1$/S$_0$ MECI, with the molecular geometry approximately halfway between QC and NBD ground-state geometries ($r_{\mathrm{cc}}\approx 2.1$ \AA, $r_{\mathrm{db}}\approx 1.4$ \AA). Notably, the plane shown in Fig.\ \ref{fig:PES} \textit{does not} contain the S$_1$ local minimum. In the electronic structure methods where this minimum does not appear, a first-order symmetric saddle-point separates the left- and right-handed variants of the S$_1$/S$_0$ CI. 

Since the S$_1$ minimum appears relatively close to the Franck-Condon region of NBD, see Fig.\ \ref{fig:geoms}, we anticipate it may significantly affect the dynamics. A deep minimum imprints a valley, attracting most of the initially excited wave packet before allowing it to proceed to the CI. Without this minimum, the potential is ridged and the wavepacket will evolve directly towards the CI. Interestingly, at the geometry of the local minimum, the S$_0$ and S$_1$ states have differing ($A_1$ and $A_2$, respectively), and thus do not interact. When displacing towards the S$_1$/S$_0$ CI (an $A_2$ distortion), both states change to the $A$ irreducible representation, leading to coupling and mixed $|2\text{ud}0\rangle$ and $|2020\rangle$ character in both states. 

The vertical excitation energies and the leading configurations at these four molecular geometries are given in Table \ref{tab:energies}. As discussed earlier, the ground state wavefunctions for NBD and QC have different leading configurations, whereas the S$_1$ state always maintains the $|2\text{ud}0\rangle$ configuration. At the MECI geometry, the S$_0$ and S$_1$ states are degenerate, and the second excited state is much lower in energy than in the NBD and QC geometries. The electronic states are thus strongly multi-configurational, with all three states having significant contributions of the $|2200\rangle$, $|2020\rangle$ and $|2\text{ud}0\rangle$ configurations.

Further, a S$_2$/S$_1$ CI appears when dynamical correlation is included. This CI epitomises the aforementioned mixing of the $|2020\rangle$ character into the S$_1$ state, but, as it is a peaked CI being approached from the lower S$_1$ surface, we do not expect the S$_2$ state to significantly affect the dynamics.\cite{galvan_role_2022} Further discussion is presented in Section 5 of the SI$^\dag$.


\subsection{Nature of the potential energy surfaces}

In the following, we discuss two key points on the potential energy surfaces. The first is the S$_1$ minimum, which only appears when using a subset of the electronic structure methods, and the second is the S$_1$/S$_0$ conical intersection, which governs decay onto the ground electronic state. While all electronic structure methods evaluated provide reasonably similar descriptions of the conical intersection, even subtle differences in the topography of the potential energy surfaces in the vicinity of a conical intersection can have consequences for the photoexcited dynamics. For simulations, it is thus crucial to describe this region as accurately as possible.

\subsubsection{The S$_1$ minimum}
\begin{figure*}
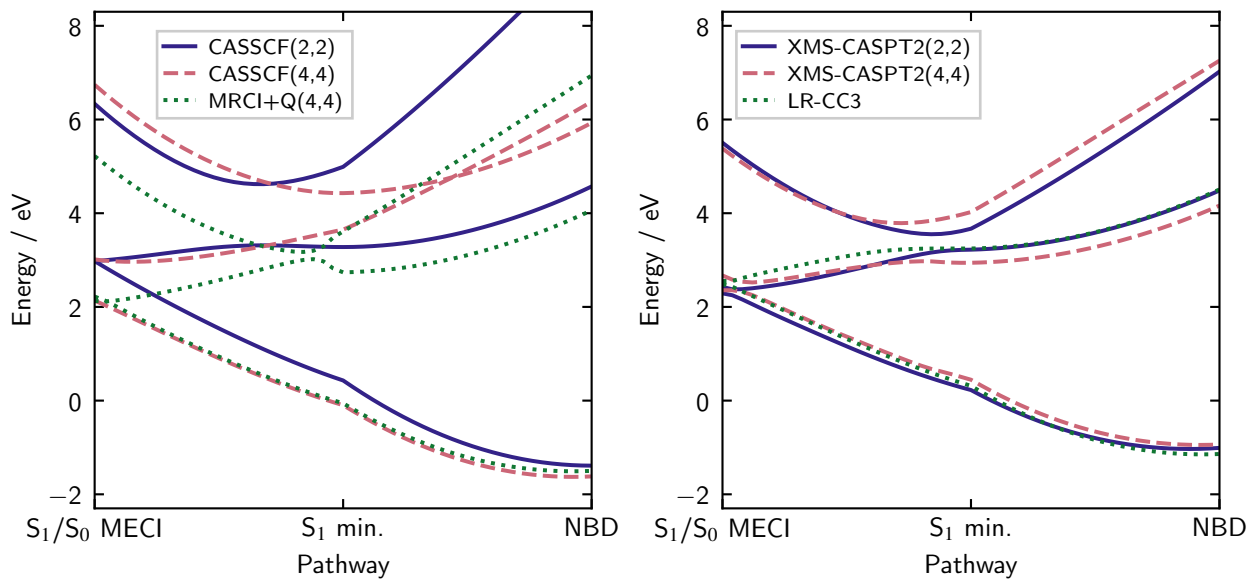

    \centering
    \input{figures/S1_1.pgf}
    \input{SI_figures/cc3_liic2.pgf}
    \caption{Energies for S$_0$, S$_1$, and S$_2$ calculated along the LIIC pathway from the S$_1$/S$_0$ MECI to the S$_1$ minimum, and finally to the NBD equilibrium geometry (see Section \ref{sec:liics_nbd} for details). All calculations employ the p-cc-(p)VDZ basis. Left: CASSCF(2,2) (purple solid), CASSCF(4,4) (rose dashed), and MRCI+Q(4,4) (green dotted). The CASSCF(4,4) is again qualitatively incorrect, whereas CASSCF(2,2) shows a qualitatively correct shape, including a barrier between the S$_1$ minimum and the S$_1$/S$_0$ MECI. The MRCI+Q(4,4) shows a small artefact around the S$_1$ minimum due to the Davidson correction and the two excited states mixing. Right: XMS-CASPT2 [(2,2) solid purple, (4,4) rose dashed] and LR-CC3 (dashed green). All three methods look very similar, but the XMS-CASPT2(2,2) shows no barrier between the S$_1$ minimum and S$_1$/S$_0$ MECI, while the other two only have a very small one.}
    \label{fig:s1_liic}
\end{figure*}

As previously mentioned, the local symmetric S$_1$ $|2\text{ud}0\rangle$ minimum geometry is not present in all methods; namely, XMS-CASPT2(2,2), CASSCF(4,4) and MRCI(4,4) do not show it\footnote[7]{We have not optimised the minimum using MRCI+Q or LR-CC3, since those methods do not have analytical gradients available.}. This is due to the increase in energy of the $|2\text{ud}0\rangle$ state, as mentioned previously. This leads to the S$_1$ state being of primary $|2020\rangle$ character, leading to a strong slope to the S$_1$/S$_0$ MECI, eliminating the possibility of a local minimum. In CASSCF(4,4) and MRCI(4,4), this simply means that a local minimum would exist on the S$_2$ surface, still with primary $|2\text{ud}0\rangle$ character. For XMS-CASPT2(2,2), the crossing of the states occurs very close to the would-be minimum, distorting the potentials and leading to no observed minimum. Further discussion of the S$_2$/S$_1$ crossing is included in Section 5 of the SI$^\dag$.

To explore this region of configuration space, we construct a different LIIC pathway, first going from the NBD ground state geometry to this local S$_1$ minimum and then onwards to the S$_1$/S$_0$ MECI geometry. 

The potential energy cuts for the previously shown methods are shown on this pathway in Fig. \ref{fig:s1_liic}. The key feature in this pathway is a barrier between the S$_1$ minimum and S$_1$/S$_0$ MECI, which must be present for an excited state minimum. Clearly, CASSCF(2,2), XMS-CASPT2(4,4) and LR-CC3 all show a small barrier, while CASSCF(4,4) and XMS-CASPT2(2,2) do not. The apparent barrier in MRCI+Q(4,4) is most likely an artefact of the optimisation procedure. CASSCF(4,4) stands out as clearly divergent, with a crossing of S$_2$ and S$_1$, showing that it lacks even the correct qualitative description. CASSCF(2,2), on the other hand, is at least qualitatively correct, although the conical intersection is higher than the other methods. LR-CC3 and XMS-CASPT2(4,4) agree closely on the shape of the potentials, with only a small energy offset --- we would expect almost identical dynamics from these methods. Finally, we note that XMS-CASPT2(2,2) S$_2$/S$_1$ crossing occurs just off the pathway shown here, but we can see that the S$_2$ state is notably closer in energy at the S$_1$ minimum than in other methods, indicating much stronger influence.

To highlight these effects, in Fig.\ \ref{fig:PES_S1} we show the S$_1$ potential energy surface in the plane defined by the NBD and S$_1$ minima and the S$_1$/S$_0$ MECI. For CASSCF(2,2) (top left), XMS-CASPT2(4,4) (bottom left) and LR-CC3 (bottom right), the overall shape of the potential energy is consistent, with a notable slope towards the S$_1$ minimum (at $(r_{\mathrm{cc}},r_{\mathrm{rh}})\approx(2.1,0)$ \AA). We can clearly see the S$_1$ minimum, with a characteristic `pinching' of the contour lines around a saddle-point between the minimum and the conical intersection. XMS-CASPT2(2,2) (Fig. \ref{fig:PES_S1}, top right) shows an entirely different shape, with a gradient pushing away from the would-be minimum and no obvious saddle-point. This is due to the crossing of the $|2\text{ud}0\rangle$ and $|2020\rangle$ states, leading to the S$_2$/S$_1$ conical intersection (see further discussion in Section 5 of the SI$^\dag$).

In summary, with both XMS-CASPT2(4,4) and LR-CC3 all showing an S$_1$ minimum, we believe the evidence leans towards the presence of a bound local minimum on the excited state, as predicted by CASSCF(2,2).

\begin{figure*}
    \centering
    \includegraphics{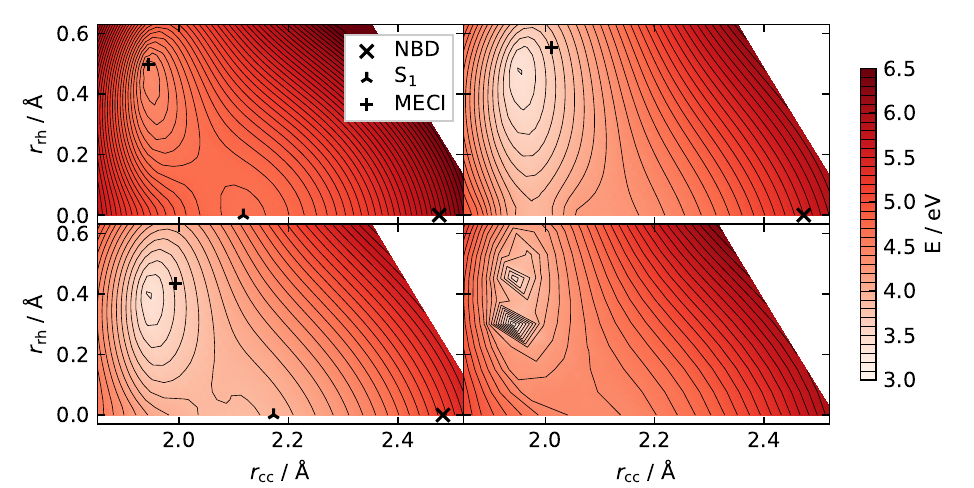}
  \caption{S$_1$ potential energies in the Cartesian plane defined by the NBD and S$_1$ minima and the S$_1$/S$_0$ MECI geometries, calculated at CASSCF(2,2)/p-cc-(p)VDZ level. Energies calculated with CASSCF(2,2) (top left), XMS-CASPT2(2,2) (top right), XMS-CASPT2(4,4) (bottom left), and LR-CC3 (bottom right). Locations of optimised geometries (if they exist) are shown for the active space methods --- we note that, for XMS-CASPT2, these are not the minimum geometries in this plane, as it is calculated using the structures at CASSCF(2,2) level. All but the XMS-CASPT2(2,2) have a similar overall shape, with a notable gradient towards the S$_1$ minimum. LR-CC3 shows instability around the conical intersection but is smooth elsewhere.}
    \label{fig:PES_S1}
\end{figure*}

\subsubsection{The S$_1$/S$_0$ conical intersection}

\begin{figure}
    \centering
    \begin{tikzpicture}[scale=0.8]
        
    \node at (0,0) {\includegraphics[width=0.7\linewidth]{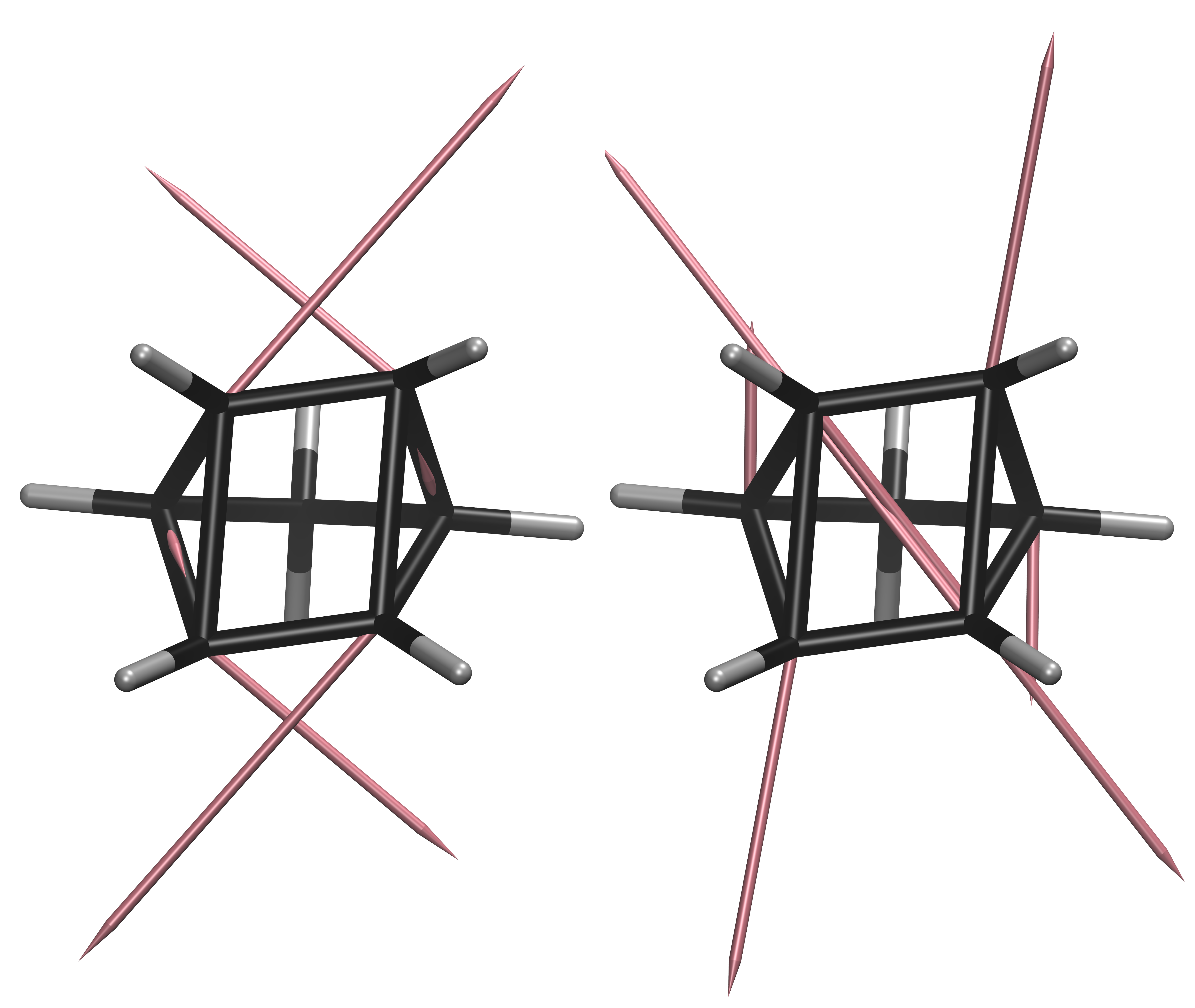}};
    \node at (-2.1,-2.2) {\Large $X$};
    \node at (2.1,-2.2) {\Large $Y$};
    \end{tikzpicture}
    \caption{Branching plane $X$ and $Y$ vectors from S$_1$/S$_0$ MECI, optimised at CASSCF(2,2)/p-cc-(p)VDZ level.\cite{fdez_galvan_analytical_2016} Displacements here lie approximately in the plane of the four-carbon ring. The $X$ vector shortens the C$_1$C$_4$ bond and lengthens the C$_1$C$_2$ bond, forming NBD in the positive direction and QC in the negative. The $Y$ coordinate controls the rhombicity, with negative $Y$ displacement forming the square four-carbon ring of NBD and QC. Hydrogen displacements are small and thus not shown.}
    \label{fig:BP_vec}
\end{figure}

\begin{table}
    \centering
    \caption{Conical intersection parameters. $P$ and $B$ parameters and carbon-carbon distances (in \AA ngstroms) for the  S$_1$/S$_0$ MECI optimised with the (2,2) and (4,4) active spaces for SA(3)-CASSCF, XMS-CASPT2, and MRCI, all with the p-cc-(p)VDZ basis. The conical intersections all have a $C_2$ optimised geometry. Branching plane energy surfaces are shown in Section 4 of the SI$^\dag$.}
    \begin{tabular}{ccccccc}
    \hline
        \textbf{Method} & ($m$,$n$) & $P$ & $B$ & $r_{\mathrm{cc}}$ & $r_{\mathrm{db}}$ & $r_{\mathrm{rh}}$  \\ \hline
        \textbf{CASSCF}             & (2,2) & 0.79 & 0.86 & 1.94 & 1.44 & $\pm$0.50\\ 
        \textbf{}                   & (4,4) & 0.82 & 1.82 & 2.03 & 1.51 & $\pm$0.75\\
        \textbf{MRCI}               & (2,2) & 0.58 & 0.83 & 1.96 & 1.45 & $\pm$0.49\\ 
        \textbf{}                   & (4,4) & 0.78 & 1.11 & 1.99 & 1.48 & $\pm$0.62\\
        \textbf{XMS-CASPT2}         & (2,2) & 0.56 & 1.17 & 2.01 & 1.49 & $\pm$0.56\\ 
        \textbf{}                   & (4,4) & 0.30 & 0.83 & 1.99 & 1.49 & $\pm$0.44\\
        \hline
    \end{tabular}
    \label{tab:MECIs}
\end{table}

The central S$_1$/S$_0$ conical intersection plays a central role in the non-radiative decay to the ground state. {As mentioned previously, all three $|2200\rangle$, $|2020\rangle$, and $|2\text{ud}0\rangle$ configurations are important in this region (the doubly-excited character is quantified\cite{do_casal_classification_2023} in Section 11 of the SI$^\dag$). To analyse the performance of the different electronic structure methods in this region, we use the `local linear approximation' of Fdez.\ Galv\'an \etal,\cite{fdez_galvan_analytical_2016} which provides a consistent scheme for analysing intersections. The orthonormalised branching plane vectors $X$ and $Y$ are shown in Fig.\ \ref{fig:BP_vec}. The $X$ is a wing-separation coordinate related to both $r_{\mathrm{cc}}$ and $r_{\mathrm{db}}$; extension along positive $X$ moves towards the NBD ground state, while negative $X$ tends towards QC. $Y$ is a rhombic distortion clearly related to $r_{\mathrm{rh}}$, with positive displacement increasing the rhombicity and negative displacement increasing the squareness of the four-carbon ring. Unsurprisingly, these vectors are reminiscent of the branching-plane vectors in ethylene dimerisation.\cite{bernardi_mechanism_1990,celani_geometry_1995,serrano-perez_extended_2012} The energy gap is smallest along the $Y$ coordinate, and thus motion along $X$ is most likely to induce non-adiabatic effects.\cite{fdez_galvan_analytical_2016} We can also mention that this analysis has previously been applied to this system,\cite{jorner_unraveling_2017,coppola_norbornadienequadricyclane_2023} albeit in the context of different electronic structure theories.

Fdez.\ Galv\'an \etal\ also introduce two parameters, $P$ and $B$, to quantify the local topography of the conical intersection.\cite{fdez_galvan_analytical_2016} These parameters are functions of the gradients and coupling of the two states and provide a convenient two-parameter representation of the conical intersection. Briefly, $P$ quantifies the overall gradient of the intersection, with $P>1$ for sloped and $P<1$ for peaked intersections. Peaked intersections are known to `funnel' the wavepacket more efficiently towards strong coupling regions and should transfer populations faster. The $B$, on the other hand, quantifies the barriers on the lower potential energy surface. When $B>1$, there are two separate minima, while for $B<1$ there is only one minimum. In principle, an intersection with two separate minima can afford two separate reaction paths and is `bifurcating' --- a crucial feature of photoswitches. Having one ground state potential energy minimum can only lead to a single outcome of the dynamics, and so is a `single-path' intersection. Values near the boundary of $P,B=1$ indicate intersections of mixed character.

Table \ref{tab:MECIs} gives these parameters for the MECI geometries for the multi-configurational methods tested here (standard MRCI is shown due to the lack of analytical gradients in MRCI+Q). All methods agree that this is a peaked conical intersection ($P<1$), which agrees well with previous studies and is concordant with the rapid dynamics seen in experimental work. The different methods, however, do not agree on the number of minima, with methods alternating between predicting bifurcating and single-path intersections. With the exception of the CASSCF(4,4), with a value of $B=1.823$, all methods converge on values around $B=1\pm0.2$, indicating that if there are two minima, one is shallow. 

\begin{figure}
    \centering
    \includegraphics[width=\linewidth]{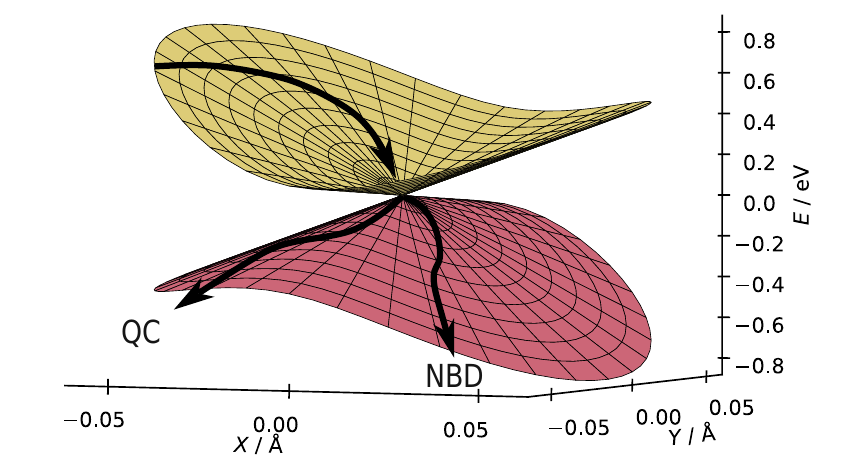}
    \caption{Local linear approximation of energies in the branching plane of the S$_1$/S$_0$ MECI, optimised at CASSCF(2,2)/p-cc-(p)VDZ level. This is a peaked bifurcating conical intersection, and the branching plane vectors are shown in Fig.\ \ref{fig:BP_vec}. Approximately, NBD is located towards positive $X$ and negative $Y$, and QC towards negative $X$ and $Y$. Two plausible reaction paths are shown, one pointing towards QC and the other towards NBD.}
    \label{fig:BP_fig}
\end{figure}

In Fig.\ \ref{fig:BP_fig}, we show the energies in the branching plane for CASSCF(2,2), with the other methods in this study shown in Section 5 of the SI$^\dag$. Here, we can see the ground state potential barrier, in line with the potential energy surface shown in Fig.\ \ref{fig:PES}. The (quasi)bifurcating nature reflects this system's utility as a photoswitch --- there are two minima, corresponding to the QC and NBD products, which both can be accessed from this intersection. In the present case, NBD has the deeper ground state minimum, while the minimum in the QC direction is shallower and more sensitive to the method used for the calculation. The borderline values of $B$ hint at the unsubstituted system's low quantum-yield; no significant potential energy well promotes the conversion of excited NBD into QC. In the context of applications, we suspect that substituted systems with both a lower $B$ value \textit{and} the deeper minimum towards QC would lead to a more efficient route for the formation of QC and, thus, a higher quantum yield. These numbers act as useful distillations of the potential energy surfaces and help assess the similarity of the different electronic structure methods. However, we must bear in mind that while the nature of the conical intersection is important to the outcome of the dynamics, the dynamics preceding the CI is likely to play an even more critical role. 

Interestingly, we notice that the earlier results by Hernandez \etal,\cite{hernandez_multiconfigurational_2023} which indicate significant formation of ground state NBD at relatively short time-scales, are based on electronic structure calculations similar to the CASSCF(4,4), aligning nicely with the pronounced single-path nature of the conical intersection observed in this particular method.

\section{Conclusions}

We have presented an extensive analysis of the multi-configurational electronic structure of the valence states of quadricyclane and norbornadiene. The previously used CASSCF(4,4) is shown to disagree with higher-level methods, including XMS-CASPT2, MRCI and LR-CC3, while the compact CASSCF(2,2) model is found to yield qualitatively correct results. Additionally, we present a small basis which provides excellent results in comparison to larger basis sets, limiting the computational time required and, thus, the computational cost of dynamics simulations. 

The effects of different electronic structure methods on the potential energy surfaces are demonstrated, specifically concerning the presence of an S$_1$ minimum and the shape of the S$_1$/S$_0$ conical intersection. The S$_1$ minimum, present in the highest-level methods, is absent in the XMS-CASPT2(2,2) calculations. On the other hand, the conical intersection is fairly consistent between low- and high-level multi-configurational methods, with only comparatively minor differences in the topography. 

From this work, three electronic structure methods suitable for dynamics are identified: CASSCF(2,2), XMS-CASPT2(2,2) and XMS-CASPT2(4,4). Clearly, further work performing the actual dynamics simulations is important to gain greater understanding of the interplay between dynamics and electronic structure, both in this system specifically and more generally. Furthermore, the erroneous CASSCF(4,4) model can be used as a control case, making it possible to assess how much the dynamics is affected by qualitatively incorrect surfaces. The effect of the conical intersection topography on the outcome of the dynamics is quite interesting, particularly regarding how dynamics prior to and at the conical intersection may affect the (short-time) quantum yields of products.

As chemical modifications via substitution are extremely important in practical applications based on the NBD/QC system, future work should aim to develop a better understanding of how steric and electronic effects due to substituent groups modify the potential energy surfaces. Furthermore, as most practical applications will be in the condensed phase, better theoretical understanding of solvent interactions and their effects on the potential energy surfaces of both native and substituted systems is essential.

Looking ahead, the models identified in this work provide an opportunity to explore how changes in barrier heights and CI topographies affect photochemical dynamics. It is striking that even in a set of nearly-correct electronic structure models, subtle differences in \textit{e.g.}\ the position and appearance of CIs appear. These can have significant consequences for the dynamics and, thus, important quantities for applications, such as decay times, photostability, and branching ratios. The non-trivial electronic structure of QC/NBD might also make this an interesting, albeit challenging, test system for emerging black-box electronic structure methods.\cite{hutton_using_2024} Finally, a fundamental understanding of how the dynamics is influenced by changes in the electronic structure should prove useful for the identification of suitable substituent groups for the QC/NBD system, leading to more efficient and effective MOST applications.

\section*{Conflicts of interest}
There are no conflicts to declare.

\section*{Author contributions}
AK and JCC jointly planned the research. JCC performed and analysed the calculations. Both authors discussed the results and wrote the manuscript.

\section*{Data availability}
Optimised geometries used in this work are available in the supplementary information (SI).

\section*{Acknowledgements}
The authors thank Roland Lindh (Uppsala) for helpful discussions at an early stage of the project and Lauren Bertram (Oxford) for useful comments on the draft manuscript. JCC acknowledges a doctoral studentship from the University of Oxford. AK acknowledges funding from the Engineering and Physical Sciences Research Council EP/V006819/2, EP/V049240/2, EP/X026698/1, and EP/X026973/1, and funding from the Leverhulme Trust, RPG-2020-208, as well as support from the U.S.\ Department of Energy, Office of Science, Basic Energy Sciences, under award number DE-SC0020276. 


\balance


\bibliography{references-2} 
\bibliographystyle{rsc_jabbrv} 

\end{document}


{\noindent\LARGE Supplementary Information\\
\Large{Electronic structure of norbornadiene and quadricyclane}}
\vspace{0.5cm}

{\noindent\large Joseph C. Cooper and Adam Kirrander\footnote{adam.kirrander@chem.ox.ac.uk}}
\vspace{0.5cm}

{\noindent Physical and Theoretical Chemistry Laboratory, Department of Chemistry, University of Oxford, South Parks Road, Oxford, UK, OX1 3QZ}

\tableofcontents
\newpage

\section{Orbitals}

\begin{figure}[H]
    \centering
    \includegraphics[width=0.6\linewidth]{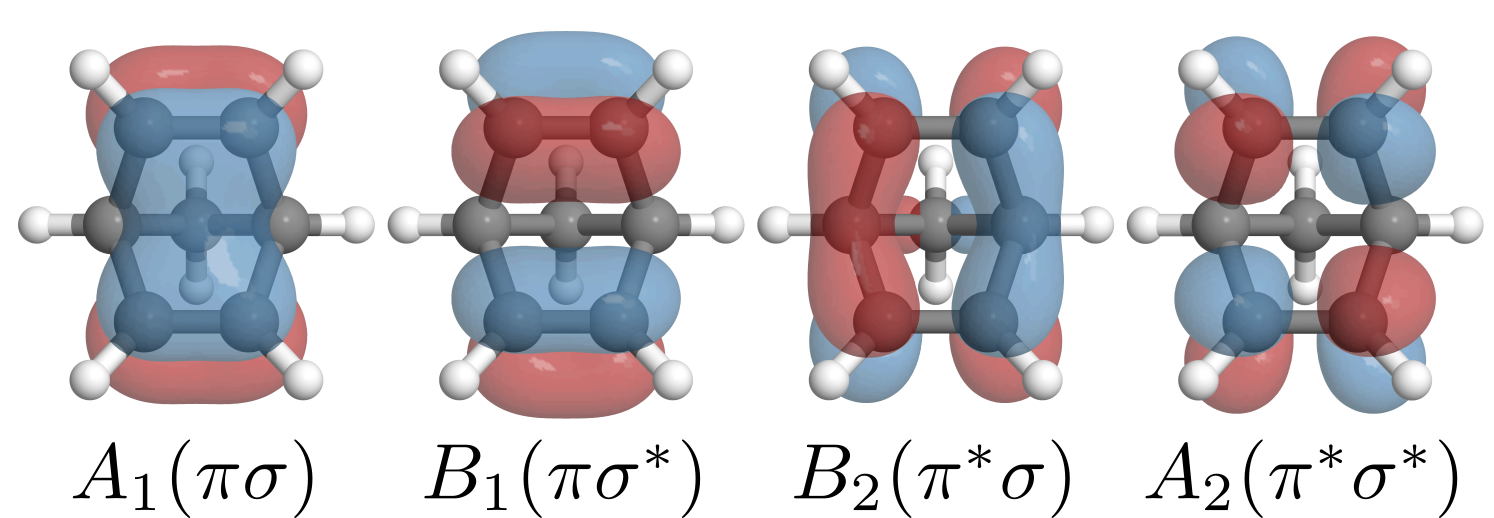}
    \caption{Orbitals (isosurface value of 0.05) for CASSCF(4,4)/p-cc-(p)VDZ at the NBD geometry. We use the same labelling as Fig.\ 2 in the main text}.
    \label{fig:44orbs}
\end{figure}

\begin{figure}[H]
    \centering
    \includegraphics[width=0.7\linewidth]{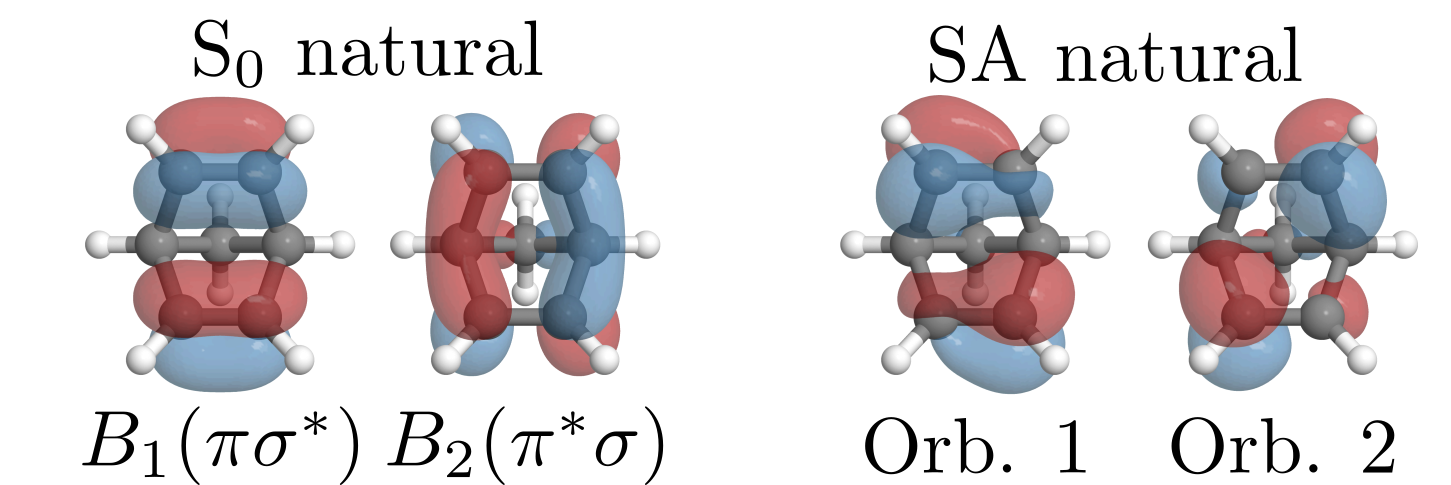}
    \caption{Orbital (iso-surface value of 0.05) for CASSCF(2,2)/p-cc-(p)VDZ at the NBD geometry. On the right, we show the state-averaged (SA) natural orbitals, which exhibit pronounced asymmetry. The state-specific orbitals for the S$_0$ state (left, S$_1$ state gives similar results), do not show this asymmetry, and look much closer to Fig.\ \ref{fig:44orbs}.}
    \label{fig:22orbs}
\end{figure}

In Figure \ref{fig:44orbs}, we show the orbitals in NBD for CASSCF(4,4), which match Fig. 2 in the main text. For the CASSCF(2,2)/p-cc-(p)VDZ calculations, as shown in Figu.\ \ref{fig:22orbs}, the state-averaged (SA) natural orbitals do not resemble the orbitals of the (4,4) active space. This is due to the effect of the S$_2$ state, which is not well described, adding asymmetry. If we look at the natural orbitals for the S$_0$ state (or the S$_1$ state), then we can see the more familiar shapes, akin to the $B_1$ and $B_2$ orbitals in the (4,4) active space. The asymmetry in the orbitals does not affect the calculations --- the two sets of orbitals are simply a rotation of each other.

\section{Geometry parameters}
\begin{table}[H]
\caption{Geometry parameters for ground state optimised geometries, calculated using CASSCF and XMS-CASPT2 for various basis sets and active spaces. All geometries are C$_{2v}$, making $r_{\mathrm{rh}}=0$. The reference calculations (MP2/aug-cc-pVQZ) are taken from our previous work.\cite{cooper_valence_2024} All distances in {\AA}ngstr{\"o}m. The (4,4) active space is unstable around the QC ground state minimum, so those results are not included.}
\centering
\begin{tabular}{lllllll}\hline\hline
             &              &            & QC    &              & NBD   &              \\ \hline
Basis        & Active space & Method     & $r_{\mathrm{cc}}$   & $r_{\mathrm{db}}$   & $r_{12}$   & $r_{14}$   \\ \hline
p-cc-(p)VDZ  & (2,2)        & CASSCF     & 1.549 & 1.530  & 2.476 & 1.330  \\
             &              & XMS-CASPT2 & 1.518 & 1.555  & 2.473 & 1.365  \\
             &              & MRCI       & 1.535 & 1.538  & 2.474 & 1.339  \\\cline{2-7} 
             & (4,4)        & CASSCF     &       &        & 2.474 & 1.352  \\
             &              & XMS-CASPT2 &       &        & 2.483 & 1.356  \\\hline
ANO-L-VTZ(p) & (2,2)        & CASSCF     & 1.552 & 1.525  & 2.470 & 1.322  \\
             &              & XMS-CASPT2 & 1.503 & 1.544  & 2.455 & 1.351  \\ \hline
aug-cc-pVQZ  & HF           & MP2        & 1.515 & 1.538  & 2.462 & 1.339  \\ \hline\hline
\end{tabular}
\label{tab:SI_gs_geoms}
\end{table}

\section{Excitation energies}

Table \ref{tab:vees} shows the vertical excitation energies for S$_1$ and S$_2$, calculated from the NBD ground state. We also show the experimental value (5.25 eV). Most correlated methods predict the excitation energy in the right ballpark, with energies in the $5.25\pm0.4$ eV range. We note that the steeply sloped nature of this state could significantly affect the excitation energies, as a small displacement of the nuclear geometry leads to a relatively large change in excitation energy. As we use the same geometries (MRCI(2,2) for correlated methods, CASSCF(2,2) for non-correlated) across all comparisons, the calculations are not at the exact minimum for each of the methods, which could affect the excitation energy. Of particular note are the values for CASSCF(4,4), which is over 2 eV higher than the predictions of the other methods, and MRCI(4,4), which fails to significantly correct for the errors in CASSCF(4,4).

\begin{table}[H]
\caption{Calculated vertical excitation energies in NBD in eV. Experimentally, the state is posited to have a vertical excitation of $\approx$5.25 eV, but the dissociative nature and lack of oscillator strength make this assignment difficult. All calculations are performed on MRCI(2,2)/p-cc-(p)VDZ ground state minimum geometry, except CASSCF calculations, which are performed on CASSCF(2,2)/p-cc-(p)VDZ geometry. The MP2/aug-cc-pV6Z value is taken from Cooper \textit{et al.}.\cite{cooper_valence_2024} Values next to CASPT2 in brackets indicate shift values, with IPEA indicating that the IPEA shift was used with the recommended value of 0.25 E$_{\mathrm{h}}$. All state-averaged calculations are performed over three states. ANO-L-VTZ(p) indicates an ANO-L-VTZP basis on the carbon atoms and an ANO-L-VTZ basis on the hydrogens.}
\centering
\label{tab:vees}
\begin{tabular}{lllll}
\hline\hline
Basis         & ($n$,$m$) & Method           & S$_1$ $|2ud0\rangle$ & S$_2$ $|2020\rangle$ \\ \hline\hline
Exp.          &              &                  & 5.25 & \\\hline
p-cc-(p)VDZ   & (2,2)        & CASSCF           & 5.96         & 11.24                       \\
              &              & MS-CASPT2 (0.2i) & 4.85         & 8.55                        \\
              &              & MS-CASPT2 (IPEA, 0.2i) &  5.21  & 8.84                        \\
              &              & XMS-CASPT2 (0.1i)& 5.40         & 7.71                        \\
              &              & XMS-CASPT2 (0.2i)& 5.41         & 7.85                        \\
              &              & XMS-CASPT2 (0.4i)& 5.43         & 8.44                        \\
              &              & MRCI             & 5.74         & 10.34                       \\
              &              & MRCI+Q (DV1)     & 5.68         & 9.89                        \\
              &              & MRCI+Q (DV2)     & 5.66         & 9.69                        \\
              &              & MRCI+Q (DV3)     & 5.62         & 9.26                        \\
              &              & MRCI+Q (Pople)   & 5.63         & 9.40                        \\
              &              & QD-NEVPT2        & 5.36         & 8.14                        \\
              &              & XMC-QDPT2        & 5.44         & 8.09                        \\
              &              & HCI(20, 80, $\epsilon_1=10^{-4}$ )& 5.83         &                             \\
              &              & SHCI(20, 80, $\epsilon_1=10^{-4}$, $\epsilon_2=10^{-7}$) & 5.72         &                             \\ \cline{2-5} 
              & (4,4)        & CASSCF           & 7.55         & 8.00                        \\
              &              & XMS-CASPT2 (0.2i)& 5.01         & 8.01                        \\
              &              & MRCI             & 6.47         & 8.13                        \\
              &              & MRCI+Q (DV1)     & 6.01         & 8.20                        \\
              &              & MRCI+Q (DV2)     & 5.83         & 8.22                        \\
              &              & MRCI+Q (DV3)     & 5.47         & 8.25                        \\
              &              & MRCI+Q (Pople)   & 5.59         & 8.24                        \\ \cline{2-5} 
              & HF           & LR-CC3           & 5.56         &                             \\ 
              &              & LR-CCSD          & 5.70         &                             \\
              &              & LR-CC2           & 5.46         &                             \\ \hline
aug-cc-pVTZ   & HF           & LR-CCSD          & 5.57         &                             \\
              &              & LR-CC2           & 5.34         &                             \\ \hline
ANO-L-VTZ(p)  & (2,2)        & CASSCF           & 5.93         & 11.19                       \\
              &              & XMS-CASPT2 (0.2i)& 5.28         & 7.66                        \\ \hline
ANO-L-VQZP    & (2,2)        & CASSCF           & 5.89         & 11.15                       \\
              &              & XMS-CASPT2 (0.2i)& 5.26         & 7.66                        \\ \hline
aug-cc-pV6Z   & HF           & ADC(2)          & 5.34  &  \\ 
\hline\hline\end{tabular}
\end{table}

\section{LIIC geometries}

\begin{figure}[H]
    \centering
    \input{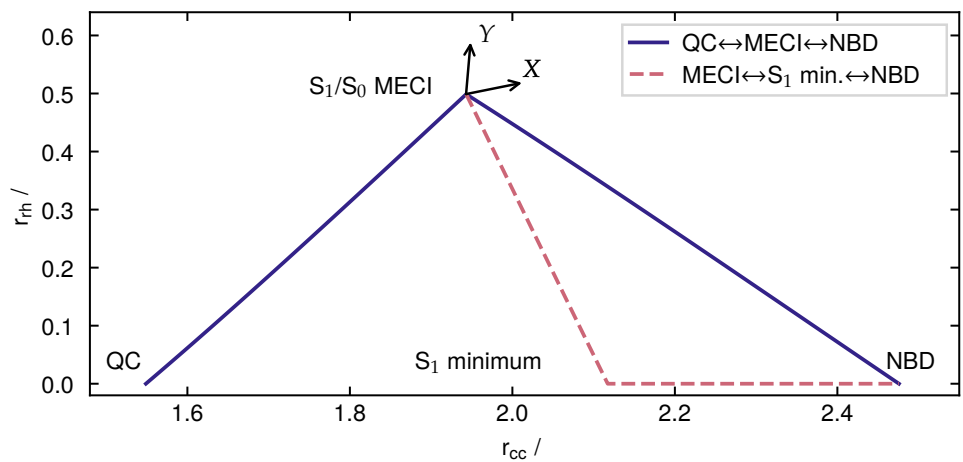}
    \caption{Carbon-carbon distance coordinates for the two LIICs used in the $(r_{\mathrm{cc}},r_{\mathrm{rh}})$-plane, with key geometries labelled. All geometries are calculated at CASSCF(2,2)/p-cc-(p)VDZ level. The projection of the two branching plane coordinates $X$ and $Y$ into the plane are also shown (cf Fig.\ 12, main text).}
    \label{fig:liic_distances}
\end{figure}

In Fig.\ \ref{fig:liic_distances}, we show the two different LIICs. The QC$\leftrightarrow$S$_1$/S$_0$ MECI$\leftrightarrow$NBD LIIC (solid purple line) travels straight between the minima and the conical intersection, conserving $C_2$ symmetry (approximately, as calculations are run without symmetry turned on). For the S$_1$/S$_0$ MECI$\leftrightarrow$S$_1$ minimum$\leftrightarrow$NBD LIIC, the first part of the pathway traverses from the conical intersection to the S$_1$ minimum, maintaining (approximate) $C_2$ symmetry. The second part of the pathway holds $r_{\mathrm{rh}}=0$, maintaining $C_{2v}$ symmetry. From here, it is clear that we do not need to plot the QC$\leftrightarrow$S$_1$/S$_0$ MECI on both pathways.

\section{Conical intersections}

\begin{figure}[H]
    \centering
    \begin{tikzpicture}
        
    \node at (0,0) {\includegraphics[width=0.5\linewidth]{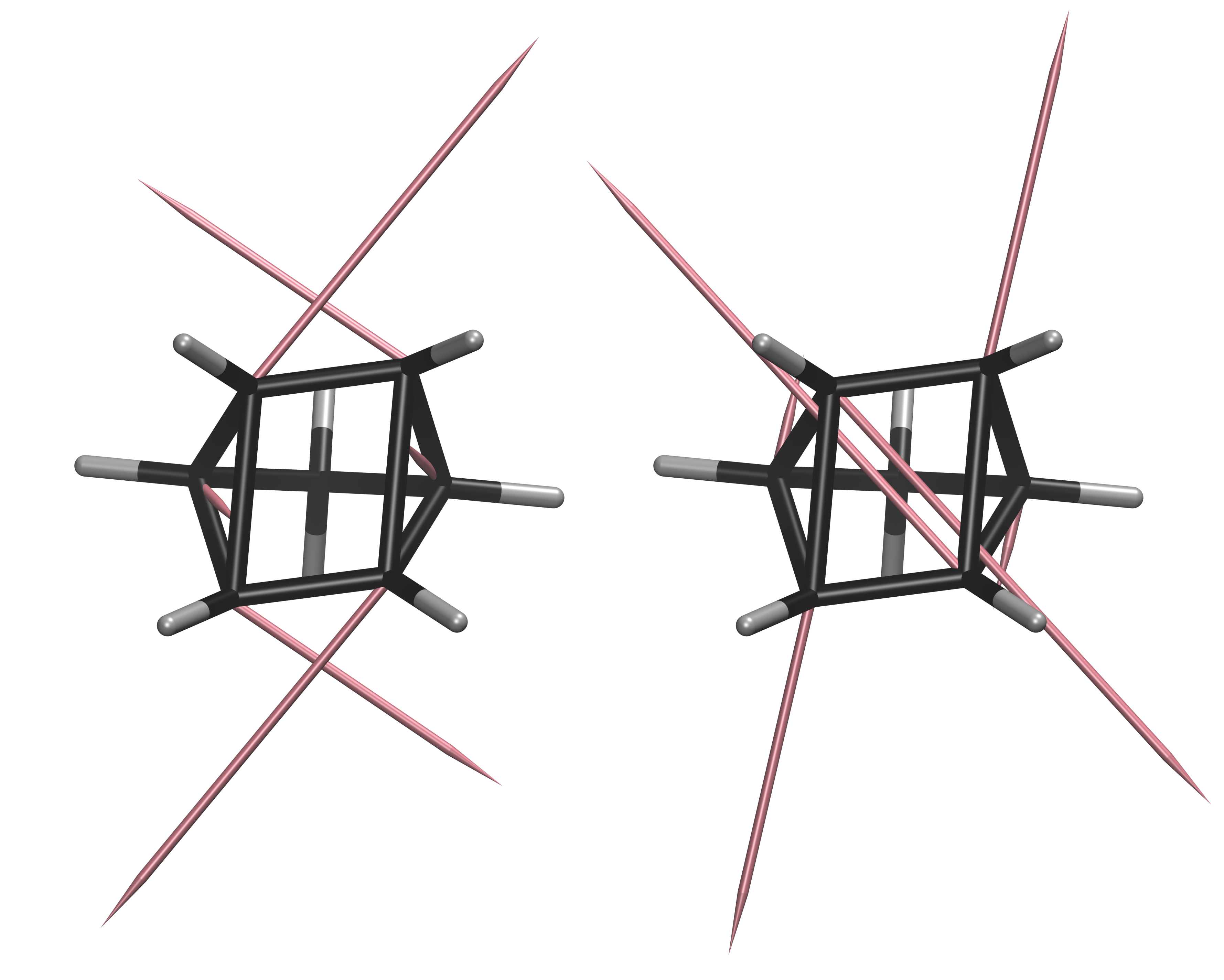}};
    \node at (-2.1,-2.2) {\Large X};
    \node at (2.1,-2.2) {\Large Y};
    \end{tikzpicture}
    \caption{Branching plane X and Y vectors from S$_1$/S$_0$ MECI, optimised at XMS-CASPT2(2,2)/p-cc-(p)VDZ level. These vectors are very similar to the CASSCF(2,2) vectors shown in the main text, and all other methods tested here.}
    \label{fig:caspt2_bp}
\end{figure}

\begin{figure}[H]
    \centering
    \includegraphics[width=0.5\linewidth]{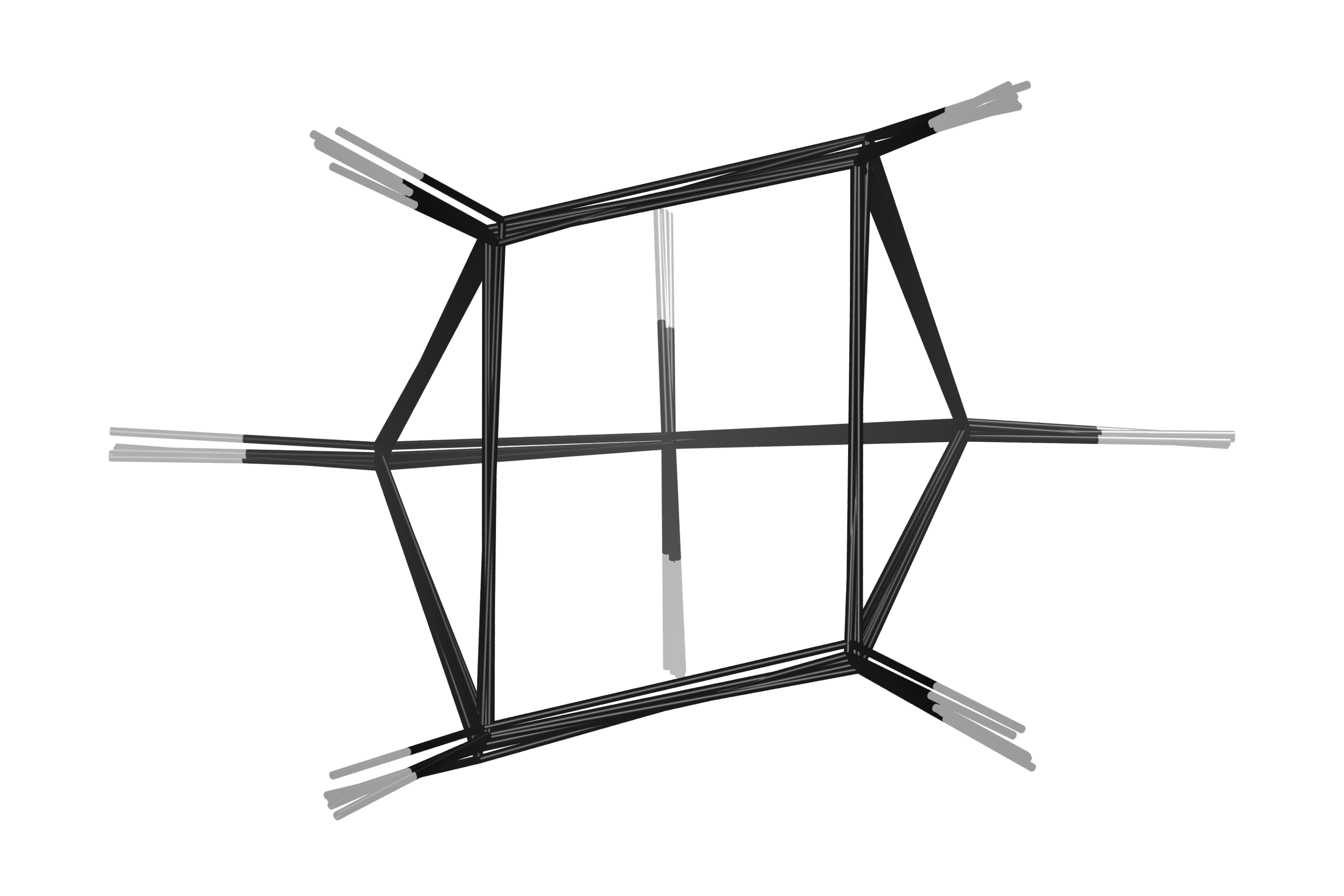}
    \caption{S$_1$/S$_0$ MECI geometries for all combinations of CASSCF, MRCI, and XMS-CASPT2, and the (2,2) and (4,4) active spaces overlaid. All geometries show approximately the same distortions from the ground state minima.}
    \label{fig:meci_overlay}
\end{figure}

Firstly, we show the branching plane vectors from XMS-CASPT2(2,2)/p-cc-(p)VDZ in Fig.\ \ref{fig:caspt2_bp}. These compare well with the CASSCF(2,2) vectors shown in the main text, with only very minor changes in the direction of the vectors. All methods also give very similar optimised geometries, as discussed in Table 2 in the main text. The geometries are shown superimposed in Fig.\ \ref{fig:meci_overlay}. There are minor differences (especially in the hydrogens attached to the four-carbon ring), but the overall features of the geometry are replicated in all methods.

\begin{figure}[H]
    \centering
    \includegraphics[width=0.8\linewidth]{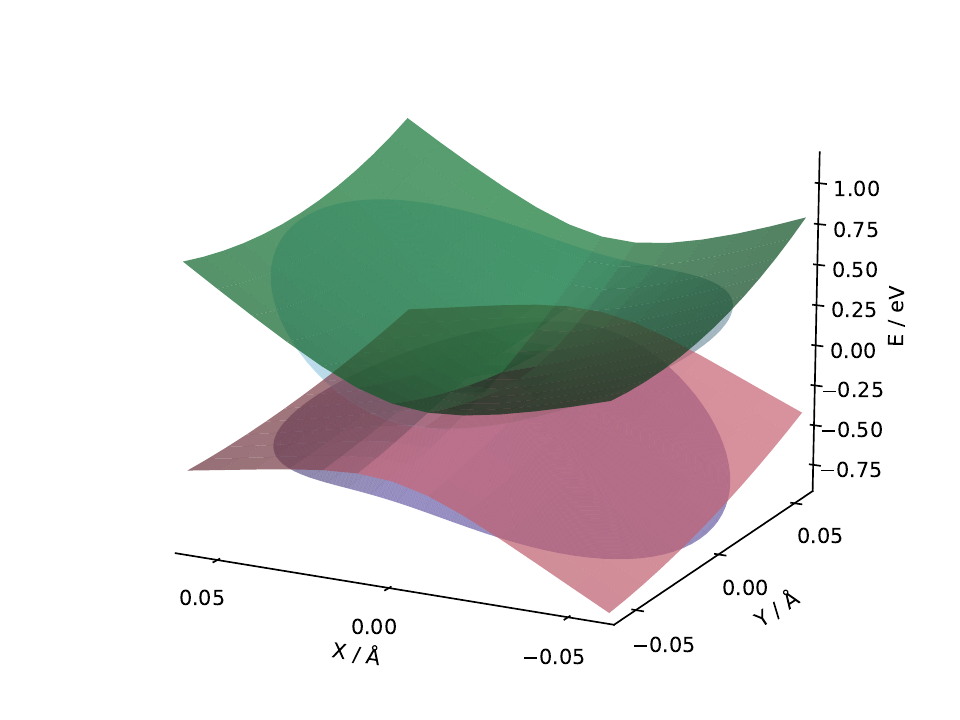}
    \caption{Appropriateness of local linear approximation in the S$_1$/S$_0$ conical intersection: Analytical local linear representation (purple and light blue surfaces) against CASSCF(2,2) (rose and green surface) calculations inside the branching plane. The agreement is excellent, demonstrating that the local linear approximation is reasonable in the plotted region. Parameters taken from SA(3)-CASSCF/p-cc-(p)VDZ optimised geometry. Square surfaces are the \textit{ab initio} results, the round surfaces the local linear representation.} 
    \label{fig:3dllr}
\end{figure}

\begin{figure}[H]
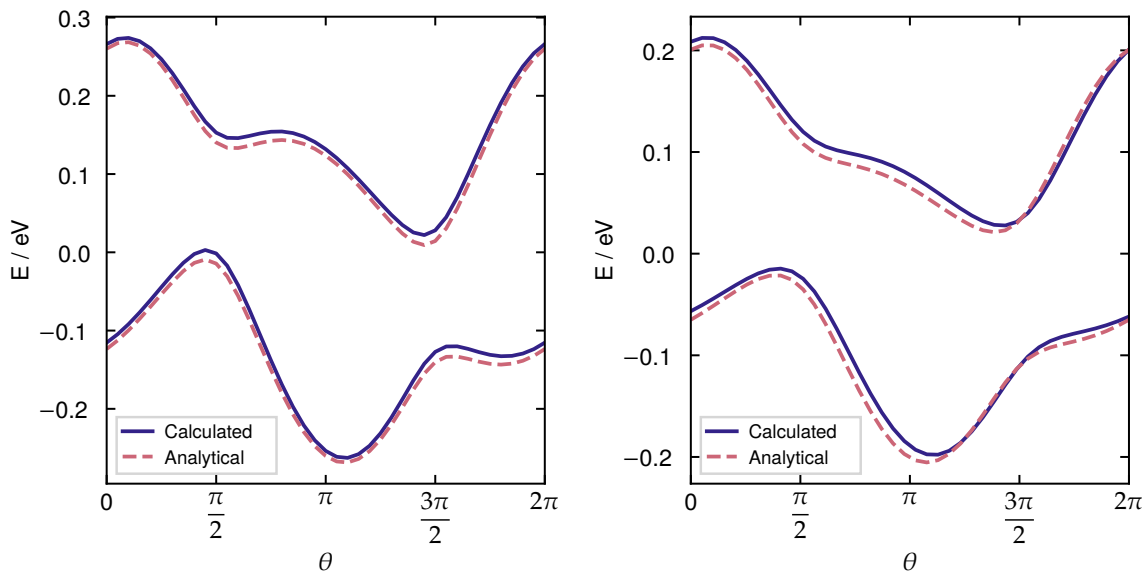

    \centering
    \input{SI_figures/ana_vs_calc.pgf}\input{SI_figures/pt2_ana_vs_calc.pgf}
    \caption{Appropriateness of local linear approximation in the S$_1$/S$_0$ conical intersection: Calculated \textit{ab initio} (solid purple) energies vs analytical local linear representation (dashed rose) as a function of polar angle $\theta$ at a distance of $0.02$ \AA\space around the S$_1$/S$_0$ MECI. Left: Energies and MECI calculated at SA(3)-CASSCF/p-cc-(p)VDZ level. Right: Energies and MECI calculated at XMS-CASPT2(2,2)/p-cc-(p)VDZ level. For both calculations, the analytical representation matches the calculated energies well, with only a slight underestimation. One can see the difference in `single-path' vs `bifurcating' intersections by the small minimum on the lower surface at $\theta\approx\frac{7\pi}{4}$. The $\theta$ is defined to be 0 along the positive $X$ direction, and $\frac{\pi}{2}$ along the positive $Y$ direction.}
    \label{fig:2dllr}
\end{figure}

The energies in the branching plane are approximated according to the formula taken from the original work of Fdez.-Galv\'an et al.,\cite{fdez_galvan_analytical_2016} $$E^{\pm}(r, \theta) = \delta_{gh}r\left(\sigma\;\mathrm{cos}(\theta-\theta_s) \pm \sqrt{1+\Delta_{gh}\;\mathrm{cos}(2\theta)}\right),$$ where $E^{\pm}$ are the energies of the two states relative to the minimum energy crossing point, $r$ and $\theta$ are the polar distance and angle, and $\delta_{gh}$, $\Delta_{gh}$, $\sigma$ and $\theta_s$ are various parameters taken from the \textit{ab initio} calculation at the optimised geometry (more details in Fdez.-Galvan et al.\cite{fdez_galvan_analytical_2016}).

For small $r$, these provide an excellent approximation of the local potential energy surfaces around the conical intersection. This can be seen clearly in Figs.\  \ref{fig:3dllr} and \ref{fig:2dllr}, which compare the energies of \textit{ab initio} calculations to the local linear representations. Figure \ref{fig:3dllr} shows the linear representation (the two circular surfaces) and the energies of ab initio calculations (the square surfaces). Clearly, the two surfaces are almost on top of one another. This might be easier to see in Fig.\ \ref{fig:2dllr}, which show the potential energies on a circular trip around the minimum energy conical intersection (at $r=0.02$ \AA). Here, the agreement between the analytical representation and the energies calculated using \textit{ab initio} methods is clearly very good.

\begin{figure}[H]
    \centering
    \includegraphics[width=\textwidth]{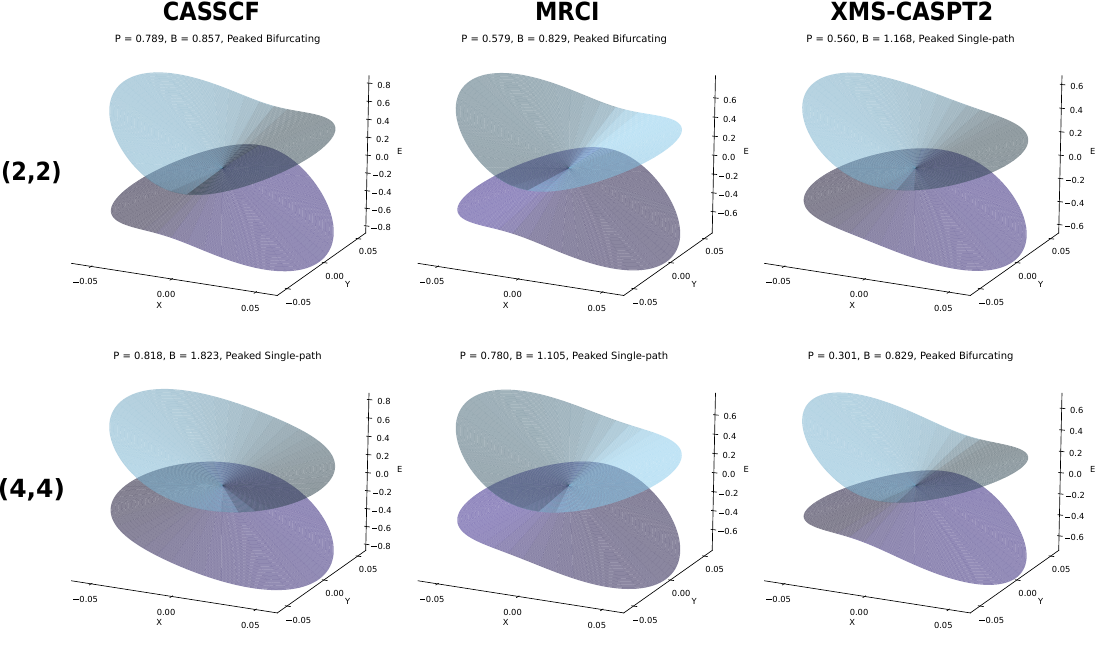}
    \caption{Local linear approximations to the energy in the branching plane around the S$_1$/S$_0$ MECI geometries, optimised for CASSCF, XMS-CASPT2, and MRCI with both the (2,2) and (4,4) active spaces (columns represent methods, rows active spaces). S$_1$ is shown in light blue, S$_0$ in purple. The values $P$ and $B$ are discussed in the main text and are found using the methods of Fdez.-Galvan \textit{et al}.. The NBD ground state minimum is roughly towards positive $X$ and negative $Y$, whereas the QC ground state minimum is towards negative $X$ and negative $Y$. To the naked eye, all methods show a similar energy landscape in the branching plane except CASSCF(4,4) (bottom left), which shows a slightly more tilted geometry with a less obvious second minimum. These plots are shown with energy in eV and distance in {\AA}ngstr{\"o}m.}
    \label{fig:all_mecis}
\end{figure}

In Fig.\ \ref{fig:all_mecis}, we show the local linear representation of the energies in the branching plane of the S$_1$/S$_0$ MECIs for all combinations of the (2,2) and (4,4) active space and SA(3)-CASSCF, XMS-CASPT2, and MRCI. Almost all methods agree about the rough shape of the intersection, with a notable ridge approximately along the $X=0$ line. The notable exception is CASSCF(4,4), which gives a more slanted overall intersection, with a shallower minimum towards the negative $X$ direction (towards QC).

\subsection{The S$_2$/S$_1$ conical intersection}

\begin{figure}[H]
    \centering
\includegraphics{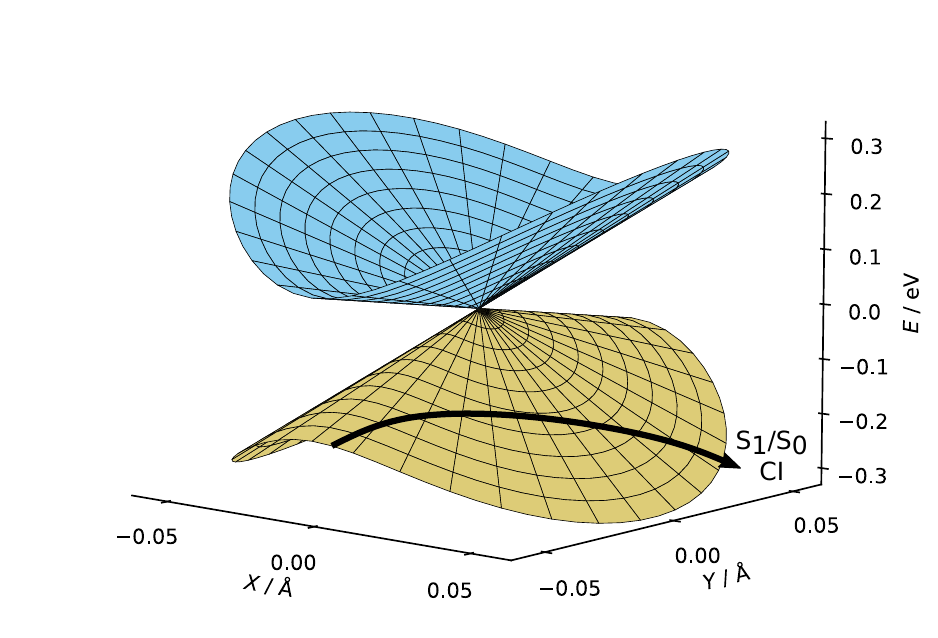}
    \caption{Energies in the branching plane of S$_2$/S$_1$ conical intersection, using the local linear representation\cite{fdez_galvan_analytical_2016}. This is a peaked, bifurcating intersection. The branching plane vectors are shown in Figure \ref{fig:S2S1BP}, with $X$ breaking the symmetry and $Y$ preserving it. The conical intersection was optimised using XMS-CASPT2(4,4)/p-cc-(p)VDZ. An illustrative trajectory is shown, which would travel from NBD to the S$_1$/S$_0$ conical intersection, missing the S$_2$/S$_1$ conical intersection}
    \label{fig:S2S1meci}
\end{figure}

\begin{figure}[H]
    \centering
    \includegraphics[width=0.8\linewidth]{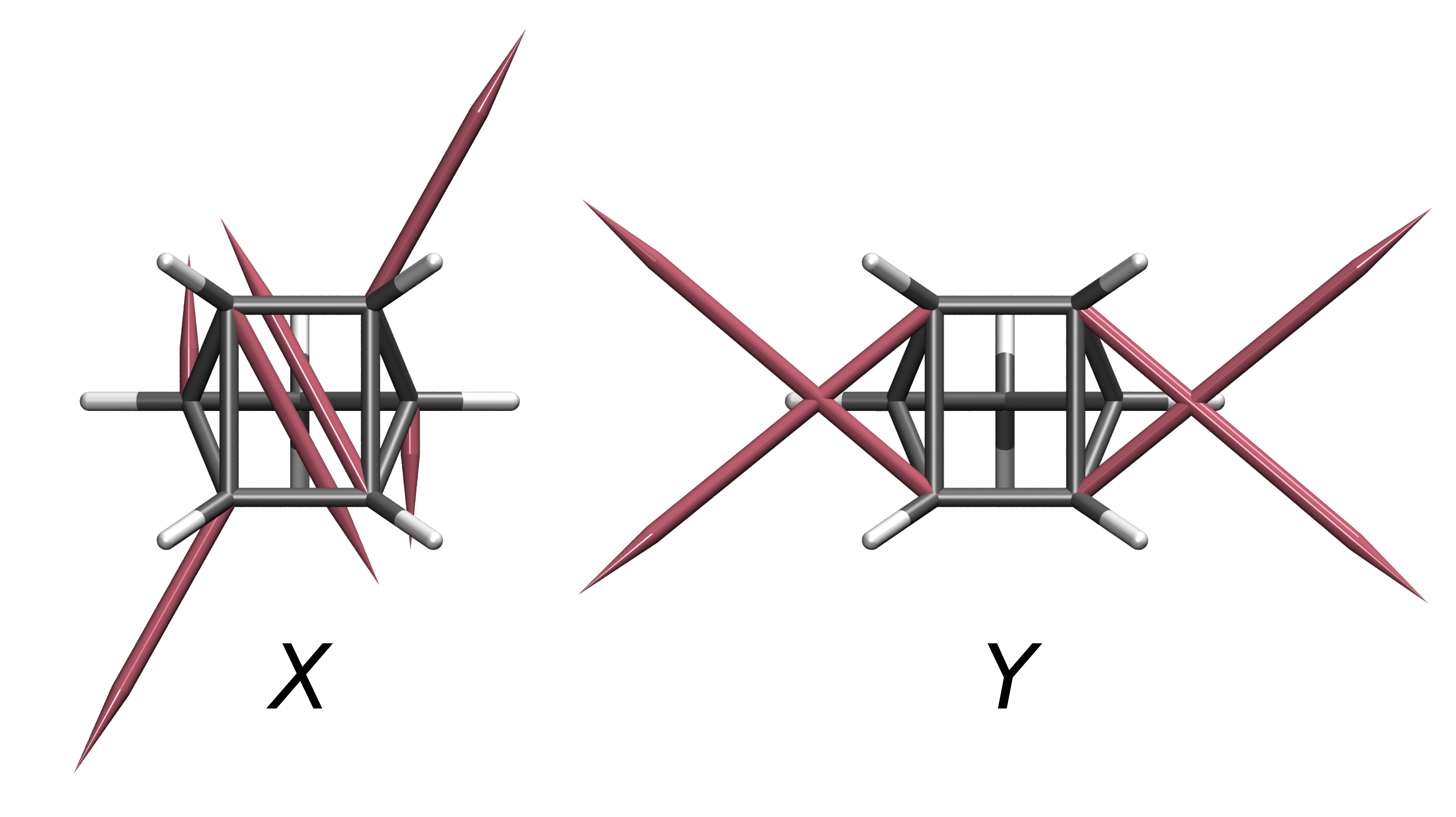}
    \caption{Branching plane vectors for the S$_2$/S$_1$ conical intersection, optimised using XMS-CASPT2(4,4)/p-cc-(p)VDZ. The $X$ vector is a rhombic displacement, which takes the molecule almost directly to the S$_1$/S$_0$ conical intersection, while the $Y$ vector is very similar to the displacement vector from NBD to QC. All vectors are almost entirely within the plane of the figure.}
    \label{fig:S2S1BP}
\end{figure}

Finally, we mentioned the presence of a S$_2$/S$_1$ conical intersection. When using XMS-CASPT2(4,4)/p-cc-(p)VDZ (it does not exist in CASSCF(2,2)), the intersection can be found at $(r_{\mathrm{cc}},r_{\mathrm{rh}},r_{\mathrm{db}})\approx(2.01,0,1.48)$ \AA, close to the S$_1$ minimum. Figure \ref{fig:S2S1meci} shows the branching plane energies for this intersection, calculated using the same method as in the main text\cite{fdez_galvan_analytical_2016}. It is clear that this is a peaked intersection with clear bifurcating character. By looking at the vectors in the branching plane (shown in Fig.\ \ref{fig:S2S1BP}), we can identify the $X$ direction as breaking the symmetry with a rhombic displacement (increasing magnitude of $r_{\mathrm{rh}}$), while the $Y$ direction is symmetric, decreasing $r_{\mathrm{cc}}$ and increasing $r_{\mathrm{db}}$. In short, the $X$ displacement moves you towards the S$_1$/S$_0$ conical intersection, while the $Y$ displacement moves you from NBD to QC. 

Using Fig.\ \ref{fig:S2S1meci}, it is clear that any dynamics which started in NBD (\textit{i.e.} coming from negative $Y$ along the lower S$_1$ surface) would almost certainly be split into two separate channels which go towards the two mirror-image copies of the S$_1$/S$_0$ conical intersection, and we expect minimal coupling in the dynamics. For clarity, we illustrate this with an example trajectory, drawn as a vector. This conical intersection is also present in the XMS-CASPT2(2,2) method, with almost identical energies and branching plane vectors. Notably, the conical intersection is more peaked, leading to less transfer to the S$_2$ state.

\section{Location of Rydberg states}

\begin{table}[H]
    \centering
    \caption{Summary of previous active spaces applied to excited states of this molecule. The nominal active space and basis set are given, with comments on which additional orbitals (all Rydberg) were included over the CASSCF(4,4) method. Coppola \textit{et al.}\cite{coppola_norbornadienequadricyclane_2023}  also performed some MS- and SS-CASPT2 calculations, and Hernandez \textit{et al.}\cite{hernandez_multiconfigurational_2023} some XMS-CASPT2.}
    \begin{tabular}{ccrr}
    \hline\hline
      Work                                                                  & Active space  &  Basis & Differences from (4,4)\\\hline
      Antol\cite{antol_photodeactivation_2013}&  (4,4)+3s     & cc-pVDZ + diffuse S & Additional auxiliary 3s\\
      Valentini \textit{et al.}\cite{valentini_selective_2020}              &  (4,8)   & aug-cc-pVDZ & 3s and three 3p\\
      Coppola   \textit{et al.}\cite{coppola_norbornadienequadricyclane_2023} &  (4,7)  & ANO-L-VDZP+1s1p1d & (4,3) with 3s and three 3p\\
      Hernandez \textit{et al.}\cite{hernandez_multiconfigurational_2023}   &  (4,7)  & ANO-S-VDZP        & 3s, 3p$_x$, 3p$_y$ \\ \hline\hline
    \end{tabular}
    \label{tab:other_studies}
\end{table}

\begin{figure}[H]
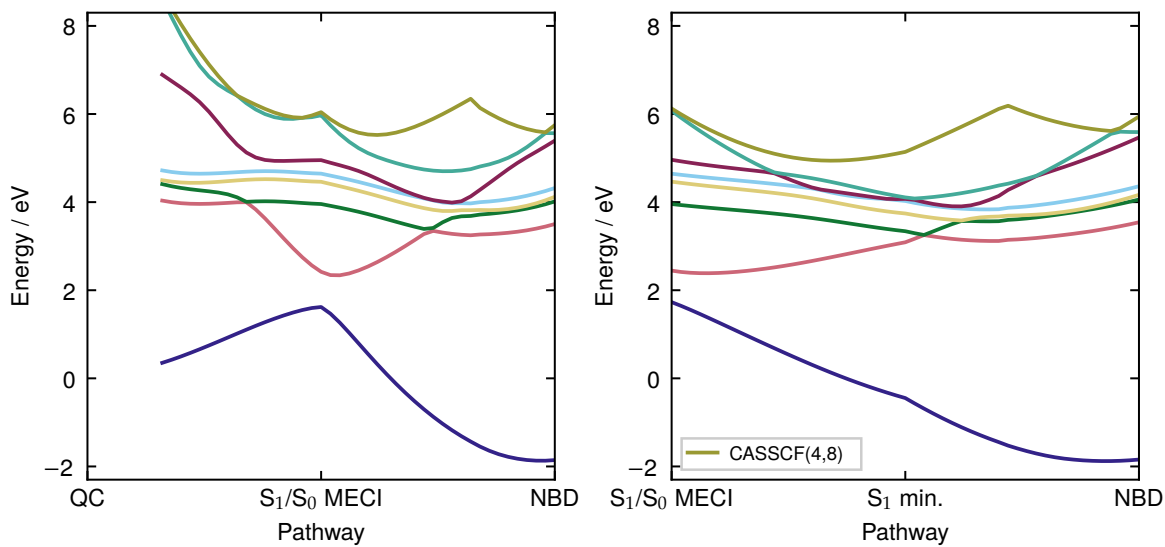

    \centering
    \input{SI_figures/ryd_casscf.pgf}\input{SI_figures/ryd_casscf_s1.pgf}
    \caption{SA(8)-CASSCF(4,8)/aug-cc-pVDZ Rydberg states. Left: QC $\leftrightarrow$ S$_1$/S$_0$ MECI $\leftrightarrow$ NBD LIIC. The important valence state (referred to as S$_1$ elsewhere) is far higher in energy, starting as the burgundy state at $\approx7.5$ eV above the ground state of NBD. It crosses with the doubly-excited state and moves through the Rydberg manifold (the flat states at $\approx4$ eV) before heading to the ground state. Right: S$_1$/S$_0$ MECI $\leftrightarrow$ S$_1$ minimum $\leftrightarrow$ NBD LIIC. Again, the important states start extremely high in energy. On the right half of this pathway, the geometry maintains $C_{2v}$ symmetry, and thus, the singly excited state crosses diabatically with the doubly-excited state. The doubly excited state crosses the Rydberg manifold around the S$_1$ minimum geometry and then mixes with the singly excited state to move closer to the intersection.}
    \label{fig:CASSCF_rydberg}
\end{figure}

As mentioned in the main text, all previous work on this system has used Rydberg states in the electronic structure. The Rydberg states are generally absent in applications, where the molecules are always in the condensed phase.

When calculating Rydberg states, dynamic correlation is essential. For example, we show SA(8)-CASSCF(4,8)/aug-cc-pVDZ, a non-dynamically correlated method very closely related to those used in previous studies.\cite{antol_photodeactivation_2013,valentini_selective_2020,hernandez_multiconfigurational_2023} A brief description of the methods used in those studies is shown in Table \ref{tab:other_studies}. The (4,8) active space contains all of the orbitals in the (4,4) active space used in this study \textit{and} additional 3s and three 3p Rydberg orbitals. We use the full aug-cc-pVDZ basis to ensure that all Rydberg states are equally described.

In Fig.\ \ref{fig:CASSCF_rydberg}, one can see that the low-lying valence state (called S$_1$ elsewhere in this work) is far above the Rydberg manifold, appearing first as the burgundy state at $\approx 7.5$ eV above the NBD ground state minimum, far away from the correct value should be $\approx 5.25$ eV. The overall shape of this state is consistent with the CASSCF(4,4) results (\textit{e.g.} in Fig.\ \ref{fig:MRCI}), showing that the description of the valence states is unaffected by the inclusion of the Rydberg states (a fact that also applies to the (2,2) active space). The CASSCF(4,8) thus describes these states as poorly as CASSCF(4,4).

Further, mixed Rydberg/valence systems are not well described by CASSCF. This is because the Rydberg states do not require as much correlation as the valence states, as the Rydberg electron is situated far away from the other electrons. CASSCF, which is not very correlated, therefore describes the Rydberg states better than the valence, which leads to the excitation energies of Rydberg states being \textit{too low}. In this CASSCF(4,8) method, we have a Rydberg manifold which starts at $\approx5.4$ eV above the NBD ground state, lower than its experimental value of closer to 6 eV.\cite{cooper_valence_2024}

The increase in energy of the valence state (caused by the poor description of the (4,4) active space) and the relative decrease in energy of the Rydberg states (by lack of dynamic correlation) causes the Rydberg states to sit far below the valence states, when they should sit far above. Any dynamics on the valence states will have to traverse the Rydberg manifold to reach the conical intersection, leading to extensive coupling into those states and, thus,  substantively different dynamics. Finally, we mention that it is also possible to not use diffuse functions in the basis, causing the Rydberg states to artificially rise in energy (perhaps even above the valence states). This is undesirable, as you are deliberately describing one part of the system (the Rydberg states) poorly in order to get a favourable outcome. Additionally, this adds a much larger amount of energy into the system, leading to different dynamics.

\begin{figure}[H]
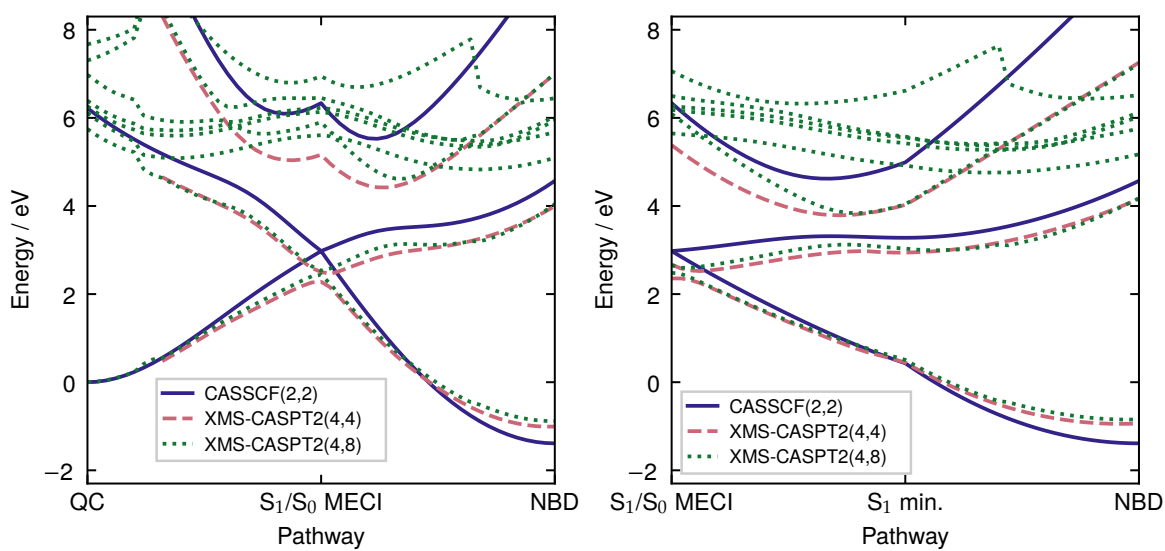

    \centering
    \input{SI_figures/rydberg_liic_norm.pgf}\input{SI_figures/rydberg_liic_s1.pgf}
    \caption{Potential energy cuts containing Rydberg states, for CASSCF(2,2) (solid purple) and XMS-CASPT2(4,4) (dashed rose), both using the p-cc-(p)VDZ basis, and XMS-CASPT2(4,8) (green dotted) using the aug-cc-pVDZ basis. Left: QC $\leftrightarrow$ S$_1$/S$_0$ MECI $\leftrightarrow$ NBD LIIC. The S$_1$ valence state (the lowest excited state in NBD) agrees between the (4,4) (without Rydberg) and (4,8) active spaces in XMS-CASPT2, indicating that the Rydberg states do not significantly affect the dynamics. This state is below the Rydberg manifold (the flat set of potentials at $\approx5$ eV). Right: S$_1$/S$_0$ MECI $\leftrightarrow$ S$_1$ minimum $\leftrightarrow$ NBD LIIC. Again, the important S$_1$ state is always below the Rydberg manifold, agreeing well between the (4,4) and (4,8) active spaces.}
    \label{fig:Rydberg_liic}
\end{figure}

We can correct these issues by adding correlation. In Figure \ref{fig:Rydberg_liic}, we show the potential energy surfaces for both LIICs considered in this study for XMS(8)-CASPT2(4,8)/aug-cc-pVDZ, which adds dynamic correlation onto the SA(8)-CASSCF(4,8) method, and compare with CASSCF(2,2) and XMS-CASPT2(4,4). As can be seen, the low-lying S$_1$ (1$A_2$) state agrees well with all methods and is far below the Rydberg manifold, the set of flat states centred at around 6 eV above the NBD minimum. The inclusion of Rydberg states thus does not significantly affect the dynamics on the low-lying S$_1$ state, as they are well separated in energy in all important geometries, and we can safely remove them from consideration.

Although unsubstituted NBD contains Rydberg states, which are primarily excited in experiments, there is Rydberg-free dynamics upon excitation at $<6$ eV. This more closely models the dynamics that would be seen in a practical application, where substituted systems lower the energy of the valence states even further, completely removing all trace of Rydberg excitation from the system.

\section{MRCI}

 The truncation of  CI calculations leads to size-consistency errors, and a multitude of corrections exist to approximate the inclusion of higher excitations.\cite{siegbahn_multiple_1978,davidson_size_1977,pople_variational_1977,szalay_multiconfiguration_2012} In MRCI calculations with multiple states, these corrections are critical, as the approximate quality of a CI calculation is related to the reference weight $c_0$, the coefficient of the wavefunction included to the reference configurations. A helpful analogy is to relate $c_0$ to the `amount of work' the MRCI calculation has to perform. A high $c_0$ indicates that the MRCI calculation has to compensate for less deficiency in the CASSCF wavefunction compared to lower $c_0$ values. As a consequence, MRCI calculations based on biased active spaces tend to remain biased. 

\begin{figure}[H]
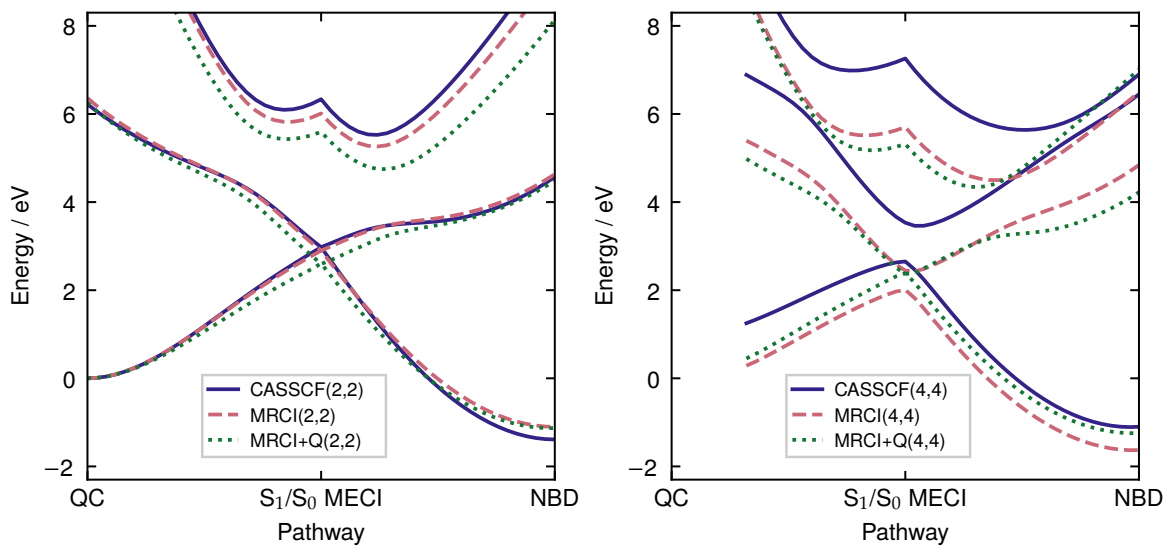

    \centering
    \input{figures/22.pgf}\input{figures/44.pgf}
    \caption{Left: SA(3)-CASSCF (solid, indigo) vs MRCI (dashed, rose) vs MRCI+Q (dotted, green), with both (2,2) (left) and (4,4) right active spaces. All calculations are performed with the p-cc-(p)VDZ basis set. For the (2,2) active space, the MRCI changes the description for S$_0$ around the NBD minimum, but qualitative agreement is seen elsewhere, even with the Davidson correction. For the (4,4) active space, the three methods do not agree even qualitatively, with a notable $\approx$ 2 eV increase in excitation energy in NBD. Notably, the MRCI+Q(4,4) calculations give a similar shape to all three (2,2) active space methods. Calculations not shown for QC, where the (4,4) active space is unstable.}
    \label{fig:MRCI}
\end{figure}

In Fig.\ \ref{fig:MRCI}, we show the CASSCF, MRCI and MRCI+Q results for the QC $\leftrightarrow$ S$_1$/S$_0$ MECI  $\leftrightarrow$  NBD LIIC for the (2,2) (left) and the (4,4) (right) active space. In the (2,2) active space, the reference weight is relatively constant for all the states, and so the MRCI and MRCI+Q agree closely on the shape of the potential energy surface. We also see that the CASSCF(2,2) results are also fairly close to the MRCI results. On the other hand, the (4,4) active space shows no signs of agreement. This is because the reference weight is much lower for S$_1$ than the other two states around NBD. This can also be seen in Table \ref{tab:vees}, where the CASSCF(4,4) S$_1$ excitation energy is far higher (7.55 eV) than the other methods. As we include dynamic correlation with MRCI, the result improves (to 6.47 eV) but is still not close to the experimental value of 5.25 eV. Only by including the Davidson corrections do we get closer, moving to 5.47 eV (DV3, see later), in acceptable agreement with experiment. 

We use the terminology of the COLUMBUS program\cite{szalay_multiconfiguration_2012,lischka_generality_2020} in defining the size-consistency corrections. For clarity, we repeat them here.

\vspace{0.5em}
DV1: $\displaystyle E_{\mathrm{DV1}} = (1-c_0^2)E_{\mathrm{corr}},$

DV2: $\displaystyle E_{\mathrm{DV2}} = \frac{(1-c_0^2)}{c_0^2}E_{\mathrm{corr}},$

DV3: $\displaystyle E_{\mathrm{DV3}} = \frac{(1-c_0^2)}{2c_0^2-1}E_{\mathrm{corr}},$

\begin{figure}[H]
    \centering
    \input{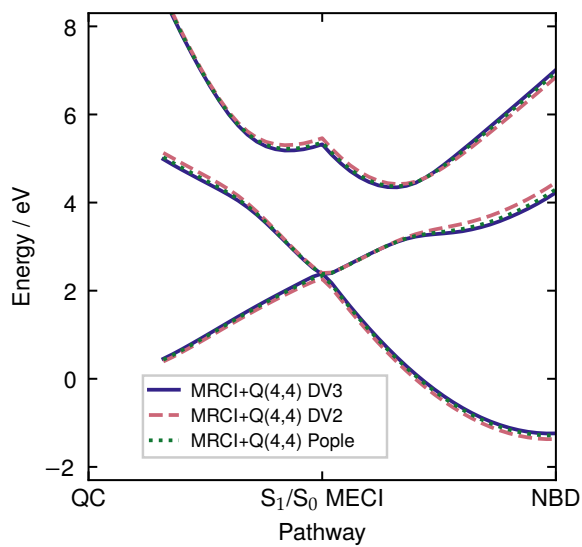}
    \caption{MRCI+Q(4,4) Davidson corrections, calculated using the p-cc-(p)VDZ basis set. The DV3 (solid purple) and Pople (dotted green) corrections agree excellently, while the DV2 slightly deviates, especially on the first excited state of NBD.}
    \label{fig:DV}
\end{figure}

Pople: $\displaystyle E_{\mathrm{Pople}} = \frac{\sqrt{N^2+2N\mathrm{tan}^2(2\theta)} - N}{2(\mathrm{sec}(2\theta)-1)}E_{\mathrm{corr}},$
\vspace{0.5em}

\noindent where $c_0$ is the reference weight, $E_{\mathrm{corr}} = E_{\mathrm{MRCI}}-E_{\mathrm{CASSCF}}$, $\theta=\mathrm{arccos}(c_0)$, and $N$ is the number of electrons. We compare DV3, DV2, and the Pople correction in Fig.\ \ref{fig:DV}. There are no huge deviations in these calculations, and all corrections closely replicate the overall shape of the potentials.

\begin{figure}[H]
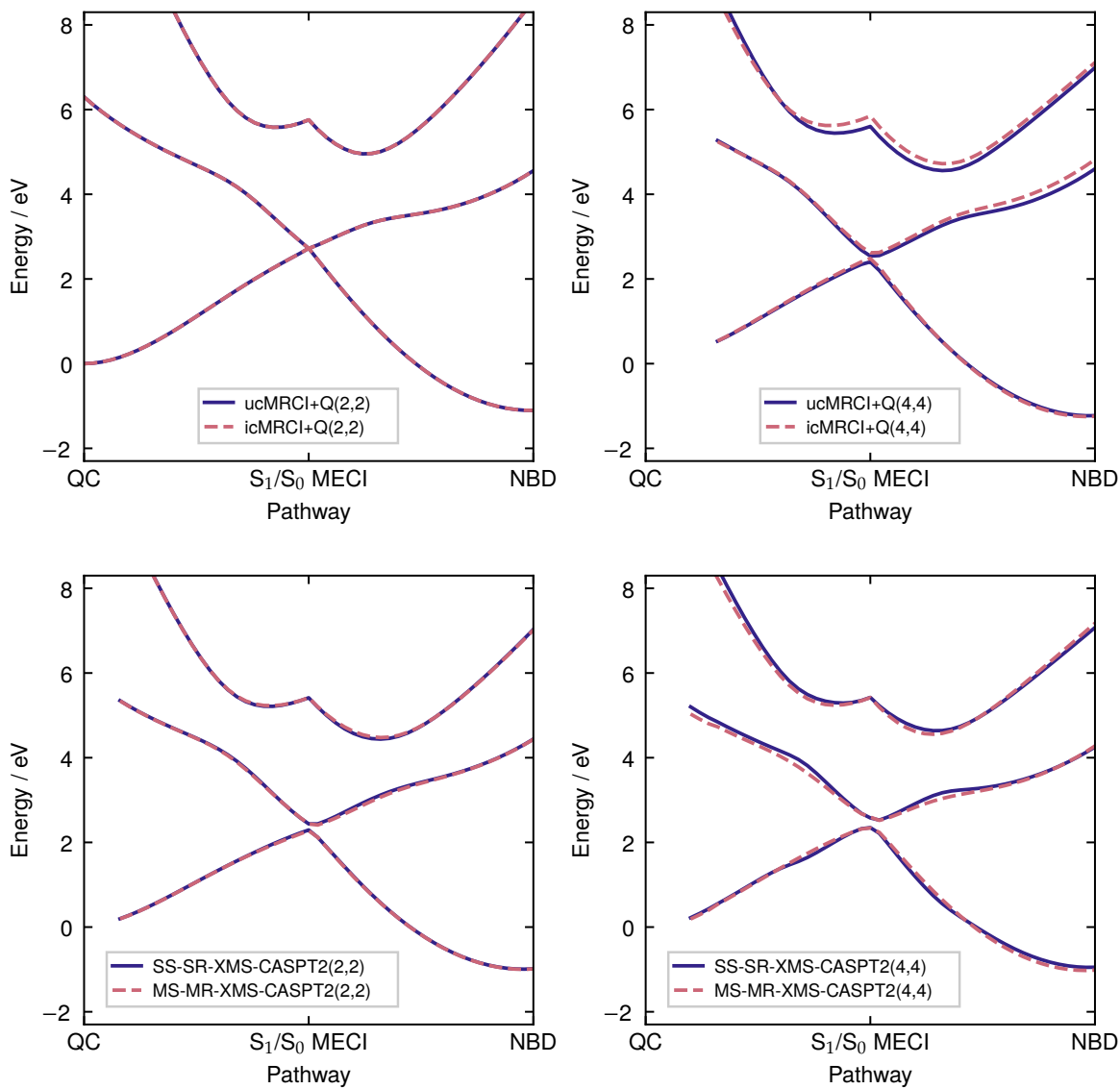

    \centering
    \input{SI_figures/22_ucic.pgf}\input{SI_figures/44_ucic.pgf}
    \input{SI_figures/22_sssr.pgf}\input{SI_figures/44_sssr.pgf}
    \caption{Multi-reference formalisms. Potential energy cuts for contracted and uncontracted MRCI schemes, for the QC $\leftrightarrow$ S$_1$/S$_0$ MECI $\leftrightarrow$ NBD LIIC. Upper left: uc-MRCI+Q(2,2) (solid purple) and ic-MRCI+Q(2,2) (dashed rose) potentials. Internal contraction makes no difference to the shape of the potential energy surfaces. Upper right: uc-MRCI+Q(4,4) (solid purple) and ic-MRCI+Q(4,4) (dashed rose) potentials. The internal contraction gives different energies from the uncontracted calculations, especially for the states with significant $|2\mathrm{ud}0\rangle$ character. Lower right: XMS-CASPT2(2,2) calculations. The contraction has a minimal effect, as seen in the equivalent MRCI calculations. Lower right: XMS-CASPT2(4,4) calculations. A small effect is seen, but here, it affects all states. Geometries are from MRCI(2,2)/p-cc-(p)VDZ LIIC. Calculations with convergence/intruder state issues are not shown. (2,2) CASPT2 calculations use a real shift of 0.2 $\mathrm{E_h}$, (4,4) use 0.4 $\mathrm{E_h}$. All calculations use p-cc-(p)VDZ basis. ucMRCI is performed in COLUMBUS 7.6, SS-SR-XMS-CASPT2 in OpenMolcas v23.2, and icMRCI and MS-MR-XMS-CASPT2 in MOLPRO 2022.2.}
    \label{fig:ucvsic}
\end{figure}

The implementation of MRCI in COLUMBUS is `uncontracted' (ucMRCI), which includes all single and double excitations from each determinant individually. Other MRCI implementations, such as those in ORCA 5.0.4 \cite{neese_orca_2020} and Molpro 2022.2 \cite{werner_molpro_2020}, use a `internally contracted' formalism (icMRCI), which only applies the excitation to the entire wavefunction. This leads to fewer variational parameters and, thus, a more approximate wavefunction. The computational saving is often large but is less obvious when dealing with smaller active spaces, as we do here. 

In this case, as shown in Fig.\ \ref{fig:ucvsic} (upper two panels), the difference is dependent on the active space. For the (2,2) active space, the uncontracted formalism gives effectively identical curves. For the (4,4) active space, the difference is much larger. Interestingly, this difference is specific to the second state, only affecting the state with large $|2\mathrm{ud}0\rangle$ character around NBD. This might be rationalised as a consequence of the poor description of the second state - the icMRCI method, which excites from the reference wavefunction, does not have the full flexibility needed to describe the final state. The (2,2) active space, which gives a more equal description of the state, gives little qualitative different between the two contraction schemes.

It should be noted that a similar problem exists in CASPT2. The split is between the `multi-state multi-reference' (MS-MR), referring to a formalism which calculates all states together, and `single-state single-reference' (SS-SR), a formalism which combines all of the states separately after calculation (which decreases computational time). Previous work has shown that these approximations are less important than they are in MRCI.\cite{sen_comprehensive_2018} We show the results for the different approximations in Fig.\ \ref{fig:ucvsic} (lower panels). In XMS-CASPT2(2,2), the effect of contraction has a far less obvious effect, while there is at least a small difference in XMS-CASPT2(4,4), in line with the results of MRCI. The SS-SR calculations come from the OpenMolcas program (as used throughout), while the MS-MR calculations come from Molpro. As the imaginary shift used previously is not implemented in Molpro, a real shift is used  (0.2 $\mathrm{E_h}$ for the (2,2) active space, 0.4 $\mathrm{E_h}$ for the (4,4)). The real shift is known to cause its own issues with intruder states, and this is seen around the ground state minimum of QC, which is not shown on the figure.

\section{SHCI and LR-CCSD}

In the main text, we compare our multi-configurational methods with LR-CC3, a very high-level method for calculating excitation energies from a coupled-cluster wavefunction. Here, we show two more non-active space methods: SHCI, which is a variant of selected configuration interaction, and LR-CCSD, which which calculates the excitation energies from a CCSD ground state. 

\begin{figure}
    \centering
    \input{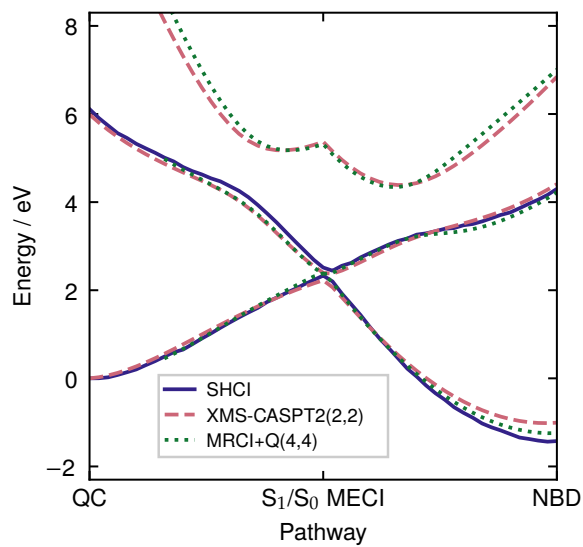}
    \caption{SHCI(2,2) (solid, purple) vs XMS-CASPT2(2,2) (dashed, rose) vs MRCI+Q(4,4) (dotted, green) calculations, all for p-cc-(p)VDZ basis. The agreement is strong across all geometries, with only a small deviation in excitation energies. All calculations on MRCI geometries (see Methods).}
    \label{fig:LIIC_final}
\end{figure}

Selected configuration interaction (CI) methods approximate full CI by attempting to include only important configurations in the expansion. Unlike traditional truncated CI, such as CISD, they do limit the degree of excitation, but rather include configurations based on an importance criterion, effectively ignoring determinants that do not contribute significantly. The most important advantage is their lack of human bias; they do not require the selection of an active space and generally provide a comparatively even-handed description of all electronic states considered. While many flavours of selected CI exist, we use heat-bath CI (HCI),\cite{holmes_excited_2017} which is known to be computationally efficient. We perform the calculations in a large active space that consists of 20 electrons in 80 orbitals since all-electron calculations are not feasible. This active space is chosen to ensure that all low-energy orbitals are included. Furthermore, the contribution from determinants not selected is estimated using a stochastic perturbation theory.\cite{holmes_excited_2017}

We only calculate the ground and first excited ($|2\mathrm{ud}0\rangle$) states using SHCI. As seen in Fig.\ \ref{fig:LIIC_final}, the SHCI results agree well with MRCI+Q(4,4), confirming that the active space procedure is reasonably unbiased. It is worth noting that as well as providing an unbiased reference, the selected CI methods are also interesting candidates for 'black-box' electronic structure methods for dynamics, with a good mixture of static and dynamic correlation and relatively straightforward gradient and non-adiabatic coupling implementations.\cite{coe_analytic_2023} Currently, their usage is limited by their computational costs, but ongoing improvements in theory and algorithms could make them ubiquitous.

\begin{figure}[H]
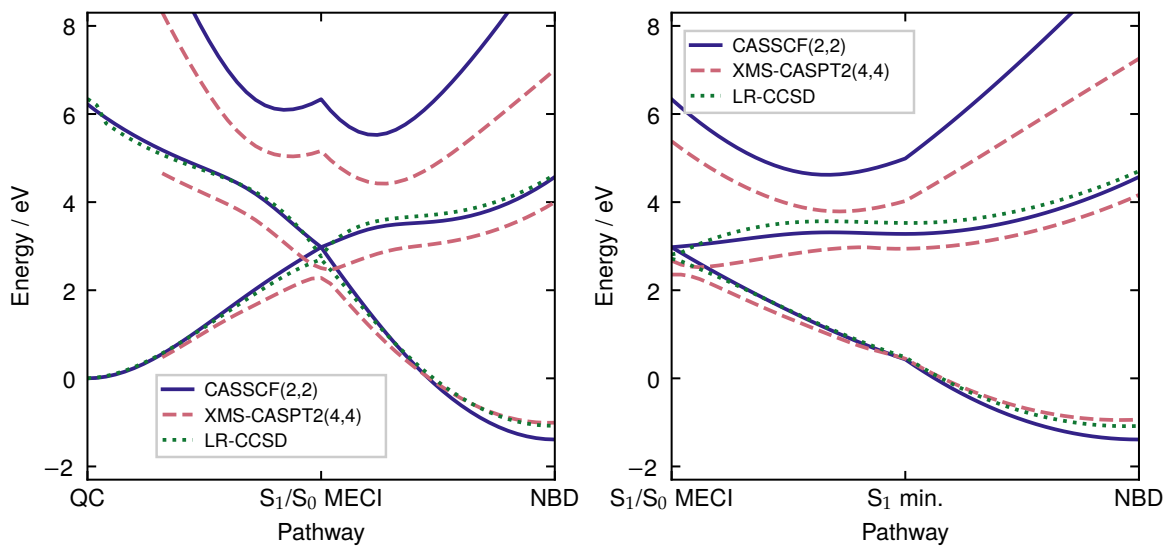

     \centering
     \input{SI_figures/lr-ccsd_l1.pgf}\input{SI_figures/lr-ccsd_l2.pgf}
     \caption{Potential energy cuts for LR-CCSD, contrasted against the CASSCF(2,2) and XMS-CASPT2(4,4). Left: QC $\leftrightarrow$ S$_1$/S$_0$ MECI $\leftrightarrow$ NBD LIIC. The LR-CCSD agrees very well with the CASSCF(2,2), but shows a slightly different shape to the XMS-CASPT2(4,4) on the excited state. Notably, the gradients around the MECI are slightly too steep. Right: S$_1$/S$_0$ MECI $\leftrightarrow$ S$_1$ minimum $\leftrightarrow$ NBD LIIC. Again, the LR-CCSD agrees fairly well with CASSCF(2,2) and XMS-CASPT2(4,4).}
     \label{fig:LIIC_22_BSE}
 \end{figure}

Finally, we show LR-CCSD.\cite{folkestad_et_2020,musial_equation--motion_2020} Unfortunately, LR-CCSD is limited to only describing singly-excited states well --- this system, which contains important doubly-excited states, is not well described by it. This can be seen in Fig.\ \ref{fig:LIIC_22_BSE}, which shows the LR-CCSD with a notably steep gradient around the intersection in both LIICs. Other single-reference correlated excited state methods (e.g. ADC(2) and CC2) only give worse results, as the double excitation crucial to the description is included at an even lower level, if at all.

\begin{figure}[H]
    \centering
    \input{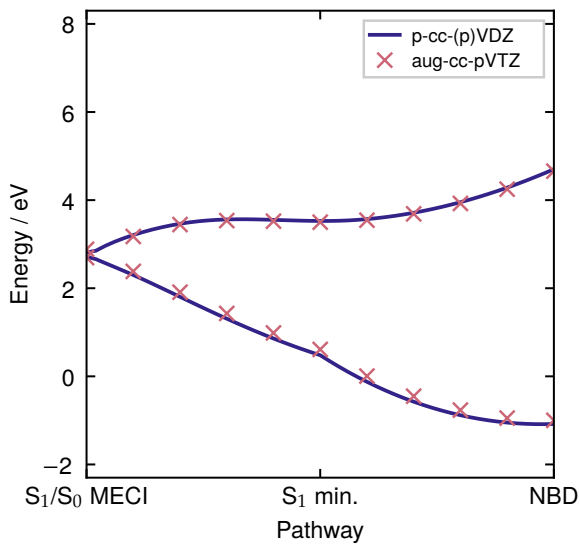}
    \caption{Basis sets for LR-CCSD, using the S$_1$/S$_0$ MECI $\leftrightarrow$ S$_1$ minimum $\leftrightarrow$ NBD LIIC, the overall agreement between the p-cc-(p)VDZ and aug-cc-pVTZ basis sets is good, indicating that the p-cc-(p)VDZ is at least somewhat converged.}
    \label{fig:ccsd_basis}
\end{figure}

As a final check, we show the LR-CCSD calculations performed with our p-cc-(p)VDZ basis and the larger aug-cc-pVTZ basis in Fig.\ \ref{fig:ccsd_basis}. We do this primarily to show that our new contraction, p-cc-(p)VDZ, is close to convergence, as coupled cluster methods are notorious for needing large basis sets to get good agreement. The shape of the potentials is consistent between the two methods, implying that our basis is appropriate. These calculations were not feasible for LR-CC3, but we assume the trend is approximately continued, and we can trust that the current potentials are close to correct.

\subsection{Computational details for SHCI}

 Stochastic heat bath configuration interaction (SHCI) calculates were performed using the PySCF\cite{sun_recent_2020}/DICE\cite{holmes_heat-bath_2016,holmes_excited_2017,sharma_semistochastic_2017,smith_cheap_2017} interface. The SHCI used the pseudo-canonical orbitals from SA(3)-CASSCF(2,2) calculations to ensure that orbitals are smooth along the LIICs, and then performs the SHCI within a (20,80) active space. Expanding the active space is not computationally feasible for the entire pathways but gave similar relative energy gaps at test points. The variational portion was converged to $\epsilon_{1}=10^{-4}$, and the stochastic perturbative correction used 30 samples of 200 variationally chosen configurations, with $\epsilon_{2} = 10^{-7}$. The stochastic portion gave errors of less than $5\times10^{-4}\;\mathrm{E_h} \approx 0.01$ eV, which suffices for comparison --- note, this is \textit{not} the error with respect to FCI, but the error in the stochastic sampling. SHCI calculations for the QC$\leftrightarrow$S$_1$/S$_0$ MECI$\leftrightarrow$NBD were performed simultaneously for three states, with the second (triplet) solution discarded, while calculations for the S$_1$/S$_0$ MECI$\leftrightarrow$S$_1$ min.$\leftrightarrow$NBD were performed with time-reversal symmetry to remove the triplet state.

\section{Other multi-reference methods}

\begin{figure}[H]
    \centering
    \input{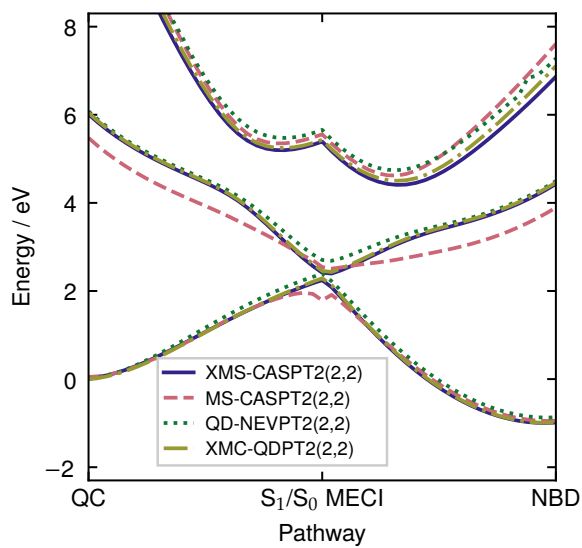}
    \caption{Potential energy curves for XMS-CASPT2 (purple solid), MS-CASPT2 (rose dashed), QD-NEVPT2 (green dotted), XMC-QDPT2 (sand dashdot lines), all based on a SA(3)-CASSCF(2,2)/p-cc-(p)VDZ reference. The QD-NEVPT2, XMC-QDPT2 and XMS-CASPT2 agree well, with only a small and roughly constant energy difference between the two curves. The MS-CASPT2, on the other hand, seems to get the overall shape of the potential energy surface wrong, at least compared to XMS-CASPT2, which compares favourably with the MRCI+Q(4,4).}
    \label{fig:SI_pt2}
\end{figure}

There are many ways to add correlation to a CASSCF calculation. We utilised MRCI and XMS-CASPT2, but here we show additional calculations performed using MS-CASPT2, QD-NEVPT2 and XMC-QDPT2 (three different flavours of quasi-degenerate multi-reference perturbation theory). We compare it to XMS-CASPT2(2,2), which gave good agreement to the MRCI+Q(4,4) calculations. The vertical excitation values are shown in Table \ref{tab:vees}, where QD-NEVPT2 and XMC-QDPT2 show very similar values to XMS-CASPT2 with the (2,2) active space. MS-CASPT2, on the other hand, shows a remarkably low S$_1$ vertical excitation energy of 4.85 eV. These trends are continued in Fig.\ \ref{fig:SI_pt2}, with QD-NEVPT2 and XMC-QDPT2 both showing exceptional agreement. MS-CASPT2, on the other hand, is remarkably lower across the potential. Similar values are also seen in the results of Coppola \textit{et al}.,\cite{coppola_norbornadienequadricyclane_2023} as well as our own previous work.\cite{borne_ultrafast_2024} We noticed that the agreement of MS-CASPT2 improved when including the IPEA shift with its standard value of 0.25 E$_{\mathrm{h}}$.

All calculations in Fig.\ \ref{fig:SI_pt2} use the p-cc-(p)VDZ basis and the (2,2) active space. The QD-FIC-NEVPT2 calculations were performed using the ORCA 5.0.4 program.\cite{neese_orca_2020} XMC-QDPT2(2,2) calculations were performed in Firefly.\cite{granovsky_firefly_nodate} Some minor deviations are seen around the NBD, but this is likely to do with both the different implementations of the level-shift (0.2$i$ in XMS-CASPT2 vs. ISA=0.02 in XMC-QDPT2). MS-CASPT2 calculations were performed in OpenMolcas.\cite{aquilante_modern_2020}

\begin{figure}[H]
    \centering
    \input{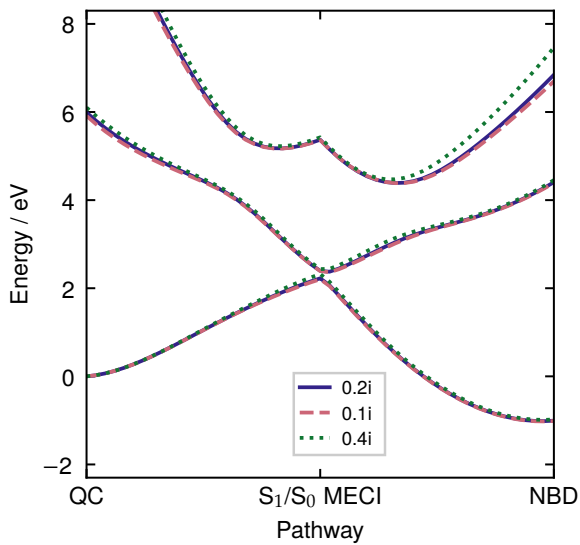}
    \caption{Potential energy cuts for XMS-CASPT2(2,2) with different shift values, all using p-cc-(p)VDZ basis. As the shift decreases, the slope of the doubly excited S$_2$ $|2020\rangle$ state around NBD gets slightly shifted. All calculations in the text use a value of 0.2$i$.}
    \label{fig:shift}
\end{figure}

Finally, in Fig.\ \ref{fig:shift}, we show the potential energy cuts using different values of the imaginary shift. Clearly, the value of the shift affects the energy of the S$_2$ state around the NBD minimum (vertical excitation energies can be found in Table \ref{tab:vees}). In general, a lower shift gives a lower excitation energy of S$_2$. Unfortunately, upon using a 0.1$i$ shift, we found minor issues with intruder states. Therefore, we use 0.2$i$ for the rest of the work as an acceptable compromise between accuracy and stability.

\section{Basis sets}

\begin{table}
\centering
\caption{Primitives and contractions for the p-cc-(p)VDZ basis set. This is a effectively a pruned aug-cc-pVDZ basis set, removing all diffuse and polarisation functions off of hydrogen atoms (a [4s$|$2p] contraction), and the S and D angular momentum diffuse functions off of carbon atoms (a [9s5p1d$|$3s3p1d] contraction). Zero values are not shown.}\label{tab:basis_contractions}
\begin{tabular}{llrrrr}
\hline\hline
\textbf{Atom} & \textbf{Ang. mom.} & \textbf{Primitives} & \textbf{Contractions} &           &   \\\hline\hline
\textbf{H}    & \textbf{S}         & 13.01               & 0.019685              &           &   \\
              &                    & 1.962               & 0.137977              &           &   \\
              &                    & 0.4446              & 0.478148              &           &   \\
              &                    & 0.122               & 0.50124               & 1         &   \\\hline
\textbf{C}    & \textbf{S}         & 6665                & 0.000692              & -0.000146 &  \phantom{-0.000146} \\
              &                    & 1000                & 0.005329              & -0.001154 &   \\
              &                    & 228                 & 0.027077              & -0.005725 &   \\
              &                    & 64.71               & 0.101718              & -0.023312 &   \\
              &                    & 21.06               & 0.27474               & -0.063955 &   \\
              &                    & 7.495               & 0.448564              & -0.149981 &   \\
              &                    & 2.797               & 0.285074              & -0.127262 &   \\
              &                    & 0.5215              & 0.015204              & 0.544529  &   \\
              &                    & 0.1596              & -0.003191             & 0.580496  & 1 \\\cline{2-6}
              & \textbf{P}         & 9.439               & 0.038109              &           &   \\
              &                    & 2.002               & 0.20948               &           &   \\
              & \textbf{}          & 0.5456              & 0.508557              &           &   \\
              &                    & 0.1517              & 0.468842              & 1         &   \\
              &                    & 0.04041             &                       &           & 1 \\\cline{2-6}
              & \textbf{D}         & 0.55                & 1                     &           &   \\\hline\hline
\end{tabular}
\end{table}

For clarity, we include a table of the basis set primitives and contractions in Table \ref{tab:basis_contractions}. The validity of this basis set can be seen by comparison to the ANO-L-VQZP basis. The ANO-L basis has a particularly large set of primitive functions, and the truncation at quadruple-zeta-plus-polarisation leads to fantastic energies and properties. A similar effect can be seen in the earlier comparison for LR-CCSD, shown in Fig.\ \ref{fig:ccsd_basis} and the associated discussion.

\begin{figure}
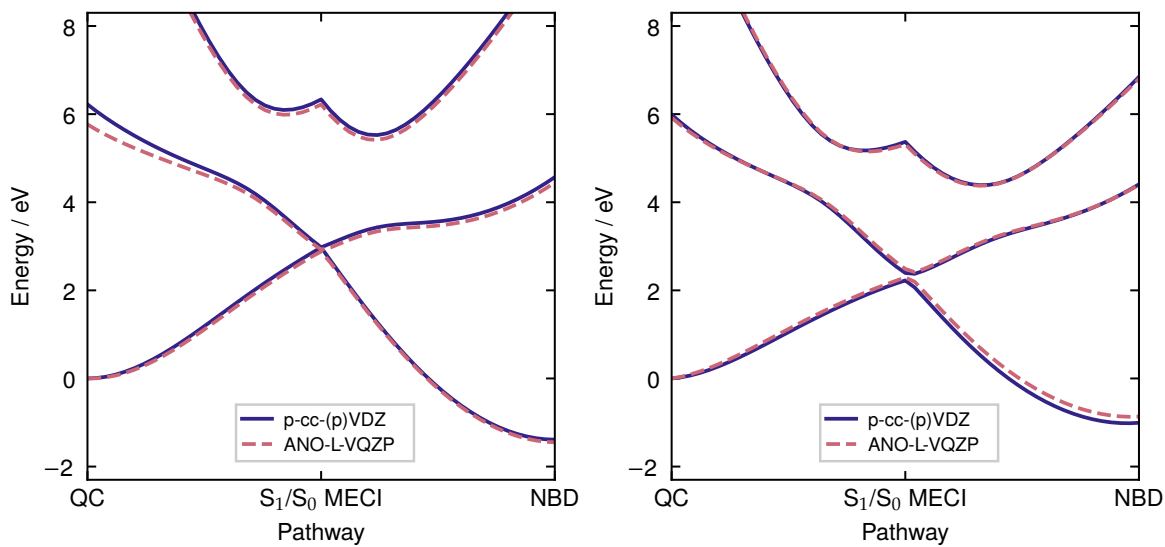

    \centering
    \input{SI_figures/cas_basis_set.pgf}\input{SI_figures/pt2_basis_set.pgf}
    \caption{Potential energy cuts calculated using p-cc-(p)VDZ basis (solid rose) vs ANO-L-VQZP (dashed green) for SA(3)-CASSCF(2,2) (left) and XMS-CASPT2 (right). CASSCF calculations performed on CASSCF geometries, and CASPT2 calculations on MRCI geometries.}
    \label{fig:casbasis}
\end{figure}

First, we can look at the potential energy surfaces in both CASSCF and XMS-CASPT2 in Fig.\ \ref{fig:casbasis}. The potential cuts show excellent agreement between the two methods, especially in the correlated XMS-CASPT2 calculations. Furthermore, we show the conical intersection parameters shown in Table \ref{tab:SI_MECIs}, which mimics Table 2 in the main text. The basis set change gives minor changes to both the structure and the potential energy surface around the intersection, but overall the methods are well converged. Finally, the structures of the ground state minima (see Table \ref{tab:SI_gs_geoms}) show close agreement, and further good agreement to a previous study using large basis sets MP2 calculations.

\begin{table}
    \centering
    \caption{Conical intersection parameters. $P$ and $B$ parameters and carbon-carbon distances (in \AA ngstr\"oms for optimised S$_1$/S$_0$ MECI for the (2,2) active space for SA(3)-CASSCF, XMS-CASPT2 (with 0.2$i$ shift), for the p-cc-(p)VDZ and ANO-L-VTZp basis set (ANO-L-VTZP on the carbons and ANO-L-VTZ on the hydrogens). The conical intersections all have a $C_2$ optimised geometry.}
    \begin{tabular}{lllllllll}
    \hline\hline
            Basis set & Method & $P$ & $B$ & $r_{\mathrm{cc}}$ & $r_{\mathrm{db}}$ & $r_{\mathrm{rh}}$\\ \hline\hline
p-cc-(p)VDZ   &    CASSCF      &  0.789 & 0.857 & 1.943 & 1.443 & 0.499  \\ 
              &    XMS-CASPT2  &  0.560 & 1.168 & 2.011 & 1.490 & 0.555  \\\hline
ANO-L-VTZ(p)  &    CASSCF      &  0.768 & 0.834 & 1.937 & 1.435 & 0.484  \\
              &    XMS-CASPT2  &  0.511 & 1.082 & 1.996 & 1.474 & 0.530  \\\hline\hline
    \end{tabular}
    \label{tab:SI_MECIs}
\end{table}

\begin{figure}[H]
    \centering
    \input{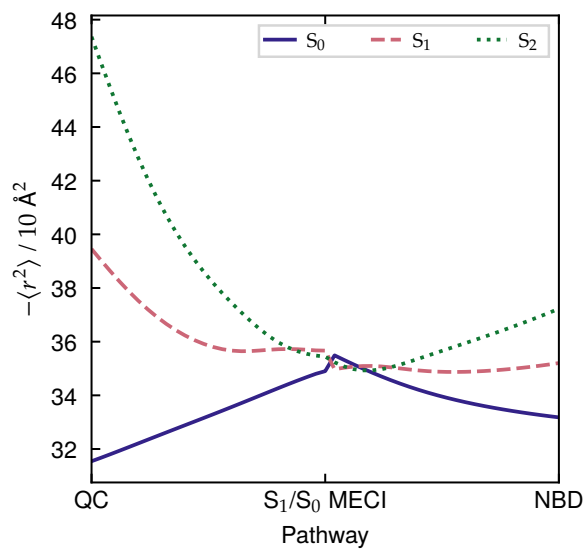}
    \caption{$-\langle r^2\rangle$  for three analysed states SA(3)-CASSCF(2,2)/p-cc-(p)VDZ calculations. Values increasing indicate increasing electronic diffuseness. The $\langle r^2\rangle$ values significantly increase for the two excited states around the QC ground state minimum. At the conical intersection, the states change character, leading to discontinuities in the curves.}
    \label{fig:r2}
\end{figure}

The primary difference between the descriptions for different basis sets is caused by the diffuse nature of the states, which is evident from the radial second moment of the charge for each state, given as $\langle r^2\rangle = \langle x^2\rangle + \langle y^2\rangle + \langle z^2\rangle$. Here, the integration is performed over both the electrons and the nuclei to cancel out effects due to the geometry change. In Fig.\ \ref{fig:r2}, the diffuseness, which increases going up the $y$-axis, for the two excited states can be seen to increase significantly around the QC ground state minimum. Here, the orbital characters change such that the $\mathrm{B_1}$ (see Fig.\ \ref{fig:44orbs}) orbital gains considerable Rydberg character. S$_2$, which is a state of $|2200\rangle$ character, is therefore even more diffuse. A smaller (but similar) effect can be seen around the NBD minimum. Incorrectly dealing with this diffuseness (\textit{e.g}.\ by using an insufficiently diffuse basis) leads to a severe overestimation of excitation energies into these states, as seen in Fig.\ \ref{fig:casbasis}.

\section{Nature of the potential energy surfaces}

\begin{figure}[H]
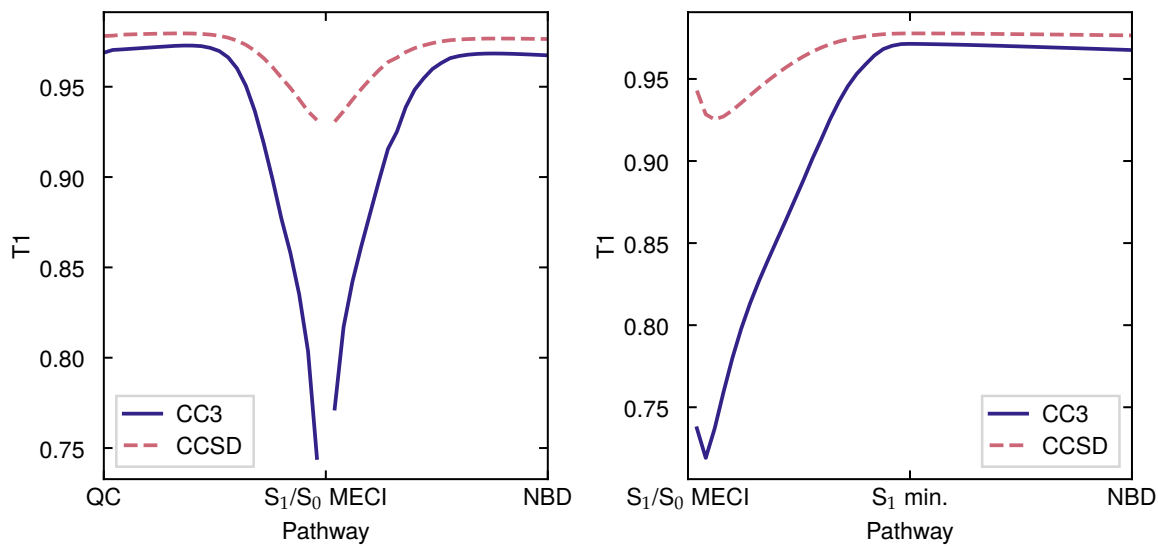

    \centering
    \input{SI_figures/T1_liic_1.pgf}\input{SI_figures/T1_liic_2.pgf}
    \caption{Fraction of single excitations (T1) for the S$_1$ LR-CC3 and LR-CCSD wavefunctions. Left: QC $\leftrightarrow$ S$_1$/S$_0$ MECI $\leftrightarrow$ NBD LIIC. The singly-excited character drops around the conical intersection, but much more noticeably for LR-CC3 than LR-CCSD. This spike at the conical intersection indicates the poor description. Right: S$_1$/S$_0$ MECI $\leftrightarrow$ S$_1$ minimum $\leftrightarrow$ NBD LIIC. Again, the doubly-excited character is highest at the conical intersection. Points not shown for the S$_1$/S$_0$ MECI, as the LR-CC wavefunction shows convergence issues.}
    \label{fig:T1}
\end{figure}

As mentioned, the S$_1$ state gains double excitation character around the S$_1$/S$_0$ conical intersection. We use the fraction of single excitations (T1), the sum of all singly-excited amplitudes. Lower values indicate more doubly-excited character.\cite{plasser_new_2014,do_casal_classification_2023}. This provides a good estimate of the doubly-excited character of the system. We show this using LR-CC3 and LR-CCSD S$_1$ states, shown in Fig.\ \ref{fig:T1}. As is clear, the doubly-excited character is very high around conical intersection. Notably, the LR-CCSD values are higher than the LR-CC3 values, in accordance with the poor description seen in Section 8. More approximate methods, such as LR-CC2 and ADC(2), would give even worse descriptions. 

\section{Summary of Calculations}

\begin{table}[H]
    \small
    \centering
    \caption{Summary of methods evaluated in this study. In order, the columns indicated: the active space, whether they are stable in QC, whether they can be used in a dynamics simulation, whether their S$_1$/S$_0$ CI has bifurcating character and whether they show a local minimum on the S$_1$ surface. N/A indicates that the calculation has not been performed.}
    \begin{tabular}{p{2.65cm}p{1.1cm}p{2.3cm}p{1.8cm}p{2.0cm}p{1.5cm}p{1.2cm}} 
    \hline\hline Method & $(m,n)$  &  Stable in QC? & Dynamics? & Bifurcating? & S$_1$ min.? & S$_2$/S$_1$? \\\hline
CASSCF      & (2,2)   & Yes & Yes  & Yes  & Yes & No   \\
            & (4,4)   & No  & Yes  & Yes  & Yes & Yes* \\\hline
XMS-CASPT2  & (2,2)   & Yes & Yes  & No   & No  & Yes  \\
            & (4,4)   & No  & Yes  & Yes  & Yes & Yes  \\\hline
MRCI        & (2,2)   & Yes & Yes  & No   & Yes & Yes  \\
            & (4,4)   & No  & Yes  & Yes  & No  & Yes  \\\hline
MRCI+Q      & (2,2)   & Yes & No   & N/A  & Yes & N/A  \\
            & (4,4)   & No  & No   & N/A  & Yes & N/A  \\\hline
LR-CC3      &  HF     & Yes & No   & N/A  & Yes & N/A  \\
LR-CCSD     &  HF     & Yes & No   & N/A  & Yes & N/A  \\
SHCI        & (20,80) & Yes & Yes  & N/A  & Yes & N/A  \\\hline\hline
    \end{tabular}
    \label{tab:my_label}
\end{table}

\section{Important geometries}

\begin{table}[H]
    \centering
    \begin{tabular}{lrrr}
    \hline
 C   & -0.00623762  &   0.00112809  &  -0.01783509 \\
 C   &  2.46952702  &   0.00113714  &  -0.01782721 \\
 C   & -0.00623241  &   0.00112590  &  -1.34758024 \\
 C   &  2.46953222  &   0.00113360  &  -1.34756813 \\
 C   &  1.23163817  &   0.79689777  &   0.43333960 \\
 C   &  1.23165004  &   0.79689305  &  -1.79874838 \\
 C   &  1.23164036  &   1.87320553  &  -0.68270700 \\
 H   & -0.69996623  &  -0.50029786  &   0.64449358 \\
 H   &  3.16324837  &  -0.50028992  &   0.64450757 \\
 H   & -0.69995108  &  -0.50030785  &  -2.00991290 \\
 H   &  3.16326333  &  -0.50029138  &  -2.00989473 \\
 H   &  1.23163506  &   1.13729348  &   1.46891791 \\
 H   &  1.23165407  &   1.13728394  &  -2.83432830 \\
 H   &  0.33495108  &   2.49860759  &  -0.68271213 \\
 H   &  2.12832761  &   2.49861092  &  -0.68270456 \\\hline
    \end{tabular}
    \caption{NBD S$_0$ geometry optimised at CASSCF(2,2)/p-cc-(p)VDZ level. Distances in \AA ngstr\"oms}
    \label{tab:my_label}
\end{table}

\begin{table}[H]
    \centering
    \begin{tabular}{lrrr}
    \hline
 C   & -0.01145559  &   0.01158941  &  -0.01135907 \\
 C   &  1.53739527  &   0.01158779  &  -0.01135237 \\
 C   & -0.01145467  &   0.01155994  &  -1.54175823 \\
 C   &  1.53740665  &   0.01158450  &  -1.54174779 \\
 C   &  0.76297273  &   1.25126614  &   0.36989437 \\
 C   &  0.76295163  &   1.25123727  &  -1.92302270 \\
 C   &  0.76296521  &   2.24291674  &  -0.77657493 \\
 H   & -0.71360035  &  -0.49523713  &   0.63679114 \\
 H   &  2.23950681  &  -0.49526863  &   0.63680775 \\
 H   & -0.71360322  &  -0.49528638  &  -2.18989435 \\
 H   &  2.23952093  &  -0.49528504  &  -2.18988826 \\
 H   &  0.76297739  &   1.57342865  &   1.40570641 \\
 H   &  0.76297060  &   1.57337379  &  -2.95884322 \\
 H   & -0.12694439  &   2.88241733  &  -0.77657997 \\
 H   &  1.65289101  &   2.88239562  &  -0.77659876  \\\hline
    \end{tabular}
    \caption{QC S$_0$ geometry optimised at CASSCF(2,2)/p-cc-(p)VDZ level. Distances in \AA ngstr\"oms}
    \label{tab:my_label}
\end{table}

\begin{table}[H]
    \centering
    \begin{tabular}{lrrr}
    \hline
 C   &  0.06659730  &  -0.02880259  &  -0.02563437 \\
 C   &  2.00952085  &   0.00421747  &  -0.01385623 \\
 C   &  0.38714943  &   0.00324735  &  -1.43196211 \\
 C   &  2.31906961  &   0.21014800  &  -1.40784916 \\
 C   &  1.07367793  &   1.00658507  &   0.49581788 \\
 C   &  1.18685157  &   1.17562053  &  -1.78739361 \\
 C   &  1.06672984  &   2.10636002  &  -0.57377451 \\
 H   & -0.20016431  &  -0.93137068  &   0.50395810 \\
 H   &  2.41088169  &  -0.78642470  &   0.60695968 \\
 H   &  0.08153018  &  -0.73587112  &  -2.16128877 \\
 H   &  2.69126986  &  -0.56875204  &  -2.05671580 \\
 H   &  1.05710172  &   1.25334693  &   1.55126345 \\
 H   &  1.16341457  &   1.56844681  &  -2.79747720 \\
 H   &  0.13354623  &   2.67323379  &  -0.55901404 \\
 H   &  1.92221352  &   2.78102518  &  -0.50047330  \\\hline
    \end{tabular}
    \caption{S$_1$/S$_0$ MECI geometry optimised at CASSCF(2,2)/p-cc-(p)VDZ level. Distances in \AA ngstr\"oms}
    \label{tab:my_label}
\end{table}

\begin{table}[H]
    \centering
    \begin{tabular}{lrrr}
    \hline
 C   &  0.17298688  &  -0.02618520  &   0.02288223 \\
 C   &  2.29030187  &  -0.02617869  &   0.02289329 \\
 C   &  0.17299268  &  -0.02618849  &  -1.38829861 \\
 C   &  2.29030760  &  -0.02618168  &  -1.38828755 \\
 C   &  1.23163931  &   0.95165983  &   0.45363687 \\
 C   &  1.23164898  &   0.95165483  &  -1.81904650 \\
 C   &  1.23164094  &   1.99160838  &  -0.68270709 \\
 H   & -0.35088109  &  -0.71008531  &   0.67492913 \\
 H   &  2.81416716  &  -0.71007700  &   0.67494457 \\
 H   & -0.35086835  &  -0.71009296  &  -2.04034685 \\
 H   &  2.81417978  &  -0.71008144  &  -2.04033145 \\
 H   &  1.23163367  &   1.28026360  &   1.48865171 \\
 H   &  1.23165256  &   1.28025404  &  -2.85406278 \\
 H   &  0.33518152  &   2.61587726  &  -0.68271194 \\
 H   &  2.12809647  &   2.61588283  &  -0.68270504 \\\hline
    \end{tabular}
    \caption{S$_1$ minimum geometry optimised at CASSCF(2,2)/p-cc-(p)VDZ level. Distances in \AA ngstr\"oms}
    \label{tab:my_label}
\end{table}

\begin{table}[H]
    \centering
    \begin{tabular}{lrrr}
    \hline
  C & -0.005267 &  0.005148 & -0.013416 \\
  C &  2.468636 &  0.005136 & -0.013417 \\
  C & -0.005319 &  0.005169 & -1.351998 \\
  C &  2.468705 &  0.005157 & -1.351966 \\
  C &  1.231578 &  0.801709 &  0.436135 \\
  C &  1.231592 &  0.801714 & -1.801534 \\
  C &  1.231671 &  1.877629 & -0.682714 \\
  H & -0.698233 & -0.503522 &  0.654364 \\
  H &  3.161583 & -0.503609 &  0.654335 \\
  H & -0.698269 & -0.503486 & -2.019780 \\
  H &  3.161652 & -0.503574 & -2.019704 \\
  H &  1.231427 &  1.145650 &  1.476688 \\
  H &  1.231455 &  1.145545 & -2.842151 \\
  H &  0.328863 &  2.504368 & -0.682730 \\
  H &  2.134465 &  2.504341 & -0.682718 \\ \hline
    \end{tabular}
    \caption{NBD S$_0$ geometry optimised at MRCI(2,2)/p-cc-(p)VDZ level. Distances in \AA ngstr\"oms}
    \label{tab:my_label}
\end{table}

\begin{table}[H]
    \centering
    \begin{tabular}{lrrr}
    \hline
  C & -0.003622 &  0.010827 & -0.007716 \\
  C &  1.531534 &  0.011746 & -0.007673 \\
  C & -0.003609 &  0.010794 & -1.545401 \\
  C &  1.531543 &  0.011762 & -1.545406 \\
  C &  0.763251 &  1.261201 &  0.371409 \\
  C &  0.763256 &  1.261178 & -1.924555 \\
  C &  0.762701 &  2.249651 & -0.776582 \\
  H & -0.702141 & -0.506174 &  0.646330 \\
  H &  2.231175 & -0.504045 &  0.646067 \\
  H & -0.702286 & -0.506115 & -2.199341 \\
  H &  2.231015 & -0.504158 & -2.199235 \\
  H &  0.762730 &  1.583453 &  1.412934 \\
  H &  0.763126 &  1.583402 & -2.966090 \\
  H & -0.133219 &  2.890123 & -0.776586 \\
  H &  1.657823 &  2.891231 & -0.776592 \\\hline
    \end{tabular}
    \caption{QC S$_0$ geometry optimised at MRCI(2,2)/p-cc-(p)VDZ level. Distances in \AA ngstr\"oms}
    \label{tab:my_label}
\end{table}

\begin{table}[H]
    \centering
    \begin{tabular}{lrrr}
    \hline
  C &  0.057015 & -0.020780 & -0.030176 \\
  C &  2.011637 &  0.010059 &  0.000095 \\
  C &  0.383132 &  0.010877 & -1.445441 \\
  C &  2.326778 &  0.218583 & -1.402547 \\
  C &  1.055737 &  1.013460 &  0.496668 \\
  C &  1.202942 &  1.184792 & -1.787452 \\
  C &  1.065984 &  2.112571 & -0.573453 \\
  H & -0.204369 & -0.930047 &  0.503658 \\
  H &  2.412487 & -0.784986 &  0.627622 \\
  H &  0.078165 & -0.731305 & -2.182146 \\
  H &  2.694483 & -0.566871 & -2.056722 \\
  H &  1.029029 &  1.267554 &  1.556603 \\
  H &  1.188653 &  1.586820 & -2.800812 \\
  H &  0.127634 &  2.683228 & -0.573734 \\
  H &  1.926344 &  2.789366 & -0.484636 \\\hline
    \end{tabular}
    \caption{S$_1$/S$_0$ MECI geometry optimised at MRCI(2,2)/p-cc-(p)VDZ level. Distances in \AA ngstr\"oms}
    \label{tab:my_label}
\end{table}

\bibliographystyle{rsc}
\bibliography{references-2}